\documentclass[sigconf,nonacm,balance=false]{acmart}

\usepackage{popets}

\usepackage{tikz}
\usepackage{amsmath}
\usepackage{multirow}
\usepackage[normalem]{ulem}

\newcommand{\trainsize}{21,588}
\newcommand{\testsize}{9252}
\newcommand{\datasetSize}{6781}

\newcommand{\numModels}{50}

\begin{document}

\title[Automating GKC-CI Privacy Policy Annotations with LLMs]{Automating Governing Knowledge Commons and Contextual Integrity (GKC-CI) Privacy Policy Annotations with Large Language Models}

\author{Jake Chanenson}
 \authornote{Both authors contributed equally to this research.}
 \email{jchanen1@uchicago.edu}
 \orcid{0000-0003-3400-925X}
 \affiliation{%
   \institution{University of Chicago}
   \city{Chicago}
   \state{IL}
   \country{USA}}

 \author{Madison Pickering}
 \authornotemark[1]
 \email{madisonp@uchicago.edu}
 \orcid{0009-0002-4000-2951}
 \affiliation{%
   \institution{University of Chicago}
   \city{Chicago}
   \state{IL}
   \country{USA}}

 \author{Noah Apthorpe}
 \email{napthorpe@colgate.edu}
 \orcid{https://orcid.org/0000-0002-9273-4268}
 \affiliation{%
   \institution{Colgate University}
   \city{Hamilton}
   \state{NY}
   \country{USA}
}

\renewcommand{\shortauthors}{Chanenson, Pickering, and Apthorpe}

\begin{abstract}
    Identifying contextual integrity (CI) and governing knowledge commons (GKC) parameters in privacy policy texts can facilitate normative privacy analysis.
    However, GKC-CI annotation has heretofore required manual or crowdsourced effort. 
    This paper demonstrates that high-accuracy GKC-CI parameter annotation of privacy policies can be performed automatically using large language models. We fine-tune \numModels{} open-source and proprietary models on \trainsize{} ground truth GKC-CI annotations from 16  privacy policies. Our best performing model
    has an accuracy of 90.65\%, which is comparable to the accuracy of experts on the same task. 
    We apply our best performing model to 456 privacy policies from a variety of online services, demonstrating the effectiveness of scaling GKC-CI annotation for privacy policy exploration and analysis. We publicly release our model training code, training and testing data, an annotation visualizer, and all annotated policies for future GKC-CI research.
\end{abstract}


\maketitle

\section{Introduction}
\label{sec:intro}
Privacy policies are notoriously complex and lengthy documents~\cite{mcdonald2008cost}.
These policies are often written in complex language or ``legalese'' to obfuscate the extent of data collection and discourage consumers from closely interrogating their privacy implications~\cite{jensen2004privacy, reidenberg2016ambiguity, amos2021privacy}. 
Most consumers therefore choose to ignore privacy policies when agreeing to online terms and services~\cite{rudolph2018users}.
Even experts have difficulty interpreting some privacy policies~\cite{reidenberg2015disagreeable}.
However, privacy policies remain essential to Internet privacy broadly and to the privacy-relevant behaviors of online services. 

The continued importance of privacy policies has motivated substantial research into structured methods of privacy policy analysis. Some of these methods seek to provide clearer or more easily digestible information to consumers or developers~\cite{sadeh2013usable, angulo2012towards, brodie2006empirical, wilson2016creation}, while others facilitate academic studies of the policies themselves, their relation to company behavior, or to privacy regulation~\cite{liu2014step, winkler2016privacy, andow2019policylint, perez2018review, andow2020actions, tesfay2018privacyguide}. Both approaches often employ \textbf{annotation}---labeling relevant parts of privacy policy texts with metadata---as a primary technique.

Early successful efforts 
involved annotating privacy policies with a large set of metadata tags~\cite{wilson2016creation}.
A more recent approach~\cite{shvartzshnaider2019going} has leveraged the theory of \textbf{contextual integrity (CI)}~\cite{nissenbaum2009privacy} to annotate privacy policies. CI annotation uses a small set of theoretically grounded tags to facilitate comparative and longitudinal analysis of data handling practices and policy ambiguities~\cite{shvartzshnaider2019going}. Research using CI as a theoretical foundation is widespread across topics in security and privacy, human-computer interaction, formal modeling, and legal analysis~\cite{brehm2023understanding, apthorpe2019evaluating, 280276, shvartzshnaider2019vaccine, wijesekera2015android, datta2011understanding}. 
Unlike differential privacy and other \textit{probabilistic} privacy definitions, CI extends a \textit{normative} notion of privacy that is a more natural fit for natural language documents like privacy policies~\cite{nissenbaum2009privacy}.

Recent work has shown that CI is even more effective if expanded using the governing knowledge commons framework (GKC)~\cite{sanfilippo2018privacy, frischmann2014governing}. GKC provides an institutional grammar for describing strategies, norms, and rules around shared knowledge resources. The unified \textbf{GKC-CI framework}~\cite{shvartzshnaider2022gkc} (Section~\ref{sec:theory}) enables straightforward identification of privacy policy ambiguities that reduce interpretability and provide excessive leeway for behavior users may consider privacy-violating. 
GKC-CI also enables normative analyses of contextual information transfers,  the rules-in-use, and the rules-on-the-books that govern data handling practices.

All previous uses of CI parameter annotation for privacy policy analysis have involved human effort by experts or crowdworkers. Manual annotation by expert researchers produces high-quality results, but the process is tedious and slow.  Crowdsourcing produces annotations more quickly, but there is a significant rate of poor-quality annotations since the annotation task is inherently nuanced ~\cite{shvartzshnaider2019going}. Combining multiple crowdsourced annotations through a voting process can improve overall performance but further increases expense, as multiple crowdworkers must be hired to annotate overlapping sections of privacy policy text~\cite{shvartzshnaider2019going}. A prior study spent approximately \$200 for crowdworker annotation of only 48 \textit{excerpts} from 16 privacy policies~\cite{shvartzshnaider2019going}. 

Furthermore, no previous work on automating privacy policy annotation with machine learning~\cite{wilson2018analyzing, harkous2018polisis, bannihatti2020finding, andow2020actions, andow2019policylint, zimmeck2019maps, story2019natural, tang2023policygpt} has been based on CI or GKC-CI. Rather, this prior work has focused on different sets of annotation tags that are not useful to the GKC or CI scholarly communities as they do not produce results relevant to GKC or CI analysis. This motivates the development of novel automated techniques for the GKC-CI annotation task.

In this paper, we train a variety of large language models (LLMs) to perform automated GKC-CI parameter annotations of privacy policies. Doing so demonstrates the feasibility of large-scale longitudinal and cross-industry analysis of privacy policies using the GKC-CI framework. Specifically, we train and evaluate \numModels{} LLMs from five different model families, ranging from open-source to proprietary models. We perform ablation studies with the LLMs, and briefly remark on some key lessons which we believe may be broadly applicable to other researchers working in this space. We then proceed to a nuanced discussion of the behavior of our best performing model, particularly with respect to the model's errors.

We observe that of the \numModels{} models we benchmark, a version of \texttt{GPT-3.5 Turbo} performs the best. We find that the model boasts a robust accuracy of 90.65\%, better than that of prior crowdsourcing approaches~\cite{shvartzshnaider2019going}. We consider an accurate annotation to be an exact string match of the human annotator's text.
In contrast to prior approaches, our best performing LLM only costs \$0.23-0.44 on average to annotate a privacy policy (depending on length) and is able to annotate an entire privacy policy in one minute or less on average (\textasciitilde5000 words a minute).  
We could annotate all policy excerpts from \cite{shvartzshnaider2019going} for only \$7 and in less than one minute.

We use our best performing model to annotate longitudinal and cross-industry policies from the Princeton-Leuven Longitudinal Corpus of Privacy Policies~\cite{amos2021privacy}. We evaluate these policies based on GKC-CI parameter counts, densities, and variances, demonstrating how GKC-CI annotation can be used to highlight policies of interest for further analysis. 
We also present a Python-based GUI that visualizes the results of GKC-CI annotation, showing privacy policy text with annotated parameters highlighted by color.

In summary, this paper presents a practical method for analyzing privacy policies that integrates machine learning, privacy law, and governance. This is the first automated method for GKC-CI parameter annotation in legal documents, which will be extremely useful to the growing GKC and CI research communities. 
This approach could also encourage collaboration and open up new research opportunities in these interconnected fields, promoting more in-depth exploration of privacy protection and compliance. More specifically,
\begin{itemize}

    \item We demonstrate that accurate GKC-CI parameter annotations of privacy policies can be performed automatically by a fine-tuned large language model (LLM), substantially improving scalability and reducing expense compared to manual and crowdsourcing approaches. Further, through training data review, we find these annotations are as accurate as those obtained from humans. 

    \item We train and publicly release the data and scripts necessary to reproduce a large language model capable of performing automatic GKC-CI annotation of privacy policies. Our automated approach has an average per-policy annotation cost between \$0.23-0.44, depending on the length of the policy.\footnote{This cost estimation range is the average cost of annotating privacy policies from both the 164 least popular and the 164 most popular websites, per their Tranco~\cite{pochat2018tranco} ranking, in the Princeton-Leuven Longitudinal Corpus of Privacy Policies~\cite{amos2021privacy}.} We note that reproducing our model incurs a one-time training cost of \$195 at time of writing.
    
    \item We perform a large-scale longitudinal and cross-industry analysis of privacy policies using our best performing model. We demonstrate that the annotations can highlight policies with atypical parameter densities and distributions that may be good candidates for future in-depth evaluation. We compile all 456 annotated policies into a GitHub repository, which we make publicly available.\footnote{\url{https://github.com/JakeC007/Automated_GKC-CI_Privacy_Policy_Annotations}} 
\end{itemize}

\section{Related Work}
\label{sec:related}
Substantial prior research has focused on systematic analyses of privacy policies. These analyses were often intended to improve consumer understanding of data handling processes and facilitate academic study of Internet privacy trends. This paper builds on this foundation, contributing to the broad goal of developing a library of effective, scalable, and inexpensive privacy policy analysis techniques suitable for a range of applications. 

\subsection{The Usable Privacy Project}

The Usable Privacy Project~\cite{sadeh2013usable} from Carnegie Mellon University is perhaps the most visibly successful application of annotation as a method for privacy policy interpretation and explanation. This project started in 2016 with a study by Wilson et al.~\cite{wilson2016creation}, that recruited law students to manually annotate privacy policies with metadata tags such as “first party collection/use,” “user choice/control,” “data retention,” and “data security.” Wilson et al. also showed~\cite{wilson2016crowdsourcing} that annotations produced by crowdworkers agreed with those of expert annotators over 80\% of the time. This showed that crowdsourcing techniques could be used to identify paragraphs describing specific data handling practices in privacy policies. 

In 2018, Wilson et al. used 115 expert-labeled policies to train logistic regression, support vector machine, and convolutional neural network models to automatically label sentences or segments of privacy policies with data practice categories~\cite{wilson2018analyzing}.
Their best models had average F1 scores of $0.66$ for policy sentences and $0.78$ for policy segments. 
These techniques have been applied to over 7000 privacy policies from 2017, with results posted on the Usable Privacy Project website to inform consumers of the wide variety of information handling practices conducted by online services.\footnote{https://explore.usableprivacy.org/}

This line of research, while independently impressive, is not relevant for the community of scholars using contextual integrity or governing knowledge commons to interpret privacy policies and other legal documents. The sets of annotation tags used for the Usable Privacy Project are not analogous to the 5-parameter information flow description fundamental to CI nor the institutional grammar fundamental to GKC. This means that a separate thread of research, described below, has been needed to develop privacy policy annotation approaches useful for GKC and CI analyses.

Unlike the work by Wilson et al.~\cite{wilson2016creation, wilson2016crowdsourcing, wilson2018analyzing} and the Usable Privacy Project, our work is based on Shvartzshnaider et al.'s GKC-CI framework~\cite{shvartzshnaider2022gkc}, which provides a theoretically grounded basis for identifying ambiguity and potential privacy-violating behavior. Our work is less focused on helping consumers understand privacy policies than~\cite{wilson2016creation, wilson2016crowdsourcing, wilson2018analyzing} and is more focused on automatable, longitudinal, and cross-industry analysis of privacy policies. Our work is also novel in its use of large language models (the current state-of-the-art in natural language processing) and achieves considerably better performance than the machine learning annotation in~\cite{wilson2018analyzing}. 

\subsection{Manual CI Annotation}
 
In 2019, Shvartzshnaider et al. \cite{shvartzshnaider2019going} used the theory of contextual integrity (CI) \cite{nissenbaum2009privacy} to inform a new approach to privacy policy annotation. 
This approach seeks to identify the five information flow parameters defined by CI (Section~\ref{sec:theory}) in privacy policy text. CI parameter annotation enables the identification of ambiguities in information transfer descriptions. 
In 2022, Shvartzshnaider et al.~\cite{shvartzshnaider2022gkc} combined contextual integrity with governing knowledge commons (GKC)~\cite{frischmann2014governing, sanfilippo2018privacy} to create a combined GKC-CI framework. GKC-CI extends the potential scope of CI annotation to eight total parameters, four from CI and four from the GKC institutional grammar (Section~\ref{sec:theory}).

While Shvartzshnaider et al.~\cite{shvartzshnaider2019going} successfully motivated CI parameter annotation for privacy policy analysis, questions of scalability remained. As with most annotation tasks, manual annotation by experts is highly accurate but tedious and slow. Shvartzshnaider et al. demonstrated that crowdsourcing could partially solve this problem but remains expensive, as 
high error rates necessitated the combination of multiple overlapping crowdsourced annotations per policy segment to increase precision. The resulting crowdsourced annotations still had a relatively high rate of false negative errors, \textit{i.e.,} parameters missed by the majority of crowdworkers.

\textbf{The scalability issues posed by crowdsourced annotation 
clearly motivate this study}, which seeks to automate CI parameter annotation through the use of large language models (LLMs). Our work is also novel in its use of the expanded eight-parameter GKC-CI labels as annotation tags (Section~\ref{sec:theory}) rather than the five-parameter CI tags used in Shvartzshnaider et al.'s original paper.

We apply our automated method to annotate a large corpus of privacy policies, including up to 20 years of longitudinal policies from 8 major technology companies (128 policies) and 328 contemporary policies from across the technology industry. 
This is orders of magnitude more privacy policies than have been manually annotated for CI research in previous work~\cite{shvartzshnaider2019going}.

\subsection{Privacy Policy Analysis With Machine Learning}

Several other studies have also applied machine learning to privacy policies, although none have been based on nor produced output useful for GKC-CI research.
In 2018, Harkous et al.~\cite{harkous2018polisis} trained a hierarchy of convolutional neural networks to build a Question-Answering system that supports free-form querying of privacy policy content. Other ML-based approaches essentially parse privacy policies for information of interest, such as by using a logistic regression model to identify opt-out statements in privacy policy text~\cite{bannihatti2020finding}. PoliCheck~\cite{andow2020actions}, an expansion of PolicyLint~\cite{andow2019policylint}, is capable of differentiating between first-party and third-party entities in flow-to-policy consistency analysis. Zimmeck et al.~\cite{zimmeck2019maps} and Story et al.~\cite{story2019natural} used support vector machines to identify non-compliance between Android application code and the applications' privacy policies.
Their approach could be used to highlight these statements for consumers to make opt-out decisions without needing to read the entire policy themselves. Our application of machine learning to GKC-CI privacy policy annotation is similarly tightly focused, but on a task that does not overlap these earlier works.

More recently, some research groups have utilized LLMs in the legal space broadly or for privacy policy analysis specifically. LLM benchmarks in the legal space include that of Dai et al.~\cite{dai2023laiw} and Fei et al.~\cite{fei2023lawbench}. Ravichander et al.~\cite{ravichander2019question} trained a BERT-based large language model to answer questions about privacy policies in a Q\&A format (not annotation) using a corpus of 1750 questions and 3500 expert answers. Other work that adapts LLMs for legal question-answering include that of Wan et al.~\cite{wan2023reformulating} and Yue et al.~\cite{yue2023disclawllm}.

Tang et al.~\cite{tang2023policygpt} used various LLMs to annotate privacy policies via prompting. However, the annotations are very simple (e.g., ``1st Party Collection'') and could likely be found using regexes. In contrast, our annotation approach uses a nuanced and theoretically grounded framework (GKC-CI) for normative privacy analysis. We believe LLMs are particularly suited for \textit{nuanced} annotations as opposed to classical neural network methods. This is because of LLMs' increased representational capacity as well as their training on Internet webscrapes, which exposes LLMs to cultural norms~\cite{vaswani2017attention, Radford2018ImprovingLU, devlin2018bert}. However, we cannot provide a comparison to other researcher's classical approaches for automated GKC-CI annotation because there are none developed in this space. To address this, we train an RNN on the GKC-CI annotation task as a baseline for our work.

\section{GKC-CI Theory}
\label{sec:theory}

The theory of contextual integrity (CI)~\cite{nissenbaum2009privacy} defines privacy as the adherence of information transfers, or ``flows,'' to sociocultural norms in specific contexts.
For example, an information flow that might be appropriate between a patient and a doctor in a medical context (e.g., about a sensitive diagnosis) might not be appropriate between that doctor and their acquaintance in a recreational context. 

CI further defines information flows as consisting of five essential parameters: 1) the \textit{sender} of the information, 2) the \textit{recipient} of the information, 3) the \textit{subject} of the information, 4) the information content or \textit{attribute}, and 5) the \textit{transmission principle} that describes how or why the information flow occurs. The CI parameter annotation task entails identifying and labeling these five parameters in descriptions of information flows. For example, the CI annotation of: ``We also collect contact information that you provide if you upload, sync or import this information from a device,'' would label ``we'' as a recipient, ``contact information'' as an attribute, ``you'' as the sender, and ``if you upload, sync or import this information from a device'' as a transmission principle (example from \cite{shvartzshnaider2019going}). 

\begin{table*}
    \centering
    \footnotesize

\begin{tabular}{@{}l|r|l|r|l@{}}
\toprule
\multicolumn{1}{c|}{Privacy Policy Sentence} &
  \multicolumn{1}{c|}{Parameter} &
  \multicolumn{1}{c|}{Annotated Text} &
  \multicolumn{1}{c|}{Parameter} &
  \multicolumn{1}{c}{Annotated Text} \\ \midrule
\textit{If you consent,} \uline{we} \textit{may} share \uline{information} about          & Aim         & to provide...our partners            & Modality  & may \\
\uline{you} with \uline{companies that aggregate it}  & Attribute   & information & Recipient & companies that aggregate it  \\
\textit{to provide analytics and measurement } & Condition & If you consent & Sender & we \\
\textit{reports to our partners.} & Consequence & N/A              & Subject   & you \\ \bottomrule
\end{tabular}

    \caption{Example GKC-CI annotation. Italics and underlining added for emphasis.}
    \label{tab:example_params}
\end{table*}

The combined GKC-CI framework~\cite{shvartzshnaider2022gkc} further extends the CI framework, enabling the evaluation of strategies, norms, and rules drawn from theories of information governance. This allows cross-disciplinary research between the GKC and CI communities and allows a broader set of research questions to be addressed than either framework alone~\cite{shvartzshnaider2022gkc}. Relevant to annotation, the combined framework divides the \textit{transmission principle} into four categories drawn from the GKC institutional grammar: 1) \textit{aims} and/or goals for specific actions, 2) \textit{conditions} indicating when, where, or how aims apply, 3) \textit{modality} operators implying pressure (deontics) or hedging, and 4) \textit{consequences}, including sanctions for noncompliance, penalties in absence of consent, and benefits for proceeding. The GKC-CI parameter annotation task is identical to the CI annotation task except that it requires identifying the eight GKC-CI parameters instead of the five CI parameters. GKC-CI annotations thus provide more nuance than CI annotations at the expense of increased annotation difficulty. An example of what different GKC-CI parameters are present in a sample sentence is shown in Table \ref{tab:example_params}. We encourage readers interested in additional information about the GKC-CI framework to refer to the original paper~\cite{shvartzshnaider2022gkc}, which explains the theoretical foundations in detail and provides a worked demonstration of GKC-CI analysis in the Internet of things context. 
\section{Methods}
\label{sec:methods}

We next describe how we trained LLMs to accurately perform GKC-CI parameter annotation for privacy policies. In doing so, we make the following contributions. 

First, we tested how various architectures and sizes of different LLMs effect performance through ablation experiments. After comparing a broad selection of models to find the most suitable candidate, we further tuned the hyperparameters of our best performing model. Details of the model training process are provided in Section~\ref{sec:methods}, and performance results are provided in Section~\ref{sec:performance}. Because there is no previous work demonstrating how NLP models may be applied for GKC-CI annotation, we believe that these results will be particularly helpful to those considering normative privacy policy analysis in this space. 

Second, we qualitatively examined the errors made by our best performing model to gain insights into the model's behavior, particularly as it relates to practical deployment. We believe that our approach to error analysis may provide a useful foundation for how others may analyze model behavior in this space. The details of our error analysis are reported in Section \ref{sec:error-analysis}.

Finally, we used our best performing model to annotate a large set of privacy policies~\cite{amos2021privacy}, which we report in Section~\ref{sec:production}, providing examples of the types of longitudinal and cross-industry analyses that can be performed via automated GKC-CI annotation.

\subsection{Training and Testing Data}
\label{Datasets}
\label{sec:data-prep}
Our ground-truth labels were obtained by manually annotating GKC-CI parameters in 16 privacy policies from popular online services and e-learning websites, the exact breakdown of which is shown in Table~\ref{tab:manual-training} of Appendix~\ref{sec:appendix_tabs}. We downloaded these privacy policies in HTML format and converted them to plain text for annotation. We used a customized version of the Brat Rapid Annotation Tool~\cite{stenetorp2012brat} to manually label all GKC-CI parameters in the policies. 
In order to achieve consistent annotations across all annotators, we used a fixed set of guidelines defining each of the GKC-CI parameters (Appendix~\ref{sec:bratt}). These guidelines were taken from~\cite{shvartzshnaider2019going} for CI parameters and~\cite{shvartzshnaider2022gkc} for GKC-CI parameters to ensure continuity with prior work.
Our ground-truth annotations included \datasetSize{} GKC-CI parameters across all 16 policies (Table~\ref{tab:manual-training}). This ground-truth annotation process took two research assistants one semester to perform, including time spent learning the task.  

In the process of annotating, we encountered several of the challenges discussed in~\cite{shvartzshnaider2019going}, including implicit parameters, ambiguous parameters, and policies not written with the CI framework in mind. We addressed these issues consistently with~\cite{shvartzshnaider2019going}. 
In general, the annotators made best judgment calls when faced with ambiguous parameters or difficult logic, consulting with the authors to ensure consistency. Importantly, we did not expect these manual annotations to be perfect. Rather, we treated them as best-effort annotations by researchers familiar with the task.

\subsection{Formatting Examples}
\label{sec:formatting}
We lightly formatted each sentence of our ground-truth annotated privacy policies as the basis of our training and testing examples. 
Each formatted example consisted of the following parts: (1) a prefix to orient the LLM to the task, (2) a sentence from a privacy policy, (3) the GKC-CI parameter of interest, and (4) text delimiters. We chose the extremely minimal prefix ``Annotate:'' to minimize the effects of prompt choice while still leveraging the training benefits of using a prompt ~\cite{webson2022promptbased, scao2021data}. We included the text delimiters because modern LLMs decide what text to generate next based on \emph{all} the text in their context window. As such, they cannot by default determine what text has been provided via the prompt and what the LLM has generated.  We chose our text delimiters based on recommendations in OpenAI's documentation, namely ``-->'' and ``x-x-x'' respectively. Two fully formatted examples are shown below:

\begin{enumerate} 
    \item Annotate: [``We also collect contact information that you provide''] Recipient--> Recipient: [``We'']x-x-x
    \item Annotate: [``We also collect contact information that you provide''] Aim-->  Aim: N/Ax-x-x
\end{enumerate}

In formatting these examples, we wanted to ensure that the models learned to find the relevant text to be annotated. This necessitated teaching the models to filter through \textit{irrelevant} text. As such, we used \textit{positive examples} (text where a parameter is present, such as example (1) above) and \textit{negative examples} (text where a given parameter is not present, such as example (2) above). The inclusion of negative examples was necessary to ensure that the models are usable in a real-world environment; not every sentence of a privacy policy will include a GKC-CI parameter. By including negative examples, our models learned to only output a parameter if one is actually present in an input sentence. We created the negative examples by taking positive examples and changing the parameter of interest to one which does not occur in the sentence. We did this for each positive example and all possible parameters not already in that example. For instance, the negative example (2) above was created by switching the parameter of interest in the positive example (1) above from ``recipient'' (which should be labeled in the example text) to ``aim'' (which should not). 

We used sentences as the atomic unit for model input because sentence divisions are natural delimiters and previous work has shown reasonable accuracy with sentence-based annotation ~\cite{wilson2018analyzing}. We observed that other natural units of separation, such as paragraphs, tended to be particularly long due to the legal nature of privacy policies. This length ultimately resulted in problems where the paragraphs were too long to fit into the context window of some of our LLMs.\footnote{Further, to our knowledge, resolving a finite-length context window with large quantities of text is an open problem in the NLP space, although there has been substantial work in the area, such as ~\cite{press2022train} and ~\cite{beltagy2020longformer}.} To ensure that the training data was consistent between LLMs, and because of the prior work above, we used sentences as the basis of our annotations. However, we also performed a supplemental analysis with GPT-3.5 to verify that we were not critically disadvantaging our models by only feeding in sentences (Appendix~\ref{sec:appendix:paragraphs}).

\subsubsection{OpenAI Models}
\label{sec:methods:formatting_examples:openai_models}
Some OpenAI chatbot models take an additional prompt, a \textit{system message}, as input. GPT-3.5 Turbo is one such model. For these models, we replaced the default system message from \emph{``You are a helpful assistant.''} to \emph{``You are an assistant that understands Helen Nissenbaum's theory of Contextual integrity (CI) and the governance of knowledge commons framework (GKC). This framework is abbreviated as GKC-CI. You reply with brief, to-the-point answers with no elaboration.''} We did so to help orient the model to our task.

Finally, we performed an experiment leveraging the fact that GPT-3.5 Turbo is designed to respond to \textit{conversational} inputs, as it is a chatbot. Specifically, we changed the prefix from \textit{``Annotate:''} to \textit{``For the following excerpt, provide the GKC-CI annotation of `<parameter>': }''. We call the model produced under this intervention \texttt{GPT-3.5 Turbo, Prompt Engineered}. Note that this is distinct from our baseline models that we prompted \textit{without fine-tuning}.

\subsection{Model Selection}
\label{sec:meth:modelselec}

\subsubsection{Baseline Models}
\label{sec:meth:modelselec:baseline}
We considered two baselines for comparison against our fine-tuned LLMs: 1) a recurrent neural network (RNN) representing classical NLP approaches and 2) prompted \textit{non-fine-tuned} LLMs.

For the RNN, we defined BOS and EOS tokens, which is standard practice for these models~\cite{pytorchrnn}. We then perform standard training for the RNN but keep the rest of the training parameters consistent with the LLMs as described in Section~\ref{sec:training}.

For the prompted non-fine-tuned LLMs, we used GPT-4, GPT-4 Turbo, and GPT-3.5 Turbo. We used a fixed prompt format where we clearly delineated instructions from other text by placing instructions within brackets, thereby providing a structured framework that helped the model distinguish between the task directives and supplementary information. We also employed n-shot learning, a common prompting technique that has been shown to improve performance as n, the number of examples, increases ~\cite{Zamfirescu2023Johnny, brown2020language_gpt3, gao2021making}. We used n=\{1, 3, 5\} where the examples were chosen randomly from the training set.

\subsubsection{Fine-Tuned Models}
\label{sec:meth:modelselec:ft}
\begin{table*}[h]
\small
\centering
\begin{tabular}{ccccccc}
    \toprule
    Model Family & Open-Source & Size(s) Considered & Architecture & Instruction-Finetuned & RLHF & Chat Variant  \\ 
    \midrule
    GPT-2 & Yes & Base (124M), XL (1.5B) & Decoder-Only & No & No & No \\ 
    Flan-T5 & Yes & Base (248M), Large (783M) & Encoder-Decoder & Yes & No & No \\
    Llama2 & Yes & 7B & Decoder-Only & No & No & No \\
    GPT-3 & No & "Davinci" & Decoder-Only & Unconfirmed & Unconfirmed & No \\
    GPT-3.5 Turbo & No & Unknown & Decoder-Only & Unconfirmed & Unconfirmed & Yes \\ 
    \bottomrule
\end{tabular}
    \caption{A summary of the models we trained and their model families.
    } 
    \label{tab:modelprops}
\end{table*}

We considered five model families of diverse size and architecture: Flan-T5, GPT-2, Llama2, GPT-3, GPT-3.5 Turbo, and GPT-4 \cite{chung2022scaling_flan, Radford2019LanguageMA_gpt2, touvron2023llama2, brown2020language_gpt3}. Their properties are summarized in Table ~\ref{tab:modelprops}. In selecting models from these families, we wanted to choose from a wide range of high-performing or particularly usable LLMs. We ultimately omitted GPT-4 from our analysis because at the time of writing, the fine-tuning API was experimental. We also omitted Llama2's chat version from our analysis because, at the time of writing, it did not appear that Meta intended it to be fine-tuned further based on its documentation~\cite{llamarecipes}. Of the models we selected, approximately half are open-source, while the GPT-3 and GPT-3.5 models are proprietary. While the exact sizes of the GPT models are not publicly released, they are likely the among the largest of the models we tested.

We now give a quick summary of how the various architectural features of the models we trained may impact performance on the annotation task. First, we varied model size. Specifically, we considered models both in their ``base'' or default size\footnote{When loading from the HuggingFace model hub} as well as a larger size, if permissible by our hardware.\footnote{We employed an NVIDIA A100 for training, which we believe to be a reasonable baseline for the amount of compute other researchers will have available.} We considered model size because the number of parameters in a model plays a large role in its performance~\cite{hoffmann2022training, kaplan2020scaling, touvron2023llama, chung2022scaling_flan}. However, smaller models may approach larger models' performance if they are given more (pre)training data. We consequently included the Llama2 model to serve as an open-source proxy for larger models' (e.g., GPT-3) performance ~\cite{touvron2023llama, touvron2023llama2}.

Most newer LLMs employ a ``decoder-only'' architecture, a change from the original design of the Transformer~\cite{vaswani2017attention}. Such models are referred to as encoder-decoder models, while models lacking an encoder block are decoder-only. Decoder-only models can only ``look'' at the preceding tokens to determine what should be generated next and tend to excel at creative or free-form generation ~\cite{holtzman2020curious}. In contrast, encoder-decoder models look at the entire input sequence to determine what should be generated. Encoder-decoder models, like Flan-T5 or BERT, tend to perform well on tasks where the output is highly scoped by the input \cite{holtzman2020curious, chung2022scaling_flan, devlin2018bert}. We included both decoder-only and encoder-decoder models in our analysis because it was unclear what advantage, if any, an encoder-decoder model might have when annotating privacy policies.

There exist a number of training paradigms which result in a model becoming more \textbf{aligned} to human intent. We specifically mean that a model is aligned if it produces outputs which are consistent with its human operator's desires (assuming the model is capable of producing those outputs)~\cite{chen2021evaluating, ouyang2022training}. Because alignment broadly reflects a model's ability to output text consistent with input tasks, we hypothesized that models which have undergone specific alignment training might perform better. We considered two alignment mechanisms: Instruction-Finetuning and Reinforcement Learning with Human Feedback (RLHF)~\cite{ouyang2022training, wei2022finetuned}. We included models in our analysis which are either confirmed or likely to have been trained according to these techniques.

Finally, we noted that some models have been released as chat models. This is relevant because 1) such models can be prompted in a different format from other non-chat models (Section~\ref{sec:methods:formatting_examples:openai_models}), and 2) these models are generally newer. While we are loath to conflate model age with performance, it is at least in the case of GPT and Llama that newer releases tend to eclipse older releases~\cite{Radford2019LanguageMA_gpt2, brown2020language_gpt3, touvron2023llama, touvron2023llama2}. We therefore included the fine-tunable chat model GPT-3.5 Turbo in our analysis.

\subsection{Model Training}
\label{sec:training}
We processed all 16 manually-annotated privacy policies into ground truth examples for model training and testing as described in Section~\ref{sec:formatting}. We randomly reserved 70\% of the manual annotations to constitute our training data (\trainsize{} examples), while the other 30\% (\testsize{} examples) were testing data.  

We used low-rank adaptation (LoRA), a type of parameter efficient fine-tuning (PEFT), as our training method  to ensure that the open-source models were trained in a way consistent with the proprietary models~\cite{hu2021lora}. PEFT describes a number of methods for training some, but not all, of a model's parameters (i.e., being ``parameter efficient'' when training). LoRA is a PEFT method that involves freezing a model's weights and inserting trainable rank decomposition matrices into the model's architecture. These low-rank matrices are then ``trained'' during fine-tuning, while the base model remains frozen.  OpenAI's business model suggests that LoRA is being employed in the place of traditional fine-tuning, as low rank matrices are very cheap to store and entire models are typically large and expensive.\footnote{We make this claim because OpenAI allows for ``fine-tuning'' through their API. Traditional fine-tuning would require making full copies of the model 
(\textit{e.g.,} GPT-3.5) for each user. Doing so would result in terabytes of space being allocated per user due to the size of OpenAI's models. Additionally, it has been observed that previously fine-tuned OpenAI models may change in performance without warning, as OpenAI routinely updates their models. This behavior would not be observed if traditional fine-tuning were occurring because each user would have their own discrete copy of the model.}

Thus, to ensure that all our model comparisons are fair, we trained all LLMs using LoRA. We kept the training parameters constant between all open-source models. We made an exception for the learning rate of Flan-T5, as the model's documentation recommends a slightly higher learning rate than the other models. This higher learning rate could have caused the model to converge faster than others. 

During training, we performed a number of ablation experiments, changing the following in a full-factorial manner for the open-source models:
\begin{enumerate}
    \item \textbf{Training Epochs}. We varied the number of training epochs for the model. Training a model for a single epoch means that the model sees every example in the training set once. Training for \emph{n} epochs means that the model sees every example in the training set \emph{n} times. Varying the number of training epochs thus influences the quantity of training tokens to which the model is exposed, as well as the number of repeated examples to which the model is exposed.
    \item \textbf{Example formatting}. We varied the examples' formatting such that all training examples were prepended and appended with the model's BOS, EOS tokens or not depending on the ablation condition. BOS (Beginning of Sentence) and EOS (End of Sentence) tokens are typically defined and used during LLM pre-training to semantically capture the start and end of a training document. We hypothesized that the inclusion or exclusion of these tokens during fine-tuning may have an effect on the model's performance. 
\end{enumerate}

OpenAI's proprietary models do not offer the same number of training options, so while were were able to experiment with different numbers of training epochs for OpenAI's proprietary models, we were unable to perform the example formatting experiment described above with the proprietary models. We ultimately benchmarked \numModels{} models distributed between open-source and OpenAI models.

\section{Model Performance}
\label{sec:performance}

\subsection{Metrics}
\label{sec:metrics}
We begin our discussion of performance by clearly defining our metrics. Specifically, our main metric is \emph{accuracy}. We consider a model's output to be correct or \textit{accurate} if it is an exact string match of the ground truth. We do not directly consider precision and recall because they are less intuitive for this task. We envision that downstream users are likely to be particularly interested in the performance per GKC-CI parameter. As such, we focus on per-parameter accuracy.

Since we want to capture nuanced model errors, we categorize each model output into exactly one of four possible results: \textit{perfect match}, \textit{superset match}, \textit{match error}, or \textit{identification error}. These categories are defined as follows:

\begin{enumerate}
    \item \textbf{Perfect Match} indicates that the model's annotation is an \textit{exact string match} with that of the human annotator. \textit{Only a Perfect Match represents a correct annotation.}
    \item \textbf{Superset Match} indicates that the model's annotation contains all the words of the human annotator. However, the model may have highlighted additional information or may have included information which does not appear as a contiguous sequence of text in the policy.
    \item \textbf{Match Error} indicates that the model agreed that a certain parameter was present but did not identify the ``correct'' annotation. This can include completions that are flat-out incorrect, completions that don't identify the correct number of instances of a parameter in the input, and completions that have identified a proper subset ($\subset$) of the correct words. 
    \item \textbf{Identification Error} occurs when the model, despite being prompted with a specific parameter (\textit{e.g.,} ``Aim''), failed to include that parameter in its completion. 
\end{enumerate}

Because any model output is categorized as belonging to one of the above categories, accuracy is defined as:
\[\text{Accuracy }= \frac{\text{Correct Outputs}}{\text{All Outputs}} = \frac{PM}{PM + SM + ME + IE}\]
where $PM$ is the number of perfect match annotations, $SM$ is the number of superset matches, $ME$ is the number of match errors, and $IE$ is the number of identification errors.

\subsection{Baseline Results}
Our baseline RNN model had a test set accuracy of only 6\% (Figure~\ref{fig:all-bench}). While the RNN handled negative examples well, it struggled with positive examples. Specifically, the RNN failed to accurately annotate \textit{any} positive examples, while all LLMs other than GPT-2 were able to successfully annotate at least one positive example.

The prompted \textit{non-fine-tuned} LLMs also performed poorly, with the best such model having a test set accuracy of  $<$20\% (Appendix~\ref{sec:appendix:baseline}). While more effective prompting strategies are likely to exist, we believe that this finding provides evidence that LLMs should not be applied off-the-shelf for GKC-CI annotation as of now.

Comparing these results to the performance of the fine-tuned LLMs below, it is evident that fine-tuned LLMs substantially outperform both classical NLP RNNs and non-fined-tuned LLMs. This justifies the time and expense needed for fine-tuning and further indicates the non-triviality of the GKC-CI annotation task.

\subsection{Fine-Tuned Results}
\begin{figure}[t]
    \centering 
    \includegraphics[width=\linewidth]{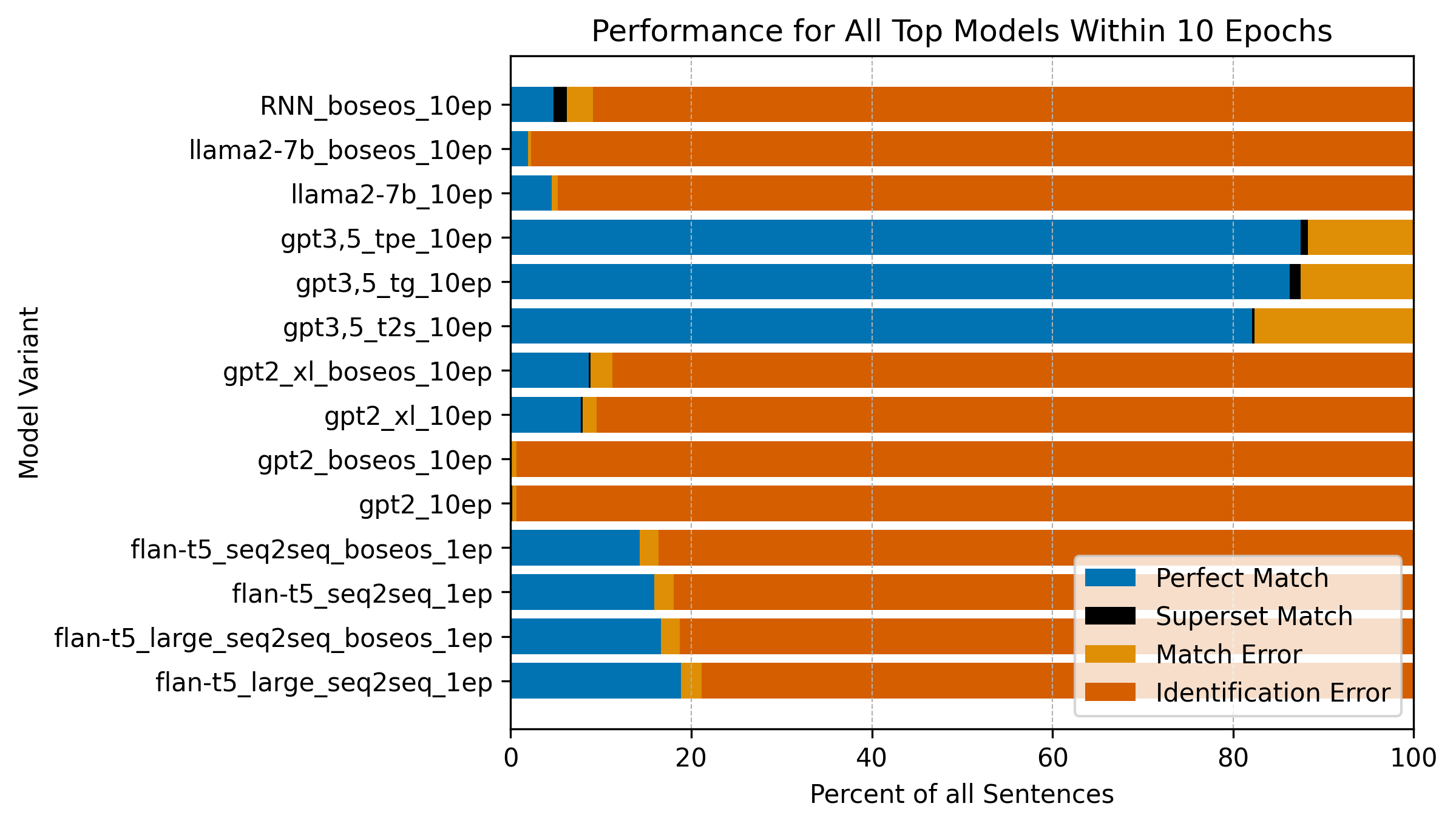}
    \caption{Test set performance of the top-performing models variants, including the RNN, with $\leq$ 10 epoch of training. \texttt{GPT3,5\_TPE} refers to the prompt-engineered version of \texttt{GPT-3.5 Turbo}, \texttt{GPT3,5\_TG} refers the generic GPT-3.5 Turbo model, and GPT3,5\_t2s refers to the joint performance of the GPT-3.5 Turbo, 2-Step models. Expanded model names in Appendix ~\ref{sec:appendix:name_mapping}.
    }
    \label{fig:all-bench}
\end{figure}

We benchmarked each of our \numModels{} models on each of the \testsize{} sentences in our test set. The results are shown in Figure~\ref{fig:all-bench}. OpenAI's proprietary models performed significantly better than any of the open-source models we tested. No open-source model that we tested had performance high enough to be considered even a poor substitute. Examination of the open-source models' outputs indicates a lack of substantial training convergence within 10 epochs for all models. Graphs of the open source models' performance at 1, 5, and 10 epochs are included in Appendix~\ref{sec:appendix:epochs}. Each of the open-source models took <12 hours to train, but the training time is extremely sensitive to specific hardware configurations. We hypothesize that given enough compute, modern open-source models may eventually reach the performance of OpenAI's proprietary models. We briefly summarize some notable lessons from working with the open-source models below. We then discuss the performance of the OpenAI models in greater detail.

\subsubsection{Open-Source Takeaways}
First, we note that within a model family, size does play a significant role in a model's ability to perform well on our annotation task. GPT-2 and Flan-T5 both failed utterly at their smaller model sizes, while their XL and Large variants performed $\approx$3\% to 15\% better. We additionally note that absolute model size in terms of parameters across model families does not appear to be a consistent indicator of performance. Llama2's 7B variety performed similarly to GPT-2's XL variety, despite being over four times the size. We mention this to caution resource-constrained researchers: \textbf{larger may not always mean better when comparing across model families.} Additionally, those models which were aligned (Flan-T5 and all GPT-3.5 models) performed the best. We \textbf{recommend compute-constrained researchers prioritize smaller, aligned models.}

Second, we observe that models appear to be vulnerable to the inclusion or exclusion of BOS, EOS tokens. Namely, we observe small performance differences between \emph{all} open-source models, but most strikingly in the Llama2 model family. We believe this is significant because it suggests that opaque defaults have observable effects on performance. Further, those effects do not appear to be consistent across model families. Specifically, as of the time of writing, HuggingFace's tokenizers \emph{all} have BOS, EOS tokens internally defined, but each model has a different default behavior when it comes to including or excluding BOS, EOS tokens. \textbf{We urge other researchers to be particularly careful of library and model defaults as they could be a potential confound in model performance.}

\begin{figure*}[ht]
    \centering
    \includegraphics[scale=0.75]{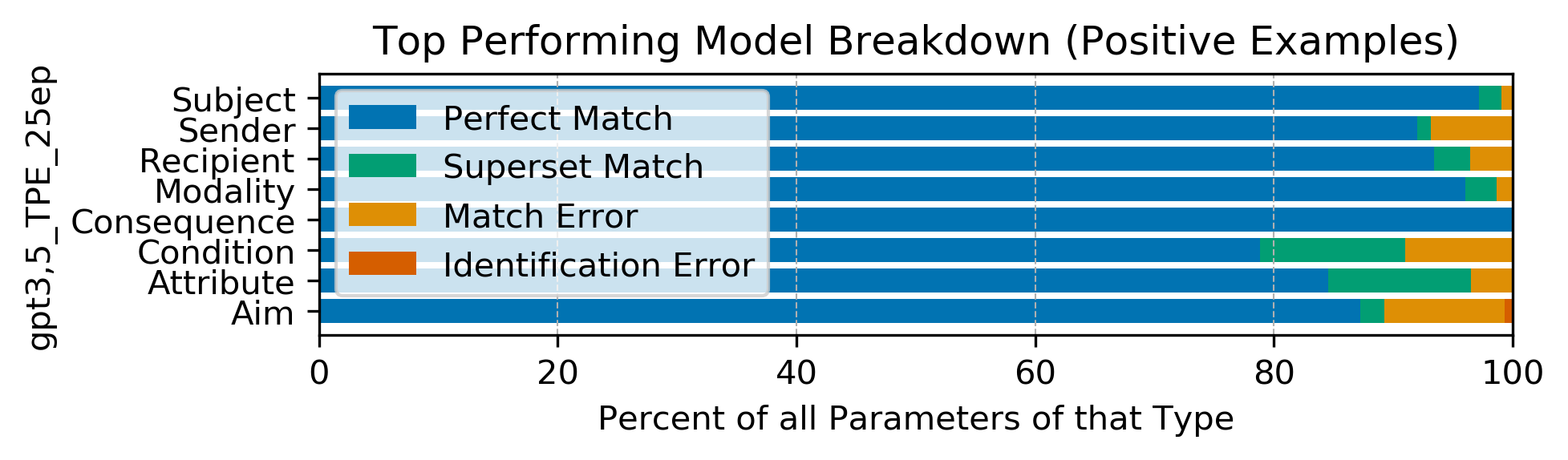}

    \caption{Performance per GKC-CI parameter for our best performing model, \texttt{GPT 3.5TPE\_25ep}.}
    \label{fig:americasnexttopmodel}
\end{figure*}

\subsubsection{OpenAI Models}
\label{sec:openai-models}

Appendix~\ref{sec:appendix:top_mods_attrs} summarizes model performance for three variants of GPT-3.5 Turbo on \textit{positive} examples, which are those examples that contain a GKC-CI parameter. We observe that these models vary substantially in their per-parameter performance. 

\texttt{GPT3.5-Turbo, two step 10 epochs} refers to a two model system where the first model determines if the text contains a GKC-CI parameter and the second model identifies the parameter. For this model, an example is a perfect match if and only if the first model and the second model classify the prompt correctly. This model performed the worst of the three GPT-3.5 Turbo variants with 30\% correct positive examples and 97\% correct negative examples. While this model demonstrated success identifying negative examples, its limited capacity to accurately identify positive examples significantly undermined its overall performance. 

\texttt{GPT3.5-Turbo, Generic 10 epochs} (\texttt{GPT 3.5TG\_10ep}) and \texttt{GPT3.5-Turbo, Prompt Engineered 10 epochs} (\texttt{GPT 3.5TPE\_10ep}) performed similarly well to each other. \texttt{GPT 3.5TG\_10ep} had 73\% correct positive examples and 88\% correct negative examples. \texttt{GPT 3.5TPE\_10ep}  had a slightly better 75\% correct positive examples and 89\% correct negative examples.

%Due to this strong performance, we further examine \texttt{GPT 3.5TPE} as our best performing model shown in Figure~\ref{fig:americasnexttopmodel}.
We speculate that the reason why \texttt{GPT 3.5TPE\_10ep} performed better than \texttt{GPT 3.5TG\_10ep} is due to the added context from prompt engineering. Specifically, prompt engineering may provide enough context for \texttt{GPT 3.5TPE\_10ep} such that the model is able to relate the annotation task to information on which it has been trained. This phenomenon has been more broadly observed in the NLP space~\cite{scao2021data, webson2022promptbased}.

\begin{figure}[h!]
    \centering
    \includegraphics[width=\linewidth]{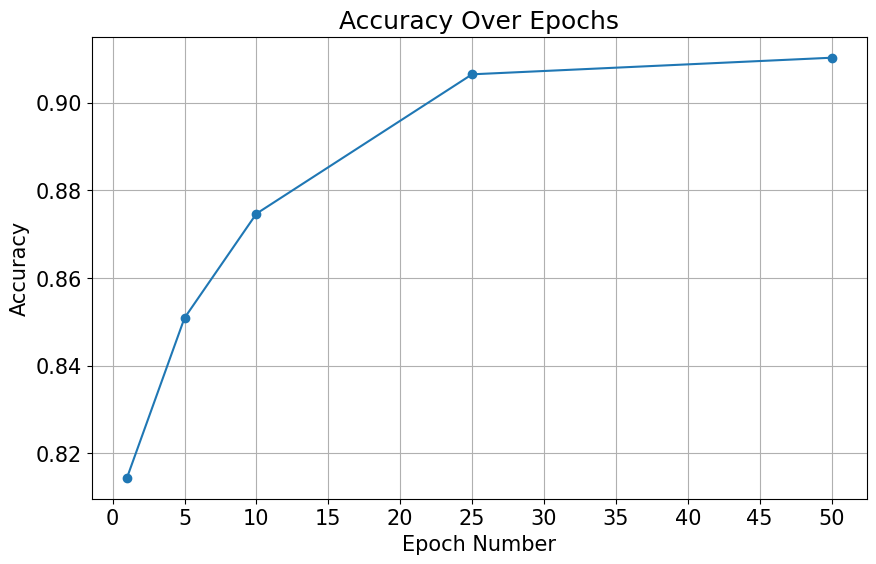}
    \caption{\texttt{GPT 3.5TPE}'s performance on the test set at 1, 5, 10, 25, and 50 epochs. Only Perfect Matches were considered to be ``correct.''}
    \label{fig:gpt3.5TPE_epochs}

    \vspace{4em}

    \includegraphics[width=\linewidth]{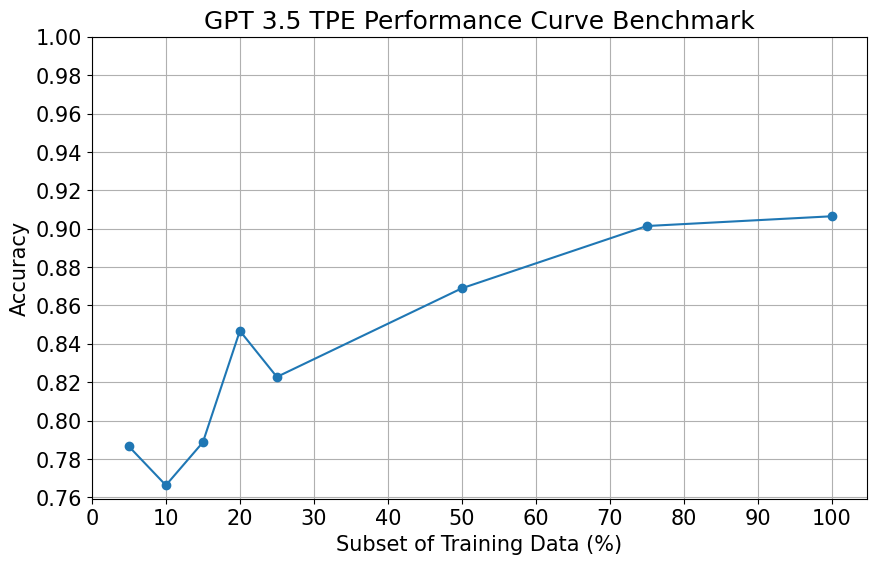}
    \caption{\texttt{GPT 3.5TPE\_25ep}'s accuracy on the test set as training data increases. Only Perfect Matches were considered to be ``correct.''} 
    \label{fig:gpt3.5TPE_trainingdata}

    \vspace{3em}
\end{figure}

We next investigated how the number of training epochs effected the performance of \texttt{GPT 3.5TPE} (Figure~\ref{fig:gpt3.5TPE_epochs}). We observed that \texttt{GPT 3.5TPE} started to have diminishing returns after 25 epochs worth of training. This behavior is typical for any model trained using gradient descent with an optimizer since both the steps grow smaller as the model approaches a minima and the learning rate decays over time. We hypothesize that the model had likely overfit after 50 epochs. We further investigated the effect of training data quantity on \texttt{GPT 3.5TPE} (Figure~\ref{fig:gpt3.5TPE_trainingdata}). The model started to have diminishing returns in terms of performance per data quantity when using 75\% of the training data, and 50\% of the training data appeared to be the necessary minimum for the model to begin to converge. However, we observed no detriment to using all available training data.

To summarize, we observed that \texttt{GPT 3.5TPE} benefited from the full training dataset and that the model was at risk of overfitting after 25 epochs. \textbf{We consequently chose \texttt{GPT 3.5TPE} as our best performing model and trained it for 25 epochs using the full training data}, which took 3 hours. \textbf{We  call this top model \texttt{GPT 3.5TPE\_25ep}}, and its performance across GKC-CI parameters in the test set is shown in Figure~\ref{fig:americasnexttopmodel}. To better understand the strengths and weaknesses of this model, we performed a qualitative analysis of the model's errors (Section~\ref{sec:error-analysis}).

Finally, to ensure that this model's performance isn't critically hindered by being trained exclusively on sentences, we fine-tuned a version of \texttt{GPT 3.5TPE} using paragraphs instead of sentences (details in Appendix ~\ref{sec:app_paragraph_lvl_annos}). The goal was to determine whether providing more context in the input would improve performance. Despite subjecting this model to the same training epoch ablation experiments as our top-performing model, it only achieved a maximum accuracy of 67.75\%, substantially lower than the 90.65\% maximum accuracy achieved with sentence-based training. For a detailed comparison of the performance between these sentence-based and paragraph-based models, please refer to Appendix~\ref{sec:appendix:paragraphs}.

\subsection{Qualitative Error Analysis}
\label{sec:error-analysis}

In order to better understand the errors made by \texttt{GPT 3.5TPE\_25ep}, we performed qualitative coding on the 37 \textit{match errors} for positive examples produced by the model, i.e., match errors where neither the ground truth text nor the model completion was ``N/A''. This served two purposes.

First, we identified cases where model outputs were mistakenly annotated as errors, specifically where the model annotation was \textit{semantically equivalent}, albeit \textit{syntactically different} from the ground truth. Second,  we conducted a detailed analysis of the model's performance, examining the factors contributing to both its strengths and weaknesses.
This provided more confidence in overall model performance. 

Identifying trends in these errors offers valuable insights into the model’s behavior. Although these insights may not  explain \textit{why} an error occurs (LLM models have notoriously poor explainability~\cite{druce2021brittleaicausalconfusion, BARREDOARRIETA202082}), they help users become aware of the model's limitations. 

\subsubsection{Qualitative Coding}
To ensure the reliability and consistency of the coding process, two expert coders initially met to collaboratively develop a comprehensive codebook consisting of ten codes: three parent codes and seven child codes. The full codebook can be seen in Appendix~\ref{sec:appendix:codebook}.

\begin{figure}[t!]
    \centering
    \includegraphics[width=7cm]{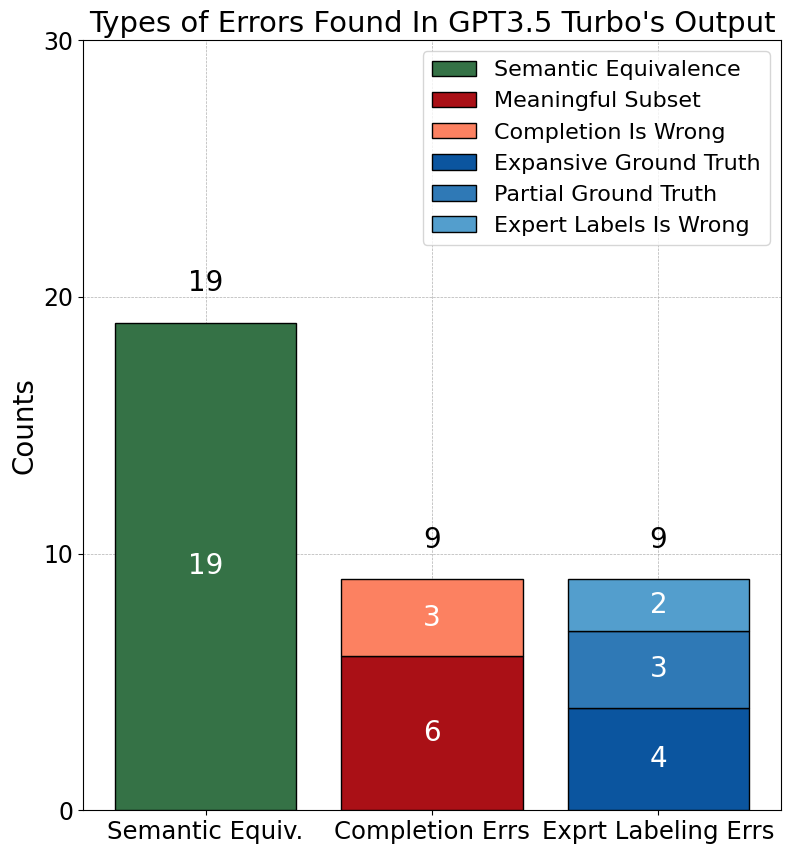}

    \caption{Breakdown by parent code of the various types of errors found from our qualitative analysis.}
    \label{fig:qual-chart}
\end{figure}

After joint codebook creation, each coder independently coded all match errors produced by \texttt{GPT 3.5TPE\_25ep}. Once the coding was complete, we computed inter-coder reliability and found a high level of agreement between the two coders with a Cohen's kappa score of 0.87~\cite{cohen1960coefficient}. The results of the qualitative coding are visualized in Figure~\ref{fig:qual-chart} and detailed further below. Note that in the following text, parent codes from the codebook are italicized, while child codes are enclosed in quotation marks

\subsubsection{Semantic Equivalence}
The child code ``Semantic Equivalence,'' the sole child code of the parent code \textit{Semantic Equivalence}, was the most prevalent code in our error analysis---accounting for 19/37 (51\%) errors. The following two examples demonstrate this type of error. The text in quotation is the expert annotation, while the underlined sections are the model’s annotation:
\begin{enumerate}
    \item Aim: ``to \uline{help us operate or administer the Services}’’
    \item Recipient: ``These \uline{Services}’’
\end{enumerate}
Note that the model's response only differed by an article or an adjective, and both are equivalently correct annotations.

\subsubsection{Incorrect Expert Annotations}
\label{sec:error-analysis:expert-errors}

The parent code \textit{Expert Labels Is Wrong} had nine examples (24\%). 
Occasional expert mis-annotations are expected for a task of this complexity. We are encouraged that there were relatively few examples under this parent code, supporting the quality of our ground truth.
Importantly, for the examples in the ``Expansive Ground Truth'' (4/9, 44\%) and ``Partial Ground Truth'' (3/9, 33\%) child codes, \textit{the model performed the task more correctly than the expert annotator}---either by omitting superfluous words included in the expert annotation or including necessary ones the expert annotator missed.

Consider the following example ``Expansive Ground Truth'' annotation. The text in quotation is the expert annotation, while the crossed out section is what the model correctly excluded: 

\begin{enumerate}
    \item Consequence: ``\sout{You can set your browser to not accept cookies, but} this may limit your ability to use the Services.''
\end{enumerate}

Conversely, the following example ``Partial Ground Truth'' annotations show the expert annotation in quotations, while the underlined words are what the model correctly choose to additionally include in its response:

\begin{enumerate}
    \item Recipient: ``trusted companies'' \uline{that work with, or on behalf of, Crowdmark to process information}
    \item Condition: ``to comply with its general obligations under the GDPR,'' \uline{in particular to process the personal data it collects in accordance with Articles 5 and 6, and to comply with Articles 13, 14, 24, 30 and 32, and to comply with any actionable rights of the data subject}
\end{enumerate}

Combined, 
``Expansive Ground Truth'' and ``Partial Ground Truth'' codes represent only 9 of the 37 errors (24\%); however,
we emphasize this finding as particularly exciting because they demonstrate that the model can identify precise annotations for the requested parameter. In other words, the model's ability has surpassed that of our trained human annotators in these situations.

\subsubsection{True Model Errors}

The parent code \textit{Completion Errors} accounted for 9 out of 37 coded examples (24\%).
Notably, a significant majority of these errors (6 out of 9, 67\%) fall under the ``Meaningful Subset'' child code. A ``Meaningful Subset`` annotation included a segment of the correct response, but missed words that altered the meaning of the annotation. 
In the following example, the text in quotation is the expert annotation while the underlined sections are the model’s annotation:

\begin{enumerate}
    \item Aim: ``solely \uline{for the purposes of providing the relevant services to Kaltura}''
    \item Attribute: ``personal data, any communications or material of any kind that you e-mail, post, or transmit through the Site, \uline{such as questions, comments, suggestions, and other data}''
\end{enumerate}

Many of these completions do encompass enough of the correct answer for someone well-versed in GKC-CI to grasp the intended annotation. However, we consider these incomplete responses to be incorrect even though the model's answer is a meaningful subset of the correct response.

Finally, the parent code \textit{Completion Is Wrong} comprises the final 3 out of 9 codes (33\%). All of these responses annotated text from within the sentence with no relation to the actual GKC-CI parameter.
For example the model incorrectly annotates the underlined part of the following sentence as an \textit{attribute}:

\begin{enumerate}
    \item Attribute: ``\uline{from the institution including the user's} identifier and organizational affiliation''
\end{enumerate}

Our qualitative analysis offers a comprehensive view of the model's performance, detailing both its strengths and limitations. We find that appreciably more than half of the purported match errors are the codes ``Semantic Equivalence,'' ``Expansive Ground Truth,'' and ``Partial Ground Truth.'' These combined child codes constitute 27 examples, resulting in a $0.29$ percent \textit{increase} in the number of correct annotations overall. This implies that our benchmarking metric of 90.65\% model accuracy serves as a conservative estimate of model performance. 

\subsection{Training Data Review}
Motivated by the nine expert annotation errors we uncovered in Section~\ref{sec:error-analysis:expert-errors}, we had two different CI experts review a random sample of 5\% of our training data. These experts agreed with the training data on 90.57\% of the sample, with the most common disagreement being the identification of occasional \textit{attribute} parameters that were missed in the original annotations. The original annotators spent several months annotating thousands of parameters, so it is unsurprising that a few were missed. Furthermore, governing knowledge commons and contextual integrity are the topic of active research and subject to different interpretive nuances by different members of the community, which is reflected in different annotations. This training data review validates our results, as our best performing model's 90.65\% accuracy is close to the inter-expert annotation agreement, suggesting that the model approaches human performance on the task.

\section{Limitations}

Despite the high accuracy of our best performing model, we point out a few limitations.
First, website privacy policies are not written with the GKC-CI framework in mind, so a perfect mapping from text to GKC-CI parameters may not be possible. We do not know the maximum performance that can be achieved by GKC-CI annotation, and it likely varies depending on the specific documents annotated. While we have created and used the largest training set for this task to date, and our analysis indicates that further increasing training set size would not substantially improve performance (Figure~\ref{fig:gpt3.5TPE_trainingdata}), it is possible that training data from a different set of policy documents might affect overall performance.

Second, the LLMs we use do not have information about the input text outside of the context window. This is unlike a human lawyer, who could review the entire document while identifying information flow descriptions. Fortunately, the information needed to identify GKC-CI parameters typically appears in the text immediately surrounding the parameter itself. The decreased accuracy we observed when using a longer context window (Section~\ref{sec:openai-models}, Appendix~\ref{sec:appendix:paragraphs}) supports our use of one-sentence contexts.

Third, we do not perform coreference resolution, i.e., identifying when different sections of text refer to the same real world entity. For example, a sentence may refer to the website user multiple times, with each instance receiving a GKC-CI parameter annotation (e.g., ``sender''). 
Although our model would accurately annotate each mention of the sender, it would not automatically recognize that these annotations refer to the same legal entity.
If that were necessary for a downstream application, another model would be required. Since coreference resolution is an active research problem in its own right, and is independent from GKC-CI parameter annotation, we leave this for future development.

\section{Example Applications}
\label{sec:production}

We applied our \texttt{GPT 3.5TPE\_25ep} model to 456 privacy policies from the Princeton-Leuven Longitudinal Corpus of Privacy Policies~\cite{amos2021privacy} to demonstrate the type of analyses enabled by GKC-CI annotation at scale. This dataset contains over 1 million privacy policies from over 100,000 companies spanning more than two decades, making it an ideal data source. However, we note that the primary contribution of our project remains the LLM training and evaluation (Sections~\ref{sec:methods}--\ref{sec:performance}). This section is not meant to provide a comprehensive analysis of policies in the Princeton-Leuven dataset. Rather, we intend the following examples to inspire future work using our LLM annotation method and the annotated policies we provide. \textbf{All 456 policies we annotated for the following analyses are publicly available at at the GitHub repository for this paper.}\footnote{\url{https://github.com/JakeC007/Automated_GKC-CI_Privacy_Policy_Annotations}} 

\subsection{Longitudinal Privacy Policy Analysis}
First, we chose 8 prominent companies and organizations\footnote{Facebook, The New York Times, GitHub, Buzzfeed, Google, Bank of America, Electronic Frontier Foundation (EFF), and the National Science Foundation (NSF)} representing a variety of sectors, including ``big tech,'' news, entertainment, finance, and government.
This variety is useful because each sector has a unique approach to data collection, user engagement, and compliance with privacy regulations. Furthermore, these websites have undergone varying levels of public scrutiny. For instance, while Facebook and Google have faced major privacy debates, leading to numerous changes in their privacy policies, entities like \textit{nsf.gov} operate under distinct governmental standards. This list also highlights geographic diversity concerning headquarters and user base, with some organizations primarily serving U.S. audiences while others have a global reach, necessitating compliance with various international privacy laws per the Brussels effect~\cite{bradford2012brussels}. 

For each of these companies and organizations, we used our model to annotate one privacy policy from every year that the company or organization appears in the dataset (128 policies total). The number of parameters in the policies of each company or organization over time are presented in Appendix~\ref{sec:long_appendix} (Figure~\ref{fig:longitudinal} and Tables~\ref{tab:longitudinal-data-1}--\ref{tab:longitudinal-data-2}).

The results provide insight into the evolution of privacy policies. 
For instance, we notice a generally increasing trend in the number of GKC-CI parameters included in privacy policies over time.
As specific examples, the privacy policies of Buzzfeed and GitHub described fewer than 60 GKC-CI parameters in their policies from 2008-2010, but now describe over 400 or 500 parameters, respectively. 
The increase in parameters in the GitHub policy from 2016 to 2019 corresponds to the acquisition of GitHub by Microsoft.

The EFF
and the NSF show similar, if less dramatic, increases in the number of parameters over time. 
This trend mirrors previously documented increases in average privacy policy length from 1996 to 2021~\cite{wagner2022privacy}, providing a sanity check for our method -- we expect longer privacy policies to include more details about information transfers. 
Indeed, the New York Times privacy policy underwent a dramatic \textit{decrease} in the number of GKC-CI parameters in 2006, corresponding to an approximately 80\% decrease in the length of the policy (23736 to 4912 words). The number of parameters then increased to above its previous maximum in 2011 when the policy length increased to 33616 words. 

We also notice that although parameter counts are generally increasing and roughly track total policy length, the relative percentages of different parameters remains nearly consistent within each company's policy.
This suggests that although companies are adding additional details to descriptions of data transfers in their privacy policies, these additions are not broadly skewed toward specific parameters.
To understand the importance of this result, consider some counterfactual examples:
If the relative percentage of \textit{aim} parameters were to have increased, it would indicate that organizations are increasingly using privacy policies to inform \textit{why} information is being collected over \textit{what} information is being collected. 
If the relative percentage of \textit{attribute} parameters were to have increased, it might indicate that organizations are collecting more data types per information transfer.
While we don't see either of these trends for these 8 companies, we examine the variance of parameter types in the privacy policies of a larger set of companies in Section~\ref{sec:cross-industry}. 

A closer look at the GitHub privacy policies from 2016 and 2019 demonstrates how GKC-CI annotation can be used to automatically identify privacy policy updates of interest. Specifically, the 2019 policy includes more \textit{aim} parameters describing new ways that the site may use collected information, such as ``to make recommendations for you, such as to suggest projects you may want to follow or contribute to'' and ``to determine your coding interests.'' Noticing a substantial increase in the number of \textit{aims} and focusing directly on those new parameters allows for quick assessment of changes in information use. Similarly, automatically identifying new or changed \textit{attributes} would allow quick assessment of new types of information collected.

Privacy policy updates happen regularly on a vast number of online services, and it is difficult for experts, to say nothing of average users, to identify which updates are salient for particular privacy concerns. We anticipate that our LLM annotation approach could be used in the future to automatically sort updates by parameter type and, combined with a visual interface, faster review of privacy policy updates.

\subsection{Cross-Industry Privacy Policy Analysis}
\label{sec:cross-industry}

We next used our fine-tuned LLM to annotate the most recent privacy policies  within the Tranco top 300~\cite{pochat2018tranco} websites that are in the Princeton-Leuven corpus (164 policies total). 

In each of the following analyses, we highlight extreme examples from across these 164 policies to demonstrate how annotation at scale facilitates directed data exploration. 
Previous work has shown that detailed analysis of individual annotated policies using the CI framework can identify specific ambiguities and 
normative shortcomings~\cite{shvartzshnaider2019going}. While deep analysis of individual policies is out of scope for this paper, the following paragraphs show how GKC-CI annotation can be used to identify policies that might be worth such detailed exploration in future work. 

\paragraph{Parameter type variance.}
We first calculated the variance in the percentages of individual parameter types across all annotated parameters in each policy. Previous work using CI annotation emphasized that descriptions of information transfers that are missing specific parameter types or that included substantially more specific parameter types (``parameter bloating'') lead to ambiguities about the actual data handling practices of the organization~\cite{shvartzshnaider2019going}. 
Since policies with a greater variance in the percentages of individual parameter types are more likely to exhibit these issues, we rank our annotated policies by this metric.

\begin{figure}[t]
    \centering
    \includegraphics[width=0.9\linewidth]{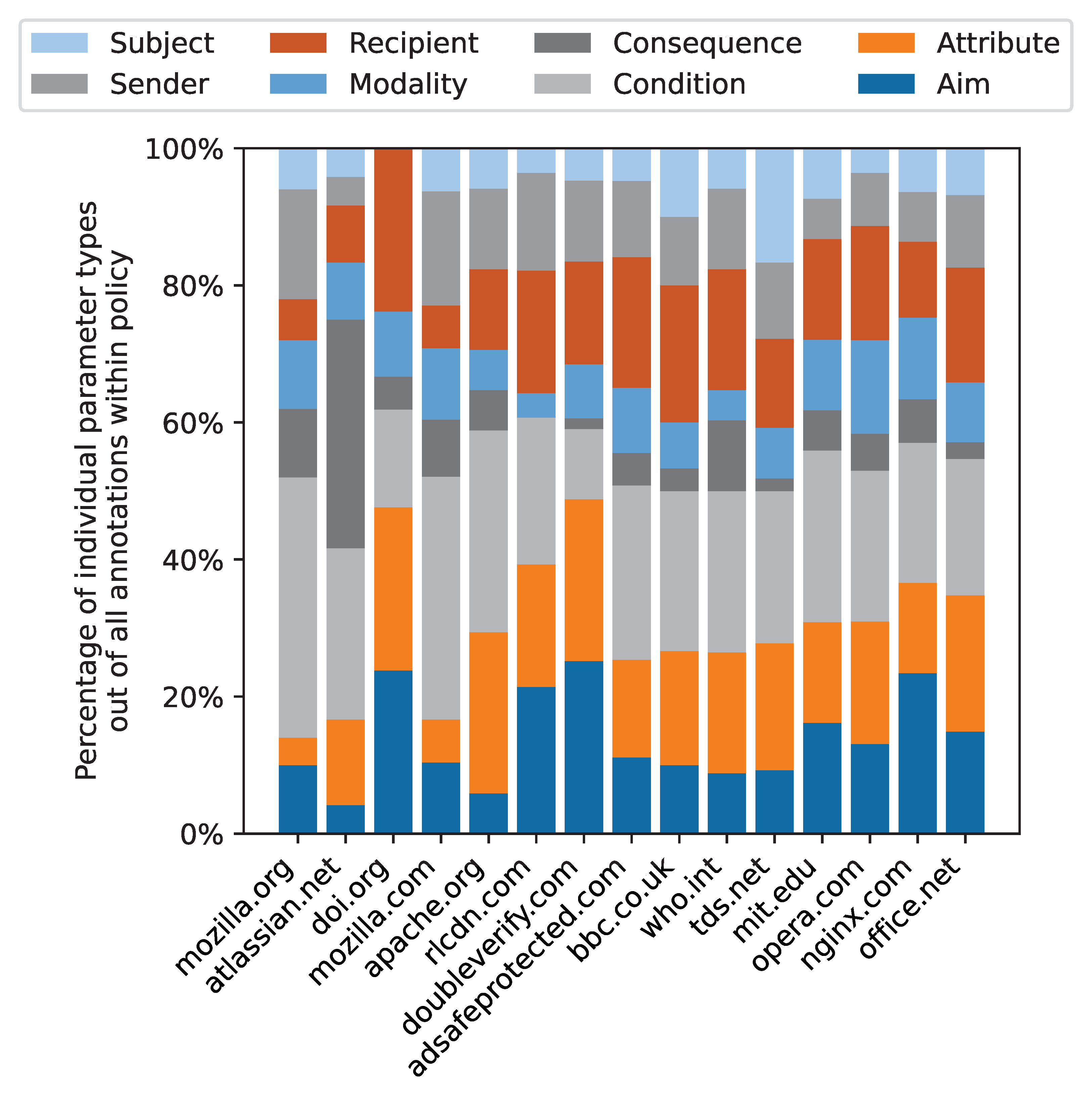}

    \caption{The 15 privacy policies with the highest variance in the percentage of individual parameter types across all parameters annotated in the policy.}
    \label{fig:paramvariance}
\end{figure}

Figure~\ref{fig:paramvariance} shows the fifteen policies with the highest variance of parameter type percentages. This includes a policy from \textit{apache.org} with relatively few \textit{aim} parameters and a policy from \textit{mozilla.org} with relatively few \textit{attribute} parameters. 
A quick review of the \textit{mozilla.org} policy shows why -- it defines all data covered by the policy in one sentence: ``For us, `personal information' means information which identifies you, like your name or email address.''
The policy does not say which particular forms of personal information are connected to the descriptions of information use in the rest of the policy, leaving it ambiguous whether e.g., what specific customer data might be shared with other entities for ``processing or providing products and services.''
While this type of high-level analysis doesn't necessarily imply the existence of policy ambiguities, it suggests that policies with high parameter type variance are promising candidates for a detailed evaluation through the lens of the GKC-CI framework. The parameter percentages for all 164 privacy policies are provided in Tables~\ref{tab:variance-data-1}--\ref{tab:variance-data-2} in Appendix~\ref{sec:variance-appendix}.

\begin{figure}[t]
    \centering
    \includegraphics[width=0.9\linewidth]{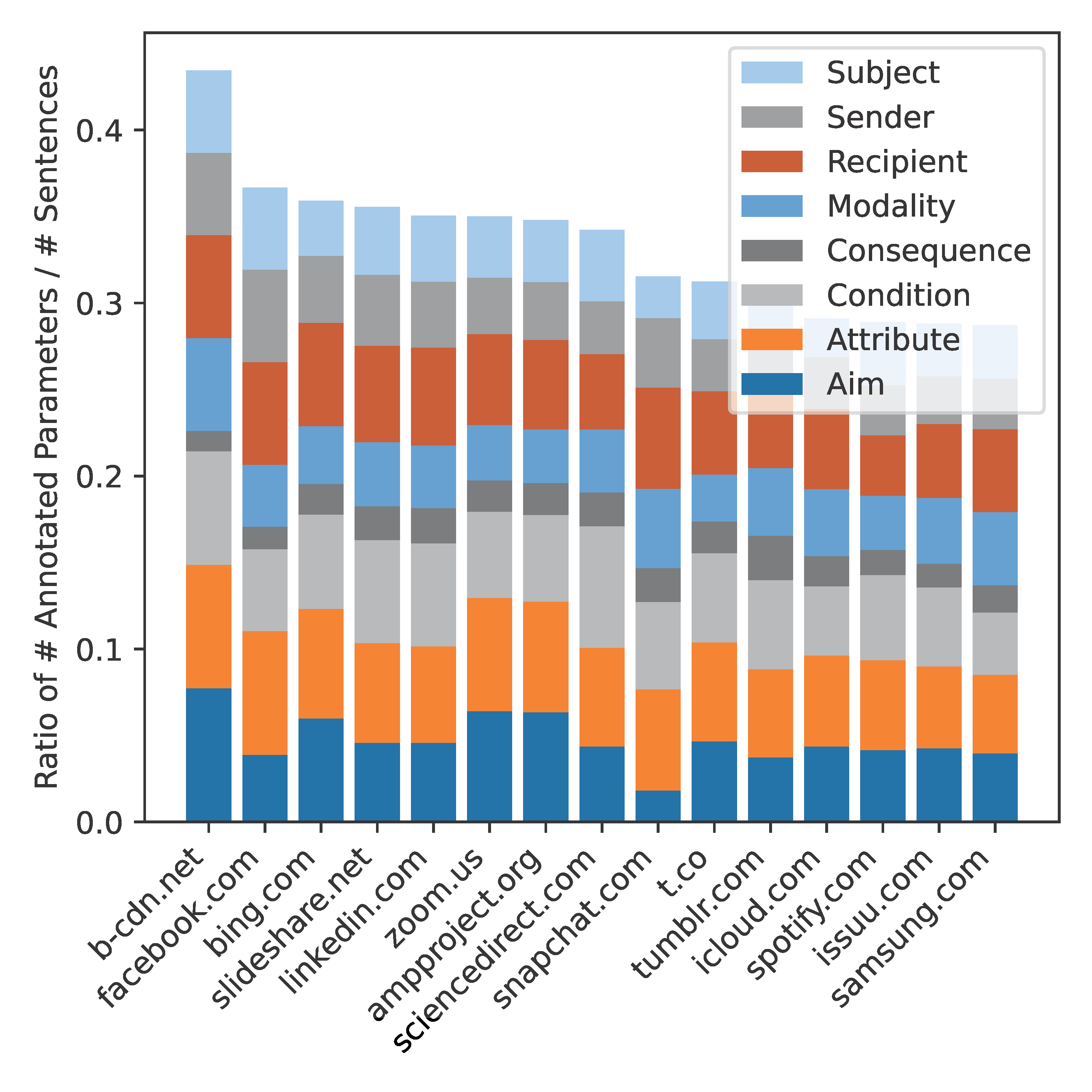}

    \caption{The 15 privacy policies with the highest ratio of GKC-CI parameters to sentences out of all 164 of the Tranco top 300 websites in the Princeton-Leuven corpus.}
    \label{fig:paramdensity}
\end{figure}

\paragraph{Parameter to sentence ratio.}
We next calculated the ratio of annotated GKC-CI parameters to the number of sentences in each policy. 
This provides a metric of the ``density'' of information transfer descriptions in the policy. 
Figure~\ref{fig:paramdensity} shows these data for the 15 privacy policies with the highest ratio of annotated parameters to sentences. The ratios for all 164 privacy policies are provided in Tables~\ref{tab:density-data-1}--\ref{tab:density-data-2} in Appendix~\ref{sec:density-appendix}.
The top 15 policies include those from content distribution networks (\textit{b-cdn.net}), web component frameworks (\textit{ampproject.org}), Facebook, Microsoft (\textit{bing.com}), Apple (\textit{icloud.com}), Zoom, and social media websites (\textit{linkedin.com}, \textit{snapchat.com}, \textit{t.co}, \textit{tumblr.com}), among others.

While these policies may exhibit parameter bloating issues due to the density of parameters, they may also be good examples of policies providing meaningful details about data handling practices. 
The first few policies on this list provide a microcosm of this variety. The \textit{b-cdn.net} policy is very minimal, but each sentence is a short, to-the-point data handling description. The Facebook policy is longer, but an earlier 2018 version was identified in \cite{shvartzshnaider2019going} as having parameter bloat.
Either way, directing future in-depth investigations toward policies with high parameter density  would provide examples of GKC-CI information flow descriptions for case studies for teaching~\cite{apthorpe2021practical} or iteration on the GKC-CI framework.

\begin{figure}[t]
    \centering
    \includegraphics[width=\linewidth]{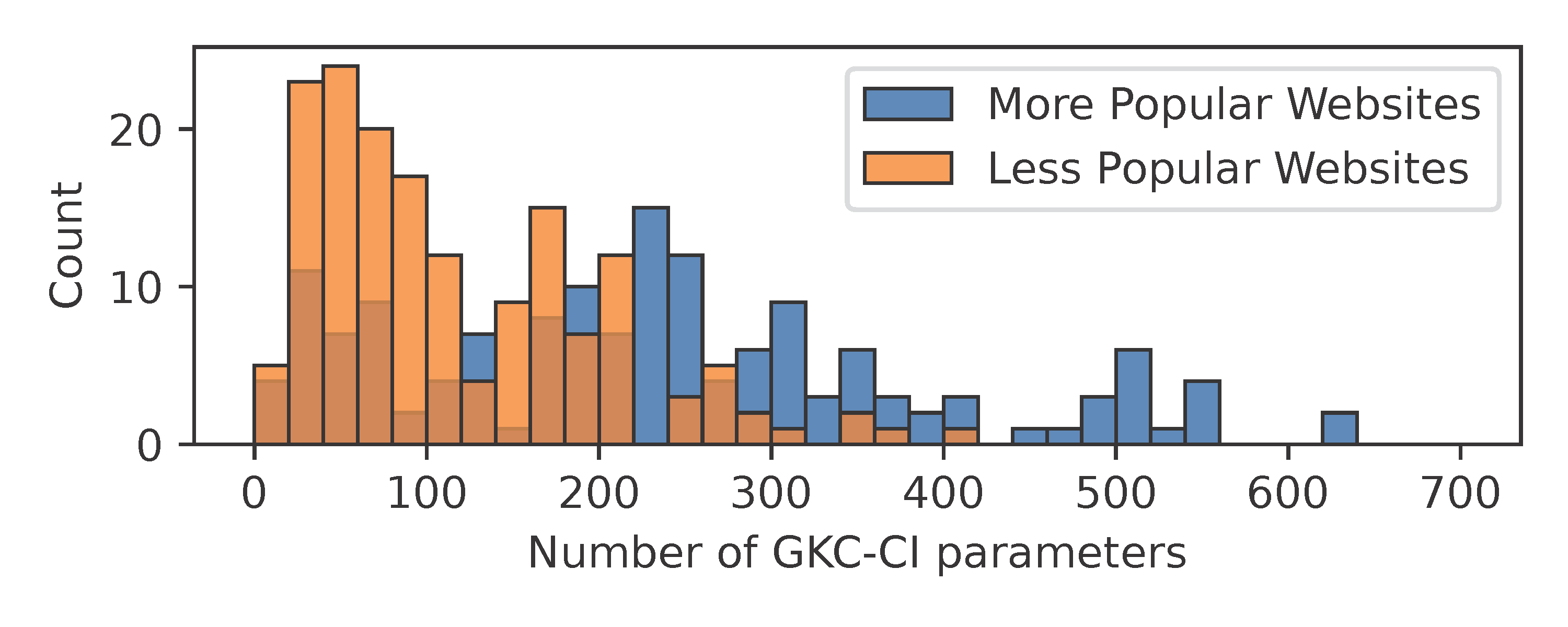}
    \includegraphics[width=\linewidth]{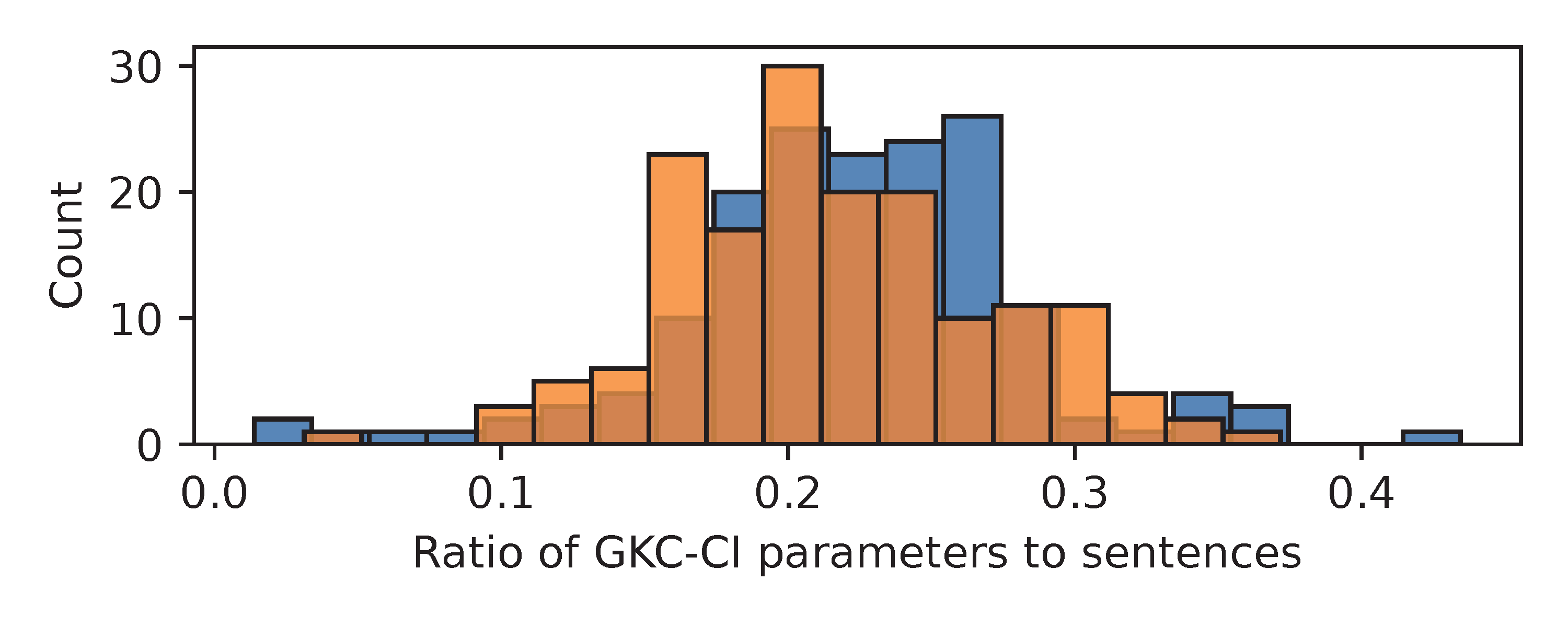}

    \caption{Comparison of all 164 websites in the Tranco top 300 versus the 164 websites lowest on the Tranco rankings in the Princeton-Leuven corpus. \textit{Top:} Distribution of the raw number of GKC-CI parameters annotated by our model. Outliers with $>700$ parameters (13 policies for the more popular websites and 1 policy for the less popular websites) are omitted. \textit{Bottom:} Distribution of the ratio of annotated GKC-CI parameters to sentence count.
    \vspace{2em}}
    \label{fig:popularitycounts}
\end{figure}

\paragraph{Website popularity.}
As a comparison against the 164 popular websites from the previous analyses, we next annotated the 164 policies from the Princeton-Leuven corpus with the \textit{lowest} Tranco rankings (999981 to 991993). 
Figure~\ref{fig:popularitycounts} compares the distributions of total parameter counts and parameter to sentence ratios between these sets of websites.

The distributions of total parameter counts are significantly different ($p < 0.005$, Mann-Whitney U test), with the more popular websites having more GKC-CI parameters in their privacy policies on average than the less popular websites (mean $412$ versus $130$). 
This result makes sense, as more popular websites are under more scrutiny about their handling of user information and therefore include more information about data practices in their privacy policies.
However, the distributions of parameter to sentence ratios are not significantly different ($p >0.05$, Mann-Whitney U test), indicating that the more popular websites are generally providing more details by adding to policy length, rather than by increasing the density of information flow descriptions.

We anticipate that automated GKC-CI parameter annotation will enable more detailed statistical analyses across privacy policies in the future. For example, one could track the changing correlation between \textit{condition} parameters in privacy policies and the text of data privacy regulations as both policies and regulation are updated -- identifying outliers for scrutiny.

\subsection{GKC-CI Annotation Visualizer}
We have developed a visualizer tool designed to enhance the interpretability of our LLM's privacy policy annotations. This tool was created to empower researchers and others using our approach to better understand the complex data flows and governing principles that underpin privacy policies by visually distinguishing between different parameter tags. By highlighting various aspects of the policy in different colors, the visualizer makes the described information flows more accessible, allowing users to quickly grasp how their data may be used and protected under the terms of the policy.

The tool functions as a local Python script that processes both the raw text of the privacy policy and the output log from the LLM. It begins by matching specific text segments from the policy with those used in the LLM prompts. The visualizer then highlights these segments in different colors based on the associated annotations, making it easy for users to see which parts of the text correspond to specific GKC-CI parameters. This highlighted text is presented in a user-friendly GUI window, where users can scroll through and interact with the augmented policy. Additionally, it generates an output file that explains the highlighted segments, helping users understand the implications of the annotations. Note that if the LLM's output has not match in the text we consider the output incorrect and do not include these instances in the visualization.

To illustrate the capabilities of the visualizer, we applied it to two versions of Facebook's privacy policy from 2015 and 2019.
Comparing visualized excerpts of these versions (Appendix~\ref{sec:appendix:viz}) effectively highlights how data practices and governing principles have evolved.
For instance,
there are more GKC-CI parameters in the 2019 excerpt than the 2015 excerpt, particularly more \textit{aims} and \textit{consequences}. The new policy does a better job informing users about the repercussions of the described data collection. 
The visualizer's color-coded presentation makes
it easier to understand how the policy has changed and will facilitate future research using GKC-CI annotation. 

\section{Future Directions}

As demonstrated in Section~\ref{sec:production}, the ability to accurately annotate privacy policies with GKC-CI parameter tags enables a variety of previously infeasible analyses. We are excited about the potential for the GKC and CI community to use our method to facilitate advances at the intersection of machine learning and privacy.

A common question posed by CI researchers is whether information flows (composed of GKC-CI parameters) are \textit{appropriate} in their respective contexts. This is a core element of the CI framework, which understands privacy breaches as inappropriate information flows that violate contextual norms~\cite{nissenbaum2009privacy}. There are many ways to discover the norms of a particular context, including surveys~\cite{apthorpe2018discovering, apthorpe2019evaluating, shvartzshnaider2016learning}, interviews~\cite{kumar2024roadmap}, focus groups~\cite{shaffer2021applying}, and textual analysis~\cite{barth2006privacy}, among others. We propose using LLM annotation as the first step in a multi-method pipeline for this task: 1) a researcher extracts GKC-CI parameters from segments of an organization's privacy policy using LLM annotation, 2) the parameters are piped into a survey template about contextual appropriateness (like that in \cite{apthorpe2018discovering}), 3) crowdsourced survey responses from relevant community members indicate (mis)alignment with existing norms. This entire process could be automated from the perspective of the researchers, enabling much larger scale CI-based audits of data handling practices across many organizations than previous possible.

We also believe that GKC-CI annotation with LLMs need not be limited to privacy policies. Many documents, from white papers to media reports, describe transfers of information. We expect that fine-tuned LLMs will be able to identify GKC-CI parameters as accurately in those documents as in privacy policies. This opens an new set of applications that we hope the community will pursue.
\section{Conclusion}
\label{sec:conclusion}
This paper demonstrates that high-accuracy annotation of contextual integrity (CI) and governing knowledge commons (GKC) parameters in privacy policies can be achieved using LLMs. We ultimately find that \texttt{GPT 3.5TPE\_25ep} had the best performance, with an accuracy of 90.65\% for exact string matches. While we find that proprietary LLMs outperformed open-source models, we report some valuable findings for researchers interested in performing LLM application studies. Namely, 1) that LLM size must be considered in context to model family; smaller, aligned models are the most economical choice, and 2) that library defaults are likely to introduce confounds and should be checked.

We demonstrate the usefulness of our fine-tuned model by annotating 456 privacy policies from the Princeton-Leuven Longitudinal Corpus of Privacy Policies~\cite{amos2021privacy}.
We show that large-scale GKC-CI annotation can be an effective tool for data exploration, highlighting changes in parameter frequency over time, policies with relatively high variances across parameter type percentages, and policies with relatively high ratios of parameters to sentences. 
This facilitates automated review of privacy policy updates to identify meaningful changes in information flow descriptions, potentially with normative implications. 
We have made our model training code, training and testing data, annotation visualizer, and all annotated policies publicly available\footnote{\url{https://github.com/JakeC007/Automated_GKC-CI_Privacy_Policy_Annotations}} in the hope that this work motivates future use of GKC-CI parameter annotation.

\begin{acks}
We thank research assistants Sophia Goffe and Wael Mohamed for their contributions. 
This research was funded by the Colgate University Faculty Research Council and the Colgate University Department of Computer Science.
Some of the results presented in this paper were obtained using the Chameleon testbed~\cite{keahey2020lessons} supported by the National Science Foundation.
\end{acks}

\bibliographystyle{ACM-Reference-Format}
\bibliography{main}


\begin{thebibliography}{92}

%%% the Web address.

\ifx \showCODEN    \undefined \def \showCODEN     #1{\unskip}     \fi
\ifx \showDOI      \undefined \def \showDOI       #1{#1}\fi
\ifx \showISBNx    \undefined \def \showISBNx     #1{\unskip}     \fi
\ifx \showISBNxiii \undefined \def \showISBNxiii  #1{\unskip}     \fi
\ifx \showISSN     \undefined \def \showISSN      #1{\unskip}     \fi
\ifx \showLCCN     \undefined \def \showLCCN      #1{\unskip}     \fi
\ifx \shownote     \undefined \def \shownote      #1{#1}          \fi
\ifx \showarticletitle \undefined \def \showarticletitle #1{#1}   \fi
\ifx \showURL      \undefined \def \showURL       {\relax}        \fi
% invisible to TeX
\providecommand\bibfield[2]{#2}
\providecommand\bibinfo[2]{#2}
\providecommand\natexlab[1]{#1}
\providecommand\showeprint[2][]{arXiv:#2}

\bibitem[Amos et~al\mbox{.}(2021)]%
        {amos2021privacy}
\bibfield{author}{\bibinfo{person}{Ryan Amos}, \bibinfo{person}{Gunes Acar}, \bibinfo{person}{Eli Lucherini}, \bibinfo{person}{Mihir Kshirsagar}, \bibinfo{person}{Arvind Narayanan}, {and} \bibinfo{person}{Jonathan Mayer}.} \bibinfo{year}{2021}\natexlab{}.
\newblock \showarticletitle{Privacy policies over time: Curation and analysis of a million-document dataset}. In \bibinfo{booktitle}{\emph{Proceedings of the Web Conference 2021}}. \bibinfo{pages}{2165--2176}.
\newblock


\bibitem[Andow et~al\mbox{.}(2019)]%
        {andow2019policylint}
\bibfield{author}{\bibinfo{person}{Benjamin Andow}, \bibinfo{person}{Samin~Yaseer Mahmud}, \bibinfo{person}{Wenyu Wang}, \bibinfo{person}{Justin Whitaker}, \bibinfo{person}{William Enck}, \bibinfo{person}{Bradley Reaves}, \bibinfo{person}{Kapil Singh}, {and} \bibinfo{person}{Tao Xie}.} \bibinfo{year}{2019}\natexlab{}.
\newblock \showarticletitle{{PolicyLint}: Investigating internal privacy policy contradictions on Google Play}. In \bibinfo{booktitle}{\emph{28th USENIX Security Symposium (USENIX Security 19)}}. \bibinfo{pages}{585--602}.
\newblock


\bibitem[Andow et~al\mbox{.}(2020)]%
        {andow2020actions}
\bibfield{author}{\bibinfo{person}{Benjamin Andow}, \bibinfo{person}{Samin~Yaseer Mahmud}, \bibinfo{person}{Justin Whitaker}, \bibinfo{person}{William Enck}, \bibinfo{person}{Bradley Reaves}, \bibinfo{person}{Kapil Singh}, {and} \bibinfo{person}{Serge Egelman}.} \bibinfo{year}{2020}\natexlab{}.
\newblock \showarticletitle{Actions speak louder than words: Entity-sensitive privacy policy and data flow analysis with {PoliCheck}}. In \bibinfo{booktitle}{\emph{29th USENIX Security Symposium (USENIX Security 20)}}. \bibinfo{pages}{985--1002}.
\newblock


\bibitem[Angulo et~al\mbox{.}(2012)]%
        {angulo2012towards}
\bibfield{author}{\bibinfo{person}{Julio Angulo}, \bibinfo{person}{Simone Fischer-H{\"u}bner}, \bibinfo{person}{Erik W{\"a}stlund}, {and} \bibinfo{person}{Tobias Pulls}.} \bibinfo{year}{2012}\natexlab{}.
\newblock \showarticletitle{Towards usable privacy policy display and management}.
\newblock \bibinfo{journal}{\emph{Information Management \& Computer Security}} (\bibinfo{year}{2012}).
\newblock


\bibitem[Apthorpe(2021)]%
        {apthorpe2021practical}
\bibfield{author}{\bibinfo{person}{Noah Apthorpe}.} \bibinfo{year}{2021}\natexlab{}.
\newblock \showarticletitle{Practical assignments for teaching contextual integrity}. In \bibinfo{booktitle}{\emph{3rd Annual Symposium on Applications of Contextual Integrity}}.
\newblock


\bibitem[Apthorpe et~al\mbox{.}(2018)]%
        {apthorpe2018discovering}
\bibfield{author}{\bibinfo{person}{Noah Apthorpe}, \bibinfo{person}{Yan Shvartzshnaider}, \bibinfo{person}{Arunesh Mathur}, \bibinfo{person}{Dillon Reisman}, {and} \bibinfo{person}{Nick Feamster}.} \bibinfo{year}{2018}\natexlab{}.
\newblock \showarticletitle{Discovering smart home internet of things privacy norms using contextual integrity}.
\newblock \bibinfo{journal}{\emph{Proceedings of the ACM on Interactive, Mobile, Wearable and Ubiquitous Technologies}} \bibinfo{volume}{2}, \bibinfo{number}{2} (\bibinfo{year}{2018}), \bibinfo{pages}{1--23}.
\newblock


\bibitem[Apthorpe et~al\mbox{.}(2019)]%
        {apthorpe2019evaluating}
\bibfield{author}{\bibinfo{person}{Noah Apthorpe}, \bibinfo{person}{Sarah Varghese}, {and} \bibinfo{person}{Nick Feamster}.} \bibinfo{year}{2019}\natexlab{}.
\newblock \showarticletitle{Evaluating the contextual integrity of privacy regulation: Parents' {IoT} toy privacy norms versus {COPPA}}. In \bibinfo{booktitle}{\emph{28th USENIX Security Symposium (USENIX Security 19)}}. \bibinfo{pages}{123--140}.
\newblock


\bibitem[Bannihatti~Kumar et~al\mbox{.}(2020)]%
        {bannihatti2020finding}
\bibfield{author}{\bibinfo{person}{Vinayshekhar Bannihatti~Kumar}, \bibinfo{person}{Roger Iyengar}, \bibinfo{person}{Namita Nisal}, \bibinfo{person}{Yuanyuan Feng}, \bibinfo{person}{Hana Habib}, \bibinfo{person}{Peter Story}, \bibinfo{person}{Sushain Cherivirala}, \bibinfo{person}{Margaret Hagan}, \bibinfo{person}{Lorrie Cranor}, \bibinfo{person}{Shomir Wilson}, {et~al\mbox{.}}} \bibinfo{year}{2020}\natexlab{}.
\newblock \showarticletitle{Finding a choice in a haystack: Automatic extraction of opt-out statements from privacy policy text}. In \bibinfo{booktitle}{\emph{Proceedings of The Web Conference 2020}}. \bibinfo{pages}{1943--1954}.
\newblock


\bibitem[{Barredo Arrieta} et~al\mbox{.}(2020)]%
        {BARREDOARRIETA202082}
\bibfield{author}{\bibinfo{person}{Alejandro {Barredo Arrieta}}, \bibinfo{person}{Natalia Díaz-Rodríguez}, \bibinfo{person}{Javier {Del Ser}}, \bibinfo{person}{Adrien Bennetot}, \bibinfo{person}{Siham Tabik}, \bibinfo{person}{Alberto Barbado}, \bibinfo{person}{Salvador Garcia}, \bibinfo{person}{Sergio Gil-Lopez}, \bibinfo{person}{Daniel Molina}, \bibinfo{person}{Richard Benjamins}, \bibinfo{person}{Raja Chatila}, {and} \bibinfo{person}{Francisco Herrera}.} \bibinfo{year}{2020}\natexlab{}.
\newblock \showarticletitle{Explainable artificial intelligence (XAI): Concepts, taxonomies, opportunities and challenges toward responsible AI}.
\newblock \bibinfo{journal}{\emph{Information Fusion}}  \bibinfo{volume}{58} (\bibinfo{year}{2020}), \bibinfo{pages}{82--115}.
\newblock
\showISSN{1566-2535}


\bibitem[Barth et~al\mbox{.}(2006)]%
        {barth2006privacy}
\bibfield{author}{\bibinfo{person}{Adam Barth}, \bibinfo{person}{Anupam Datta}, \bibinfo{person}{John~C Mitchell}, {and} \bibinfo{person}{Helen Nissenbaum}.} \bibinfo{year}{2006}\natexlab{}.
\newblock \showarticletitle{Privacy and contextual integrity: Framework and applications}. In \bibinfo{booktitle}{\emph{2006 IEEE Symposium on Security and Privacy (S\&P'06)}}. IEEE, \bibinfo{pages}{15--pp}.
\newblock


\bibitem[Beltagy et~al\mbox{.}(2020)]%
        {beltagy2020longformer}
\bibfield{author}{\bibinfo{person}{Iz Beltagy}, \bibinfo{person}{Matthew~E Peters}, {and} \bibinfo{person}{Arman Cohan}.} \bibinfo{year}{2020}\natexlab{}.
\newblock \showarticletitle{Longformer: The long-document transformer}.
\newblock \bibinfo{journal}{\emph{arXiv preprint arXiv:2004.05150}} (\bibinfo{year}{2020}).
\newblock


\bibitem[Benthall(2022)]%
        {280276}
\bibfield{author}{\bibinfo{person}{Sebastian Benthall}.} \bibinfo{year}{2022}\natexlab{}.
\newblock \showarticletitle{Integrating differential privacy and contextual integrity}. \bibinfo{publisher}{USENIX Association}, \bibinfo{address}{Santa Clara, CA}.
\newblock


\bibitem[Bradford(2012)]%
        {bradford2012brussels}
\bibfield{author}{\bibinfo{person}{Anu Bradford}.} \bibinfo{year}{2012}\natexlab{}.
\newblock \showarticletitle{The {Brussels} effect}.
\newblock \bibinfo{journal}{\emph{Northwestern University Law Review}}  \bibinfo{volume}{107} (\bibinfo{year}{2012}), \bibinfo{pages}{1}.
\newblock


\bibitem[Brehm and Shvartzhnaider(2023)]%
        {brehm2023understanding}
\bibfield{author}{\bibinfo{person}{Karoline Brehm} {and} \bibinfo{person}{Yan Shvartzhnaider}.} \bibinfo{year}{2023}\natexlab{}.
\newblock \showarticletitle{Understanding privacy in virtual reality classrooms: A contextual integrity perspective}.
\newblock \bibinfo{journal}{\emph{IEEE Security \& Privacy}} (\bibinfo{year}{2023}).
\newblock


\bibitem[Brodie et~al\mbox{.}(2006)]%
        {brodie2006empirical}
\bibfield{author}{\bibinfo{person}{Carolyn~A Brodie}, \bibinfo{person}{Clare-Marie Karat}, {and} \bibinfo{person}{John Karat}.} \bibinfo{year}{2006}\natexlab{}.
\newblock \showarticletitle{An empirical study of natural language parsing of privacy policy rules using the SPARCLE policy workbench}. In \bibinfo{booktitle}{\emph{Proceedings of the Second Symposium on Usable Privacy and Security}}. \bibinfo{pages}{8--19}.
\newblock


\bibitem[Brown et~al\mbox{.}(2020)]%
        {brown2020language_gpt3}
\bibfield{author}{\bibinfo{person}{Tom Brown}, \bibinfo{person}{Benjamin Mann}, \bibinfo{person}{Nick Ryder}, \bibinfo{person}{Melanie Subbiah}, \bibinfo{person}{Jared~D Kaplan}, \bibinfo{person}{Prafulla Dhariwal}, \bibinfo{person}{Arvind Neelakantan}, \bibinfo{person}{Pranav Shyam}, \bibinfo{person}{Girish Sastry}, \bibinfo{person}{Amanda Askell}, {et~al\mbox{.}}} \bibinfo{year}{2020}\natexlab{}.
\newblock \showarticletitle{Language models are few-shot learners}.
\newblock \bibinfo{journal}{\emph{Advances in Neural Information Processing Systems}}  \bibinfo{volume}{33} (\bibinfo{year}{2020}), \bibinfo{pages}{1877--1901}.
\newblock


\bibitem[Cengage(2022)]%
        {cengage2022}
\bibfield{author}{\bibinfo{person}{Cengage}.} \bibinfo{year}{2022}\natexlab{}.
\newblock \bibinfo{howpublished}{\url{https://www.cengagegroup.com/privacy/notice/}}.
\newblock


\bibitem[Chen et~al\mbox{.}(2021)]%
        {chen2021evaluating}
\bibfield{author}{\bibinfo{person}{Mark Chen}, \bibinfo{person}{Jerry Tworek}, \bibinfo{person}{Heewoo Jun}, \bibinfo{person}{Qiming Yuan}, \bibinfo{person}{Henrique~Ponde de Oliveira~Pinto}, \bibinfo{person}{Jared Kaplan}, \bibinfo{person}{Harri Edwards}, \bibinfo{person}{Yuri Burda}, \bibinfo{person}{Nicholas Joseph}, \bibinfo{person}{Greg Brockman}, \bibinfo{person}{Alex Ray}, \bibinfo{person}{Raul Puri}, \bibinfo{person}{Gretchen Krueger}, \bibinfo{person}{Michael Petrov}, \bibinfo{person}{Heidy Khlaaf}, \bibinfo{person}{Girish Sastry}, \bibinfo{person}{Pamela Mishkin}, \bibinfo{person}{Brooke Chan}, \bibinfo{person}{Scott Gray}, \bibinfo{person}{Nick Ryder}, \bibinfo{person}{Mikhail Pavlov}, \bibinfo{person}{Alethea Power}, \bibinfo{person}{Lukasz Kaiser}, \bibinfo{person}{Mohammad Bavarian}, \bibinfo{person}{Clemens Winter}, \bibinfo{person}{Philippe Tillet}, \bibinfo{person}{Felipe~Petroski Such}, \bibinfo{person}{Dave Cummings}, \bibinfo{person}{Matthias Plappert}, \bibinfo{person}{Fotios Chantzis},
  \bibinfo{person}{Elizabeth Barnes}, \bibinfo{person}{Ariel Herbert-Voss}, \bibinfo{person}{William~Hebgen Guss}, \bibinfo{person}{Alex Nichol}, \bibinfo{person}{Alex Paino}, \bibinfo{person}{Nikolas Tezak}, \bibinfo{person}{Jie Tang}, \bibinfo{person}{Igor Babuschkin}, \bibinfo{person}{Suchir Balaji}, \bibinfo{person}{Shantanu Jain}, \bibinfo{person}{William Saunders}, \bibinfo{person}{Christopher Hesse}, \bibinfo{person}{Andrew~N. Carr}, \bibinfo{person}{Jan Leike}, \bibinfo{person}{Josh Achiam}, \bibinfo{person}{Vedant Misra}, \bibinfo{person}{Evan Morikawa}, \bibinfo{person}{Alec Radford}, \bibinfo{person}{Matthew Knight}, \bibinfo{person}{Miles Brundage}, \bibinfo{person}{Mira Murati}, \bibinfo{person}{Katie Mayer}, \bibinfo{person}{Peter Welinder}, \bibinfo{person}{Bob McGrew}, \bibinfo{person}{Dario Amodei}, \bibinfo{person}{Sam McCandlish}, \bibinfo{person}{Ilya Sutskever}, {and} \bibinfo{person}{Wojciech Zaremba}.} \bibinfo{year}{2021}\natexlab{}.
\newblock \showarticletitle{Evaluating large language models trained on code}.
\newblock \bibinfo{journal}{\emph{arXiv preprint arXiv:2107.03374}} (\bibinfo{year}{2021}).
\newblock


\bibitem[Chung et~al\mbox{.}(2022)]%
        {chung2022scaling_flan}
\bibfield{author}{\bibinfo{person}{Hyung~Won Chung}, \bibinfo{person}{Le Hou}, \bibinfo{person}{Shayne Longpre}, \bibinfo{person}{Barret Zoph}, \bibinfo{person}{Yi Tay}, \bibinfo{person}{William Fedus}, \bibinfo{person}{Yunxuan Li}, \bibinfo{person}{Xuezhi Wang}, \bibinfo{person}{Mostafa Dehghani}, \bibinfo{person}{Siddhartha Brahma}, \bibinfo{person}{Albert Webson}, \bibinfo{person}{Shixiang~Shane Gu}, \bibinfo{person}{Zhuyun Dai}, \bibinfo{person}{Mirac Suzgun}, \bibinfo{person}{Xinyun Chen}, \bibinfo{person}{Aakanksha Chowdhery}, \bibinfo{person}{Alex Castro-Ros}, \bibinfo{person}{Marie Pellat}, \bibinfo{person}{Kevin Robinson}, \bibinfo{person}{Dasha Valter}, \bibinfo{person}{Sharan Narang}, \bibinfo{person}{Gaurav Mishra}, \bibinfo{person}{Adams Yu}, \bibinfo{person}{Vincent Zhao}, \bibinfo{person}{Yanping Huang}, \bibinfo{person}{Andrew Dai}, \bibinfo{person}{Hongkun Yu}, \bibinfo{person}{Slav Petrov}, \bibinfo{person}{Ed~H. Chi}, \bibinfo{person}{Jeff Dean}, \bibinfo{person}{Jacob Devlin},
  \bibinfo{person}{Adam Roberts}, \bibinfo{person}{Denny Zhou}, \bibinfo{person}{Quoc~V. Le}, {and} \bibinfo{person}{Jason Wei}.} \bibinfo{year}{2022}\natexlab{}.
\newblock \showarticletitle{Scaling instruction-finetuned language models}.
\newblock \bibinfo{journal}{\emph{arXiv preprint arXiv:2210.11416}} (\bibinfo{year}{2022}).
\newblock


\bibitem[Cohen(1960)]%
        {cohen1960coefficient}
\bibfield{author}{\bibinfo{person}{Jacob Cohen}.} \bibinfo{year}{1960}\natexlab{}.
\newblock \showarticletitle{A coefficient of agreement for nominal scales}.
\newblock \bibinfo{journal}{\emph{Educational and Psychological Measurement}} \bibinfo{volume}{20}, \bibinfo{number}{1} (\bibinfo{year}{1960}), \bibinfo{pages}{37--46}.
\newblock


\bibitem[Crowdmark(2022)]%
        {crowdmark2022}
\bibfield{author}{\bibinfo{person}{Crowdmark}.} \bibinfo{year}{2022}\natexlab{}.
\newblock \bibinfo{howpublished}{\url{https://crowdmark.com/privacy/}}.
\newblock


\bibitem[Dai et~al\mbox{.}(2023)]%
        {dai2023laiw}
\bibfield{author}{\bibinfo{person}{Yongfu Dai}, \bibinfo{person}{Duanyu Feng}, \bibinfo{person}{Jimin Huang}, \bibinfo{person}{Haochen Jia}, \bibinfo{person}{Qianqian Xie}, \bibinfo{person}{Yifang Zhang}, \bibinfo{person}{Weiguang Han}, \bibinfo{person}{Wei Tian}, {and} \bibinfo{person}{Hao Wang}.} \bibinfo{year}{2023}\natexlab{}.
\newblock \showarticletitle{LAiW: A Chinese legal large language models benchmark (a technical report)}.
\newblock \bibinfo{journal}{\emph{arXiv preprint arXiv:2310.05620}} (\bibinfo{year}{2023}).
\newblock


\bibitem[Datta et~al\mbox{.}(2011)]%
        {datta2011understanding}
\bibfield{author}{\bibinfo{person}{Anupam Datta}, \bibinfo{person}{Jeremiah Blocki}, \bibinfo{person}{Nicolas Christin}, \bibinfo{person}{Henry DeYoung}, \bibinfo{person}{Deepak Garg}, \bibinfo{person}{Limin Jia}, \bibinfo{person}{Dilsun Kaynar}, {and} \bibinfo{person}{Arunesh Sinha}.} \bibinfo{year}{2011}\natexlab{}.
\newblock \showarticletitle{Understanding and protecting privacy: Formal semantics and principled audit mechanisms}. In \bibinfo{booktitle}{\emph{International Conference on Information Systems Security}}. Springer, \bibinfo{pages}{1--27}.
\newblock


\bibitem[Devlin et~al\mbox{.}(2018)]%
        {devlin2018bert}
\bibfield{author}{\bibinfo{person}{Jacob Devlin}, \bibinfo{person}{Ming-Wei Chang}, \bibinfo{person}{Kenton Lee}, {and} \bibinfo{person}{Kristina Toutanova}.} \bibinfo{year}{2018}\natexlab{}.
\newblock \showarticletitle{Bert: Pre-training of deep bidirectional transformers for language understanding}.
\newblock \bibinfo{journal}{\emph{arXiv preprint arXiv:1810.04805}} (\bibinfo{year}{2018}).
\newblock


\bibitem[Dropbox(2022)]%
        {dropbox2022}
\bibfield{author}{\bibinfo{person}{Dropbox}.} \bibinfo{year}{2022}\natexlab{}.
\newblock \bibinfo{howpublished}{\url{https://www.dropbox.com/privacy}}.
\newblock


\bibitem[Druce et~al\mbox{.}(2021)]%
        {druce2021brittleaicausalconfusion}
\bibfield{author}{\bibinfo{person}{Jeff Druce}, \bibinfo{person}{James Niehaus}, \bibinfo{person}{Vanessa Moody}, \bibinfo{person}{David Jensen}, {and} \bibinfo{person}{Michael~L Littman}.} \bibinfo{year}{2021}\natexlab{}.
\newblock \showarticletitle{Brittle AI, causal confusion, and bad mental models: Challenges and successes in the XAI program}.
\newblock \bibinfo{journal}{\emph{arXiv preprint arXiv:2106.05506}} (\bibinfo{year}{2021}).
\newblock


\bibitem[Facebook(2022)]%
        {facebook2022}
\bibfield{author}{\bibinfo{person}{Facebook}.} \bibinfo{year}{2022}\natexlab{}.
\newblock \bibinfo{howpublished}{\url{https://m.facebook.com/privacy/explanation/}}.
\newblock


\bibitem[{Facebook Research}(2023)]%
        {llamarecipes}
\bibfield{author}{\bibinfo{person}{{Facebook Research}}.} \bibinfo{year}{2023}\natexlab{}.
\newblock \bibinfo{title}{Llama Recipes}.
\newblock \bibinfo{howpublished}{\url{https://github.com/facebookresearch/llama-recipes}}.
\newblock


\bibitem[Fei et~al\mbox{.}(2023)]%
        {fei2023lawbench}
\bibfield{author}{\bibinfo{person}{Zhiwei Fei}, \bibinfo{person}{Xiaoyu Shen}, \bibinfo{person}{Dawei Zhu}, \bibinfo{person}{Fengzhe Zhou}, \bibinfo{person}{Zhuo Han}, \bibinfo{person}{Songyang Zhang}, \bibinfo{person}{Kai Chen}, \bibinfo{person}{Zongwen Shen}, {and} \bibinfo{person}{Jidong Ge}.} \bibinfo{year}{2023}\natexlab{}.
\newblock \showarticletitle{LawBench: Benchmarking legal knowledge of large language models}.
\newblock \bibinfo{journal}{\emph{arXiv preprint arXiv:2309.16289}} (\bibinfo{year}{2023}).
\newblock


\bibitem[Frischmann et~al\mbox{.}(2014)]%
        {frischmann2014governing}
\bibfield{author}{\bibinfo{person}{Brett~M Frischmann}, \bibinfo{person}{Michael~J Madison}, {and} \bibinfo{person}{Katherine~Jo Strandburg}.} \bibinfo{year}{2014}\natexlab{}.
\newblock \bibinfo{booktitle}{\emph{Governing knowledge commons}}.
\newblock \bibinfo{publisher}{Oxford University Press}.
\newblock


\bibitem[Gao et~al\mbox{.}(2021)]%
        {gao2021making}
\bibfield{author}{\bibinfo{person}{Tianyu Gao}, \bibinfo{person}{Adam Fisch}, {and} \bibinfo{person}{Danqi Chen}.} \bibinfo{year}{2021}\natexlab{}.
\newblock \showarticletitle{Making pre-trained language models better few-shot learners}.
\newblock \bibinfo{journal}{\emph{arXiv preprint arXiv:2012.15723}} (\bibinfo{year}{2021}).
\newblock


\bibitem[Gradescope(2022)]%
        {gradescope2022}
\bibfield{author}{\bibinfo{person}{Gradescope}.} \bibinfo{year}{2022}\natexlab{}.
\newblock \bibinfo{howpublished}{\url{https://www.gradescope.com/privacy}}.
\newblock


\bibitem[Harkous et~al\mbox{.}(2018)]%
        {harkous2018polisis}
\bibfield{author}{\bibinfo{person}{Hamza Harkous}, \bibinfo{person}{Kassem Fawaz}, \bibinfo{person}{R{\'e}mi Lebret}, \bibinfo{person}{Florian Schaub}, \bibinfo{person}{Kang~G Shin}, {and} \bibinfo{person}{Karl Aberer}.} \bibinfo{year}{2018}\natexlab{}.
\newblock \showarticletitle{Polisis: Automated analysis and presentation of privacy policies using deep learning}. In \bibinfo{booktitle}{\emph{27th USENIX Security Symposium (USENIX Security 18)}}. \bibinfo{pages}{531--548}.
\newblock


\bibitem[Hoffmann et~al\mbox{.}(2022)]%
        {hoffmann2022training}
\bibfield{author}{\bibinfo{person}{Jordan Hoffmann}, \bibinfo{person}{Sebastian Borgeaud}, \bibinfo{person}{Arthur Mensch}, \bibinfo{person}{Elena Buchatskaya}, \bibinfo{person}{Trevor Cai}, \bibinfo{person}{Eliza Rutherford}, \bibinfo{person}{Diego de Las~Casas}, \bibinfo{person}{Lisa~Anne Hendricks}, \bibinfo{person}{Johannes Welbl}, \bibinfo{person}{Aidan Clark}, \bibinfo{person}{Tom Hennigan}, \bibinfo{person}{Eric Noland}, \bibinfo{person}{Katie Millican}, \bibinfo{person}{George van~den Driessche}, \bibinfo{person}{Bogdan Damoc}, \bibinfo{person}{Aurelia Guy}, \bibinfo{person}{Simon Osindero}, \bibinfo{person}{Karen Simonyan}, \bibinfo{person}{Erich Elsen}, \bibinfo{person}{Jack~W. Rae}, \bibinfo{person}{Oriol Vinyals}, {and} \bibinfo{person}{Laurent Sifre}.} \bibinfo{year}{2022}\natexlab{}.
\newblock \showarticletitle{Training compute-optimal large language models}.
\newblock \bibinfo{journal}{\emph{arXiv preprint arXiv:2203.15556}} (\bibinfo{year}{2022}).
\newblock


\bibitem[Holtzman et~al\mbox{.}(2020)]%
        {holtzman2020curious}
\bibfield{author}{\bibinfo{person}{Ari Holtzman}, \bibinfo{person}{Jan Buys}, \bibinfo{person}{Li Du}, \bibinfo{person}{Maxwell Forbes}, {and} \bibinfo{person}{Yejin Choi}.} \bibinfo{year}{2020}\natexlab{}.
\newblock \showarticletitle{The curious case of neural text degeneration}.
\newblock \bibinfo{journal}{\emph{arXiv preprint arXiv:1904.09751}} (\bibinfo{year}{2020}).
\newblock


\bibitem[Honorlock(2022)]%
        {honorlock2022}
\bibfield{author}{\bibinfo{person}{Honorlock}.} \bibinfo{year}{2022}\natexlab{}.
\newblock \bibinfo{howpublished}{\url{https://honorlock.com/wp-content/uploads/2021/05/May2021_Honorlock_App_Privacy_Policy.docx.pdf}}.
\newblock


\bibitem[Hu et~al\mbox{.}(2021)]%
        {hu2021lora}
\bibfield{author}{\bibinfo{person}{Edward~J Hu}, \bibinfo{person}{Yelong Shen}, \bibinfo{person}{Phillip Wallis}, \bibinfo{person}{Zeyuan Allen-Zhu}, \bibinfo{person}{Yuanzhi Li}, \bibinfo{person}{Shean Wang}, \bibinfo{person}{Lu Wang}, {and} \bibinfo{person}{Weizhu Chen}.} \bibinfo{year}{2021}\natexlab{}.
\newblock \showarticletitle{LoRA: Low-rank adaptation of large language models}.
\newblock \bibinfo{journal}{\emph{arXiv preprint arXiv:2106.09685}} (\bibinfo{year}{2021}).
\newblock


\bibitem[Jensen and Potts(2004)]%
        {jensen2004privacy}
\bibfield{author}{\bibinfo{person}{Carlos Jensen} {and} \bibinfo{person}{Colin Potts}.} \bibinfo{year}{2004}\natexlab{}.
\newblock \showarticletitle{Privacy policies as decision-making tools: an evaluation of online privacy notices}. In \bibinfo{booktitle}{\emph{Proceedings of the SIGCHI conference on Human Factors in Computing Systems}}. \bibinfo{pages}{471--478}.
\newblock


\bibitem[Kaplan et~al\mbox{.}(2020)]%
        {kaplan2020scaling}
\bibfield{author}{\bibinfo{person}{Jared Kaplan}, \bibinfo{person}{Sam McCandlish}, \bibinfo{person}{Tom Henighan}, \bibinfo{person}{Tom~B Brown}, \bibinfo{person}{Benjamin Chess}, \bibinfo{person}{Rewon Child}, \bibinfo{person}{Scott Gray}, \bibinfo{person}{Alec Radford}, \bibinfo{person}{Jeffrey Wu}, {and} \bibinfo{person}{Dario Amodei}.} \bibinfo{year}{2020}\natexlab{}.
\newblock \showarticletitle{Scaling laws for neural language models}.
\newblock \bibinfo{journal}{\emph{arXiv preprint arXiv:2001.08361}} (\bibinfo{year}{2020}).
\newblock


\bibitem[Keahey et~al\mbox{.}(2020)]%
        {keahey2020lessons}
\bibfield{author}{\bibinfo{person}{Kate Keahey}, \bibinfo{person}{Jason Anderson}, \bibinfo{person}{Zhuo Zhen}, \bibinfo{person}{Pierre Riteau}, \bibinfo{person}{Paul Ruth}, \bibinfo{person}{Dan Stanzione}, \bibinfo{person}{Mert Cevik}, \bibinfo{person}{Jacob Colleran}, \bibinfo{person}{Haryadi~S. Gunawi}, \bibinfo{person}{Cody Hammock}, \bibinfo{person}{Joe Mambretti}, \bibinfo{person}{Alexander Barnes}, \bibinfo{person}{Fran\c{c}ois Halbach}, \bibinfo{person}{Alex Rocha}, {and} \bibinfo{person}{Joe Stubbs}.} \bibinfo{year}{2020}\natexlab{}.
\newblock \showarticletitle{Lessons learned from the chameleon testbed}.
\newblock In \bibinfo{booktitle}{\emph{Proceedings of the 2020 USENIX Annual Technical Conference (USENIX ATC '20)}}. \bibinfo{publisher}{USENIX Association}.
\newblock


\bibitem[Kultura2022(2022)]%
        {kultura2022}
\bibfield{author}{\bibinfo{person}{Kultura2022}.} \bibinfo{year}{2022}\natexlab{}.
\newblock \bibinfo{howpublished}{\url{https://corp.kaltura.com/legal/privacy/privacy-policy/}}.
\newblock


\bibitem[Kumar et~al\mbox{.}(2024)]%
        {kumar2024roadmap}
\bibfield{author}{\bibinfo{person}{Priya~C Kumar}, \bibinfo{person}{Michael Zimmer}, {and} \bibinfo{person}{Jessica Vitak}.} \bibinfo{year}{2024}\natexlab{}.
\newblock \showarticletitle{A roadmap for applying the contextual integrity framework in qualitative privacy research}.
\newblock \bibinfo{journal}{\emph{Proceedings of the ACM on Human-Computer Interaction}} \bibinfo{volume}{8}, \bibinfo{number}{CSCW1} (\bibinfo{year}{2024}), \bibinfo{pages}{1--29}.
\newblock


\bibitem[LinkedIn(2022)]%
        {linkedin2022}
\bibfield{author}{\bibinfo{person}{LinkedIn}.} \bibinfo{year}{2022}\natexlab{}.
\newblock \bibinfo{howpublished}{\url{https://www.linkedin.com/legal/privacy-policy}}.
\newblock


\bibitem[Liu et~al\mbox{.}(2014)]%
        {liu2014step}
\bibfield{author}{\bibinfo{person}{Fei Liu}, \bibinfo{person}{Rohan Ramanath}, \bibinfo{person}{Norman Sadeh}, {and} \bibinfo{person}{Noah~A Smith}.} \bibinfo{year}{2014}\natexlab{}.
\newblock \showarticletitle{A step towards usable privacy policy: Automatic alignment of privacy statements}. In \bibinfo{booktitle}{\emph{Proceedings of COLING 2014, the 25th International Conference on Computational Linguistics: Technical Papers}}. \bibinfo{pages}{884--894}.
\newblock


\bibitem[Matlab(2022)]%
        {matlab2022}
\bibfield{author}{\bibinfo{person}{Matlab}.} \bibinfo{year}{2022}\natexlab{}.
\newblock \bibinfo{howpublished}{\url{https://www.mathworks.com/company/aboutus/policies_statements/privacy-policy.html}}.
\newblock


\bibitem[McDonald and Cranor(2008)]%
        {mcdonald2008cost}
\bibfield{author}{\bibinfo{person}{Aleecia~M McDonald} {and} \bibinfo{person}{Lorrie~Faith Cranor}.} \bibinfo{year}{2008}\natexlab{}.
\newblock \showarticletitle{The cost of reading privacy policies}.
\newblock \bibinfo{journal}{\emph{I/S: A Journal of Law and Policy for the Information Society}}  \bibinfo{volume}{4} (\bibinfo{year}{2008}), \bibinfo{pages}{543}.
\newblock


\bibitem[Niantic(2022)]%
        {niantic2022}
\bibfield{author}{\bibinfo{person}{Niantic}.} \bibinfo{year}{2022}\natexlab{}.
\newblock \bibinfo{howpublished}{\url{https://nianticlabs.com/privacy/en/}}.
\newblock


\bibitem[Nissenbaum(2009)]%
        {nissenbaum2009privacy}
\bibfield{author}{\bibinfo{person}{Helen Nissenbaum}.} \bibinfo{year}{2009}\natexlab{}.
\newblock \bibinfo{booktitle}{\emph{Privacy in Context}}.
\newblock \bibinfo{publisher}{Stanford University Press}.
\newblock


\bibitem[Ouyang et~al\mbox{.}(2022)]%
        {ouyang2022training}
\bibfield{author}{\bibinfo{person}{Long Ouyang}, \bibinfo{person}{Jeff Wu}, \bibinfo{person}{Xu Jiang}, \bibinfo{person}{Diogo Almeida}, \bibinfo{person}{Carroll~L. Wainwright}, \bibinfo{person}{Pamela Mishkin}, \bibinfo{person}{Chong Zhang}, \bibinfo{person}{Sandhini Agarwal}, \bibinfo{person}{Katarina Slama}, \bibinfo{person}{Alex Ray}, \bibinfo{person}{John Schulman}, \bibinfo{person}{Jacob Hilton}, \bibinfo{person}{Fraser Kelton}, \bibinfo{person}{Luke Miller}, \bibinfo{person}{Maddie Simens}, \bibinfo{person}{Amanda Askell}, \bibinfo{person}{Peter Welinder}, \bibinfo{person}{Paul Christiano}, \bibinfo{person}{Jan Leike}, {and} \bibinfo{person}{Ryan Lowe}.} \bibinfo{year}{2022}\natexlab{}.
\newblock \showarticletitle{Training language models to follow instructions with human feedback}.
\newblock \bibinfo{journal}{\emph{Advances in Neural Information Processing Systems}}  \bibinfo{volume}{35} (\bibinfo{year}{2022}), \bibinfo{pages}{27730--27744}.
\newblock


\bibitem[Packback(2022)]%
        {packback2022}
\bibfield{author}{\bibinfo{person}{Packback}.} \bibinfo{year}{2022}\natexlab{}.
\newblock \bibinfo{howpublished}{\url{https://www.packback.co/site/privacy/}}.
\newblock


\bibitem[Panopto(2022)]%
        {panopto2022}
\bibfield{author}{\bibinfo{person}{Panopto}.} \bibinfo{year}{2022}\natexlab{}.
\newblock \bibinfo{howpublished}{\url{https://www.panopto.com/privacy/}}.
\newblock


\bibitem[Perez et~al\mbox{.}(2018)]%
        {perez2018review}
\bibfield{author}{\bibinfo{person}{Alfredo~J Perez}, \bibinfo{person}{Sherali Zeadally}, {and} \bibinfo{person}{Jonathan Cochran}.} \bibinfo{year}{2018}\natexlab{}.
\newblock \showarticletitle{A review and an empirical analysis of privacy policy and notices for consumer Internet of things}.
\newblock \bibinfo{journal}{\emph{Security and Privacy}} \bibinfo{volume}{1}, \bibinfo{number}{3} (\bibinfo{year}{2018}), \bibinfo{pages}{e15}.
\newblock


\bibitem[Pochat et~al\mbox{.}(2018)]%
        {pochat2018tranco}
\bibfield{author}{\bibinfo{person}{Victor~Le Pochat}, \bibinfo{person}{Tom Van~Goethem}, \bibinfo{person}{Samaneh Tajalizadehkhoob}, \bibinfo{person}{Maciej Korczy{\'n}ski}, {and} \bibinfo{person}{Wouter Joosen}.} \bibinfo{year}{2018}\natexlab{}.
\newblock \showarticletitle{Tranco: A research-oriented top sites ranking hardened against manipulation}.
\newblock \bibinfo{journal}{\emph{arXiv preprint arXiv:1806.01156}} (\bibinfo{year}{2018}).
\newblock


\bibitem[Press et~al\mbox{.}(2022)]%
        {press2022train}
\bibfield{author}{\bibinfo{person}{Ofir Press}, \bibinfo{person}{Noah~A Smith}, {and} \bibinfo{person}{Mike Lewis}.} \bibinfo{year}{2022}\natexlab{}.
\newblock \showarticletitle{Train short, test long: Attention with linear biases enables input length extrapolation}.
\newblock \bibinfo{journal}{\emph{arXiv preprint arXiv:2108.12409}} (\bibinfo{year}{2022}).
\newblock


\bibitem[Proctorio(2022)]%
        {proctorio2022}
\bibfield{author}{\bibinfo{person}{Proctorio}.} \bibinfo{year}{2022}\natexlab{}.
\newblock \bibinfo{howpublished}{\url{https://proctorio.com/privacy}}.
\newblock


\bibitem[Radford and Narasimhan(2018)]%
        {Radford2018ImprovingLU}
\bibfield{author}{\bibinfo{person}{Alec Radford} {and} \bibinfo{person}{Karthik Narasimhan}.} \bibinfo{year}{2018}\natexlab{}.
\newblock \showarticletitle{Improving language understanding by generative pre-training}.
\newblock
\urldef\tempurl%
\url{https://api.semanticscholar.org/CorpusID:49313245}
\showURL{%
\tempurl}


\bibitem[Radford et~al\mbox{.}(2019)]%
        {Radford2019LanguageMA_gpt2}
\bibfield{author}{\bibinfo{person}{Alec Radford}, \bibinfo{person}{Jeff Wu}, \bibinfo{person}{Rewon Child}, \bibinfo{person}{David Luan}, \bibinfo{person}{Dario Amodei}, {and} \bibinfo{person}{Ilya Sutskever}.} \bibinfo{year}{2019}\natexlab{}.
\newblock \showarticletitle{Language models are unsupervised multitask learners}.
\newblock
\urldef\tempurl%
\url{https://api.semanticscholar.org/CorpusID:160025533}
\showURL{%
\tempurl}


\bibitem[Ravichander et~al\mbox{.}(2019)]%
        {ravichander2019question}
\bibfield{author}{\bibinfo{person}{Abhilasha Ravichander}, \bibinfo{person}{Alan~W Black}, \bibinfo{person}{Shomir Wilson}, \bibinfo{person}{Thomas Norton}, {and} \bibinfo{person}{Norman Sadeh}.} \bibinfo{year}{2019}\natexlab{}.
\newblock \showarticletitle{Question answering for privacy policies: Combining computational and legal perspectives}.
\newblock \bibinfo{journal}{\emph{arXiv preprint arXiv:1911.00841}} (\bibinfo{year}{2019}).
\newblock


\bibitem[Reidenberg et~al\mbox{.}(2016)]%
        {reidenberg2016ambiguity}
\bibfield{author}{\bibinfo{person}{Joel~R Reidenberg}, \bibinfo{person}{Jaspreet Bhatia}, \bibinfo{person}{Travis~D Breaux}, {and} \bibinfo{person}{Thomas~B Norton}.} \bibinfo{year}{2016}\natexlab{}.
\newblock \showarticletitle{Ambiguity in privacy policies and the impact of regulation}.
\newblock \bibinfo{journal}{\emph{The Journal of Legal Studies}} \bibinfo{volume}{45}, \bibinfo{number}{S2} (\bibinfo{year}{2016}), \bibinfo{pages}{S163--S190}.
\newblock


\bibitem[Reidenberg et~al\mbox{.}(2015)]%
        {reidenberg2015disagreeable}
\bibfield{author}{\bibinfo{person}{Joel~R Reidenberg}, \bibinfo{person}{Travis Breaux}, \bibinfo{person}{Lorrie~Faith Cranor}, \bibinfo{person}{Brian French}, \bibinfo{person}{Amanda Grannis}, \bibinfo{person}{James~T Graves}, \bibinfo{person}{Fei Liu}, \bibinfo{person}{Aleecia McDonald}, \bibinfo{person}{Thomas~B Norton}, {and} \bibinfo{person}{Rohan Ramanath}.} \bibinfo{year}{2015}\natexlab{}.
\newblock \showarticletitle{Disagreeable privacy policies: Mismatches between meaning and users' understanding}.
\newblock \bibinfo{journal}{\emph{Berkeley Technology Law Journal}}  \bibinfo{volume}{30} (\bibinfo{year}{2015}), \bibinfo{pages}{39}.
\newblock


\bibitem[Robertson(2024)]%
        {pytorchrnn}
\bibfield{author}{\bibinfo{person}{Sean Robertson}.} \bibinfo{year}{2024}\natexlab{}.
\newblock
\newblock
\urldef\tempurl%
\url{https://pytorch.org/tutorials/intermediate/char_rnn_generation_tutorial.html}
\showURL{%
\tempurl}


\bibitem[Rudolph et~al\mbox{.}(2018)]%
        {rudolph2018users}
\bibfield{author}{\bibinfo{person}{Manuel Rudolph}, \bibinfo{person}{Denis Feth}, {and} \bibinfo{person}{Svenja Polst}.} \bibinfo{year}{2018}\natexlab{}.
\newblock \showarticletitle{Why users ignore privacy policies--a survey and intention model for explaining user privacy behavior}. In \bibinfo{booktitle}{\emph{International Conference on Human-Computer Interaction}}. Springer, \bibinfo{pages}{587--598}.
\newblock


\bibitem[Sadeh et~al\mbox{.}(2013)]%
        {sadeh2013usable}
\bibfield{author}{\bibinfo{person}{Norman Sadeh}, \bibinfo{person}{Alessandro Acquisti}, \bibinfo{person}{Travis~D Breaux}, \bibinfo{person}{Lorrie~Faith Cranor}, \bibinfo{person}{Aleecia~M McDonald}, \bibinfo{person}{Joel~R Reidenberg}, \bibinfo{person}{Noah~A Smith}, \bibinfo{person}{Fei Liu}, \bibinfo{person}{N~Cameron Russell}, \bibinfo{person}{Florian Schaub}, {et~al\mbox{.}}} \bibinfo{year}{2013}\natexlab{}.
\newblock \showarticletitle{The usable privacy policy project}.
\newblock In \bibinfo{booktitle}{\emph{Technical Report, CMU-ISR-13-119}}. \bibinfo{publisher}{Carnegie Mellon University}.
\newblock


\bibitem[Sanfilippo et~al\mbox{.}(2018)]%
        {sanfilippo2018privacy}
\bibfield{author}{\bibinfo{person}{Madelyn Sanfilippo}, \bibinfo{person}{Brett Frischmann}, {and} \bibinfo{person}{Katherine Standburg}.} \bibinfo{year}{2018}\natexlab{}.
\newblock \showarticletitle{Privacy as commons: Case evaluation through the governing knowledge commons framework}.
\newblock \bibinfo{journal}{\emph{Journal of Information Policy}} \bibinfo{volume}{8}, \bibinfo{number}{1} (\bibinfo{year}{2018}), \bibinfo{pages}{116--166}.
\newblock


\bibitem[Scao and Rush(2021)]%
        {scao2021data}
\bibfield{author}{\bibinfo{person}{Teven~Le Scao} {and} \bibinfo{person}{Alexander~M Rush}.} \bibinfo{year}{2021}\natexlab{}.
\newblock \showarticletitle{How many data points is a prompt worth?}
\newblock \bibinfo{journal}{\emph{arXiv preprint arXiv:2103.08493}} (\bibinfo{year}{2021}).
\newblock


\bibitem[Shaffer(2021)]%
        {shaffer2021applying}
\bibfield{author}{\bibinfo{person}{Gwen Shaffer}.} \bibinfo{year}{2021}\natexlab{}.
\newblock \showarticletitle{Applying a contextual integrity framework to privacy policies for smart technologies}.
\newblock \bibinfo{journal}{\emph{Journal of Information Policy}}  \bibinfo{volume}{11} (\bibinfo{year}{2021}), \bibinfo{pages}{222--265}.
\newblock


\bibitem[Shvartzshnaider et~al\mbox{.}(2019a)]%
        {shvartzshnaider2019going}
\bibfield{author}{\bibinfo{person}{Yan Shvartzshnaider}, \bibinfo{person}{Noah Apthorpe}, \bibinfo{person}{Nick Feamster}, {and} \bibinfo{person}{Helen Nissenbaum}.} \bibinfo{year}{2019}\natexlab{a}.
\newblock \showarticletitle{Going against the (appropriate) flow: A contextual integrity approach to privacy policy analysis}. In \bibinfo{booktitle}{\emph{Proceedings of the AAAI Conference on Human Computation and Crowdsourcing}}, Vol.~\bibinfo{volume}{7}. \bibinfo{pages}{162--170}.
\newblock


\bibitem[Shvartzshnaider et~al\mbox{.}(2019b)]%
        {shvartzshnaider2019vaccine}
\bibfield{author}{\bibinfo{person}{Yan Shvartzshnaider}, \bibinfo{person}{Zvonimir Pavlinovic}, \bibinfo{person}{Ananth Balashankar}, \bibinfo{person}{Thomas Wies}, \bibinfo{person}{Lakshminarayanan Subramanian}, \bibinfo{person}{Helen Nissenbaum}, {and} \bibinfo{person}{Prateek Mittal}.} \bibinfo{year}{2019}\natexlab{b}.
\newblock \showarticletitle{Vaccine: Using contextual integrity for data leakage detection}. In \bibinfo{booktitle}{\emph{The World Wide Web Conference}}. \bibinfo{pages}{1702--1712}.
\newblock


\bibitem[Shvartzshnaider et~al\mbox{.}(2022)]%
        {shvartzshnaider2022gkc}
\bibfield{author}{\bibinfo{person}{Yan Shvartzshnaider}, \bibinfo{person}{Madelyn~Rose Sanfilippo}, {and} \bibinfo{person}{Noah Apthorpe}.} \bibinfo{year}{2022}\natexlab{}.
\newblock \showarticletitle{GKC-CI: A unifying framework for contextual norms and information governance}.
\newblock \bibinfo{journal}{\emph{Journal of the Association for Information Science and Technology}} (\bibinfo{year}{2022}).
\newblock


\bibitem[Shvartzshnaider et~al\mbox{.}(2016)]%
        {shvartzshnaider2016learning}
\bibfield{author}{\bibinfo{person}{Yan Shvartzshnaider}, \bibinfo{person}{Schrasing Tong}, \bibinfo{person}{Thomas Wies}, \bibinfo{person}{Paula Kift}, \bibinfo{person}{Helen Nissenbaum}, \bibinfo{person}{Lakshminarayanan Subramanian}, {and} \bibinfo{person}{Prateek Mittal}.} \bibinfo{year}{2016}\natexlab{}.
\newblock \showarticletitle{Learning privacy expectations by crowdsourcing contextual informational norms}. In \bibinfo{booktitle}{\emph{Proceedings of the AAAI Conference on Human Computation and Crowdsourcing}}, Vol.~\bibinfo{volume}{4}. \bibinfo{pages}{209--218}.
\newblock


\bibitem[Stenetorp et~al\mbox{.}(2012)]%
        {stenetorp2012brat}
\bibfield{author}{\bibinfo{person}{Pontus Stenetorp}, \bibinfo{person}{Sampo Pyysalo}, \bibinfo{person}{Goran Topi{\'c}}, \bibinfo{person}{Tomoko Ohta}, \bibinfo{person}{Sophia Ananiadou}, {and} \bibinfo{person}{Jun’ichi Tsujii}.} \bibinfo{year}{2012}\natexlab{}.
\newblock \showarticletitle{BRAT: A web-based tool for NLP-assisted text annotation}. In \bibinfo{booktitle}{\emph{Proceedings of the Demonstrations at the 13th Conference of the European Chapter of the Association for Computational Linguistics}}. \bibinfo{pages}{102--107}.
\newblock


\bibitem[Story et~al\mbox{.}(2019)]%
        {story2019natural}
\bibfield{author}{\bibinfo{person}{Peter Story}, \bibinfo{person}{Sebastian Zimmeck}, \bibinfo{person}{Abhilasha Ravichander}, \bibinfo{person}{Daniel Smullen}, \bibinfo{person}{Ziqi Wang}, \bibinfo{person}{Joel Reidenberg}, \bibinfo{person}{N~Cameron Russell}, {and} \bibinfo{person}{Norman Sadeh}.} \bibinfo{year}{2019}\natexlab{}.
\newblock \showarticletitle{Natural language processing for mobile app privacy compliance}. In \bibinfo{booktitle}{\emph{AAAI Spring Symposium on Privacy-Enhancing Artificial Intelligence and Language Technologies}}, Vol.~\bibinfo{volume}{2}. \bibinfo{pages}{4}.
\newblock


\bibitem[Stripe(2022)]%
        {stripe2022}
\bibfield{author}{\bibinfo{person}{Stripe}.} \bibinfo{year}{2022}\natexlab{}.
\newblock \bibinfo{howpublished}{\url{https://stripe.com/privacy}}.
\newblock


\bibitem[Tang et~al\mbox{.}(2023)]%
        {tang2023policygpt}
\bibfield{author}{\bibinfo{person}{Chenhao Tang}, \bibinfo{person}{Zhengliang Liu}, \bibinfo{person}{Chong Ma}, \bibinfo{person}{Zihao Wu}, \bibinfo{person}{Yiwei Li}, \bibinfo{person}{Wei Liu}, \bibinfo{person}{Dajiang Zhu}, \bibinfo{person}{Quanzheng Li}, \bibinfo{person}{Xiang Li}, \bibinfo{person}{Tianming Liu}, {and} \bibinfo{person}{Lei Fan}.} \bibinfo{year}{2023}\natexlab{}.
\newblock \showarticletitle{PolicyGPT: Automated analysis of privacy policies with large language models}.
\newblock \bibinfo{journal}{\emph{arXiv preprint arXiv:2309.10238}} (\bibinfo{year}{2023}).
\newblock


\bibitem[Tesfay et~al\mbox{.}(2018)]%
        {tesfay2018privacyguide}
\bibfield{author}{\bibinfo{person}{Welderufael~B Tesfay}, \bibinfo{person}{Peter Hofmann}, \bibinfo{person}{Toru Nakamura}, \bibinfo{person}{Shinsaku Kiyomoto}, {and} \bibinfo{person}{Jetzabel Serna}.} \bibinfo{year}{2018}\natexlab{}.
\newblock \showarticletitle{PrivacyGuide: Towards an implementation of the EU GDPR on internet privacy policy evaluation}. In \bibinfo{booktitle}{\emph{Proceedings of the Fourth ACM International Workshop on Security and Privacy Analytics}}. \bibinfo{pages}{15--21}.
\newblock


\bibitem[{The New York Times}(2022)]%
        {nytimes2022}
\bibfield{author}{\bibinfo{person}{{The New York Times}}.} \bibinfo{year}{2022}\natexlab{}.
\newblock \bibinfo{howpublished}{\url{https://www.nytimes.com/privacy/privacy-policy}}.
\newblock


\bibitem[Touvron et~al\mbox{.}(2023a)]%
        {touvron2023llama}
\bibfield{author}{\bibinfo{person}{Hugo Touvron}, \bibinfo{person}{Thibaut Lavril}, \bibinfo{person}{Gautier Izacard}, \bibinfo{person}{Xavier Martinet}, \bibinfo{person}{Marie-Anne Lachaux}, \bibinfo{person}{Timothée Lacroix}, \bibinfo{person}{Baptiste Rozière}, \bibinfo{person}{Naman Goyal}, \bibinfo{person}{Eric Hambro}, \bibinfo{person}{Faisal Azhar}, \bibinfo{person}{Aurelien Rodriguez}, \bibinfo{person}{Armand Joulin}, \bibinfo{person}{Edouard Grave}, {and} \bibinfo{person}{Guillaume Lample}.} \bibinfo{year}{2023}\natexlab{a}.
\newblock \showarticletitle{LLaMA: Open and efficient foundation language models}.
\newblock \bibinfo{journal}{\emph{arXiv preprint arXiv:2302.13971}} (\bibinfo{year}{2023}).
\newblock


\bibitem[Touvron et~al\mbox{.}(2023b)]%
        {touvron2023llama2}
\bibfield{author}{\bibinfo{person}{Hugo Touvron}, \bibinfo{person}{Louis Martin}, \bibinfo{person}{Kevin Stone}, \bibinfo{person}{Peter Albert}, \bibinfo{person}{Amjad Almahairi}, \bibinfo{person}{Yasmine Babaei}, \bibinfo{person}{Nikolay Bashlykov}, \bibinfo{person}{Soumya Batra}, \bibinfo{person}{Prajjwal Bhargava}, \bibinfo{person}{Shruti Bhosale}, \bibinfo{person}{Dan Bikel}, \bibinfo{person}{Lukas Blecher}, \bibinfo{person}{Cristian~Canton Ferrer}, \bibinfo{person}{Moya Chen}, \bibinfo{person}{Guillem Cucurull}, \bibinfo{person}{David Esiobu}, \bibinfo{person}{Jude Fernandes}, \bibinfo{person}{Jeremy Fu}, \bibinfo{person}{Wenyin Fu}, \bibinfo{person}{Brian Fuller}, \bibinfo{person}{Cynthia Gao}, \bibinfo{person}{Vedanuj Goswami}, \bibinfo{person}{Naman Goyal}, \bibinfo{person}{Anthony Hartshorn}, \bibinfo{person}{Saghar Hosseini}, \bibinfo{person}{Rui Hou}, \bibinfo{person}{Hakan Inan}, \bibinfo{person}{Marcin Kardas}, \bibinfo{person}{Viktor Kerkez}, \bibinfo{person}{Madian Khabsa},
  \bibinfo{person}{Isabel Kloumann}, \bibinfo{person}{Artem Korenev}, \bibinfo{person}{Punit~Singh Koura}, \bibinfo{person}{Marie-Anne Lachaux}, \bibinfo{person}{Thibaut Lavril}, \bibinfo{person}{Jenya Lee}, \bibinfo{person}{Diana Liskovich}, \bibinfo{person}{Yinghai Lu}, \bibinfo{person}{Yuning Mao}, \bibinfo{person}{Xavier Martinet}, \bibinfo{person}{Todor Mihaylov}, \bibinfo{person}{Pushkar Mishra}, \bibinfo{person}{Igor Molybog}, \bibinfo{person}{Yixin Nie}, \bibinfo{person}{Andrew Poulton}, \bibinfo{person}{Jeremy Reizenstein}, \bibinfo{person}{Rashi Rungta}, \bibinfo{person}{Kalyan Saladi}, \bibinfo{person}{Alan Schelten}, \bibinfo{person}{Ruan Silva}, \bibinfo{person}{Eric~Michael Smith}, \bibinfo{person}{Ranjan Subramanian}, \bibinfo{person}{Xiaoqing~Ellen Tan}, \bibinfo{person}{Binh Tang}, \bibinfo{person}{Ross Taylor}, \bibinfo{person}{Adina Williams}, \bibinfo{person}{Jian~Xiang Kuan}, \bibinfo{person}{Puxin Xu}, \bibinfo{person}{Zheng Yan}, \bibinfo{person}{Iliyan Zarov}, \bibinfo{person}{Yuchen
  Zhang}, \bibinfo{person}{Angela Fan}, \bibinfo{person}{Melanie Kambadur}, \bibinfo{person}{Sharan Narang}, \bibinfo{person}{Aurelien Rodriguez}, \bibinfo{person}{Robert Stojnic}, \bibinfo{person}{Sergey Edunov}, {and} \bibinfo{person}{Thomas Scialom}.} \bibinfo{year}{2023}\natexlab{b}.
\newblock \showarticletitle{Llama 2: Open foundation and fine-tuned chat models}.
\newblock \bibinfo{journal}{\emph{arXiv preprint arXiv:2307.09288}} (\bibinfo{year}{2023}).
\newblock


\bibitem[Turnitin(2022)]%
        {turnitin2022}
\bibfield{author}{\bibinfo{person}{Turnitin}.} \bibinfo{year}{2022}\natexlab{}.
\newblock \bibinfo{howpublished}{\url{https://help.turnitin.com/Privacy_and_Security/Privacy_and_Security.htm\#Privacy_Policy}}.
\newblock


\bibitem[Vaswani et~al\mbox{.}(2017)]%
        {vaswani2017attention}
\bibfield{author}{\bibinfo{person}{Ashish Vaswani}, \bibinfo{person}{Noam Shazeer}, \bibinfo{person}{Niki Parmar}, \bibinfo{person}{Jakob Uszkoreit}, \bibinfo{person}{Llion Jones}, \bibinfo{person}{Aidan~N Gomez}, \bibinfo{person}{{\L}ukasz Kaiser}, {and} \bibinfo{person}{Illia Polosukhin}.} \bibinfo{year}{2017}\natexlab{}.
\newblock \showarticletitle{Attention is all you need}.
\newblock \bibinfo{journal}{\emph{Advances in Neural Information Processing Systems}}  \bibinfo{volume}{30} (\bibinfo{year}{2017}).
\newblock


\bibitem[Wagner(2022)]%
        {wagner2022privacy}
\bibfield{author}{\bibinfo{person}{Isabel Wagner}.} \bibinfo{year}{2022}\natexlab{}.
\newblock \showarticletitle{Privacy policies across the ages: Content and readability of privacy policies 1996--2021}.
\newblock \bibinfo{journal}{\emph{arXiv preprint arXiv:2201.08739}} (\bibinfo{year}{2022}).
\newblock


\bibitem[Wan et~al\mbox{.}(2023)]%
        {wan2023reformulating}
\bibfield{author}{\bibinfo{person}{Zhen Wan}, \bibinfo{person}{Yating Zhang}, \bibinfo{person}{Yexiang Wang}, \bibinfo{person}{Fei Cheng}, {and} \bibinfo{person}{Sadao Kurohashi}.} \bibinfo{year}{2023}\natexlab{}.
\newblock \showarticletitle{Reformulating domain adaptation of large language models as adapt-retrieve-revise}.
\newblock \bibinfo{journal}{\emph{arXiv preprint arXiv:2310.03328}} (\bibinfo{year}{2023}).
\newblock


\bibitem[Webson and Pavlick(2022)]%
        {webson2022promptbased}
\bibfield{author}{\bibinfo{person}{Albert Webson} {and} \bibinfo{person}{Ellie Pavlick}.} \bibinfo{year}{2022}\natexlab{}.
\newblock \showarticletitle{Do prompt-based models really understand the meaning of their prompts?}
\newblock \bibinfo{journal}{\emph{arXiv preprint arXiv:2109.01247}} (\bibinfo{year}{2022}).
\newblock


\bibitem[Wei et~al\mbox{.}(2022)]%
        {wei2022finetuned}
\bibfield{author}{\bibinfo{person}{Jason Wei}, \bibinfo{person}{Maarten Bosma}, \bibinfo{person}{Vincent~Y. Zhao}, \bibinfo{person}{Kelvin Guu}, \bibinfo{person}{Adams~Wei Yu}, \bibinfo{person}{Brian Lester}, \bibinfo{person}{Nan Du}, \bibinfo{person}{Andrew~M. Dai}, {and} \bibinfo{person}{Quoc~V. Le}.} \bibinfo{year}{2022}\natexlab{}.
\newblock \showarticletitle{Finetuned language models are zero-shot learners}.
\newblock \bibinfo{journal}{\emph{arXiv preprint arXiv:2109.01652}} (\bibinfo{year}{2022}).
\newblock


\bibitem[Wijesekera et~al\mbox{.}(2015)]%
        {wijesekera2015android}
\bibfield{author}{\bibinfo{person}{Primal Wijesekera}, \bibinfo{person}{Arjun Baokar}, \bibinfo{person}{Ashkan Hosseini}, \bibinfo{person}{Serge Egelman}, \bibinfo{person}{David Wagner}, {and} \bibinfo{person}{Konstantin Beznosov}.} \bibinfo{year}{2015}\natexlab{}.
\newblock \showarticletitle{Android permissions remystified: A field study on contextual integrity}. In \bibinfo{booktitle}{\emph{24th USENIX Security Symposium (USENIX Security 15)}}. \bibinfo{pages}{499--514}.
\newblock


\bibitem[Wilson et~al\mbox{.}(2016a)]%
        {wilson2016creation}
\bibfield{author}{\bibinfo{person}{Shomir Wilson}, \bibinfo{person}{Florian Schaub}, \bibinfo{person}{Aswarth~Abhilash Dara}, \bibinfo{person}{Frederick Liu}, \bibinfo{person}{Sushain Cherivirala}, \bibinfo{person}{Pedro~Giovanni Leon}, \bibinfo{person}{Mads~Schaarup Andersen}, \bibinfo{person}{Sebastian Zimmeck}, \bibinfo{person}{Kanthashree~Mysore Sathyendra}, \bibinfo{person}{N~Cameron Russell}, {et~al\mbox{.}}} \bibinfo{year}{2016}\natexlab{a}.
\newblock \showarticletitle{The creation and analysis of a website privacy policy corpus}. In \bibinfo{booktitle}{\emph{Proceedings of the 54th Annual Meeting of the Association for Computational Linguistics (Volume 1: Long Papers)}}. \bibinfo{pages}{1330--1340}.
\newblock


\bibitem[Wilson et~al\mbox{.}(2018)]%
        {wilson2018analyzing}
\bibfield{author}{\bibinfo{person}{Shomir Wilson}, \bibinfo{person}{Florian Schaub}, \bibinfo{person}{Frederick Liu}, \bibinfo{person}{Kanthashree~Mysore Sathyendra}, \bibinfo{person}{Daniel Smullen}, \bibinfo{person}{Sebastian Zimmeck}, \bibinfo{person}{Rohan Ramanath}, \bibinfo{person}{Peter Story}, \bibinfo{person}{Fei Liu}, \bibinfo{person}{Norman Sadeh}, {et~al\mbox{.}}} \bibinfo{year}{2018}\natexlab{}.
\newblock \showarticletitle{Analyzing privacy policies at scale: From crowdsourcing to automated annotations}.
\newblock \bibinfo{journal}{\emph{ACM Transactions on the Web (TWEB)}} \bibinfo{volume}{13}, \bibinfo{number}{1} (\bibinfo{year}{2018}), \bibinfo{pages}{1--29}.
\newblock


\bibitem[Wilson et~al\mbox{.}(2016b)]%
        {wilson2016crowdsourcing}
\bibfield{author}{\bibinfo{person}{Shomir Wilson}, \bibinfo{person}{Florian Schaub}, \bibinfo{person}{Rohan Ramanath}, \bibinfo{person}{Norman Sadeh}, \bibinfo{person}{Fei Liu}, \bibinfo{person}{Noah~A Smith}, {and} \bibinfo{person}{Frederick Liu}.} \bibinfo{year}{2016}\natexlab{b}.
\newblock \showarticletitle{Crowdsourcing annotations for websites' privacy policies: Can it really work?}. In \bibinfo{booktitle}{\emph{Proceedings of the 25th International Conference on World Wide Web}}. \bibinfo{pages}{133--143}.
\newblock


\bibitem[Winkler and Zeadally(2016)]%
        {winkler2016privacy}
\bibfield{author}{\bibinfo{person}{Stephanie Winkler} {and} \bibinfo{person}{Sherali Zeadally}.} \bibinfo{year}{2016}\natexlab{}.
\newblock \showarticletitle{Privacy policy analysis of popular web platforms}.
\newblock \bibinfo{journal}{\emph{IEEE Technology and Society Magazine}} \bibinfo{volume}{35}, \bibinfo{number}{2} (\bibinfo{year}{2016}), \bibinfo{pages}{75--85}.
\newblock


\bibitem[Yue et~al\mbox{.}(2023)]%
        {yue2023disclawllm}
\bibfield{author}{\bibinfo{person}{Shengbin Yue}, \bibinfo{person}{Wei Chen}, \bibinfo{person}{Siyuan Wang}, \bibinfo{person}{Bingxuan Li}, \bibinfo{person}{Chenchen Shen}, \bibinfo{person}{Shujun Liu}, \bibinfo{person}{Yuxuan Zhou}, \bibinfo{person}{Yao Xiao}, \bibinfo{person}{Song Yun}, \bibinfo{person}{Xuanjing Huang}, {and} \bibinfo{person}{Zhongyu Wei}.} \bibinfo{year}{2023}\natexlab{}.
\newblock \showarticletitle{DISC-LawLLM: Fine-tuning Large Language Models for intelligent legal services}.
\newblock \bibinfo{journal}{\emph{arXiv preprint arXiv:2309.11325}} (\bibinfo{year}{2023}).
\newblock


\bibitem[Zamfirescu-Pereira et~al\mbox{.}(2023)]%
        {Zamfirescu2023Johnny}
\bibfield{author}{\bibinfo{person}{JD Zamfirescu-Pereira}, \bibinfo{person}{Richmond~Y Wong}, \bibinfo{person}{Bjoern Hartmann}, {and} \bibinfo{person}{Qian Yang}.} \bibinfo{year}{2023}\natexlab{}.
\newblock \showarticletitle{Why Johnny can’t prompt: How non-AI experts try (and fail) to design LLM prompts}. In \bibinfo{booktitle}{\emph{Proceedings of the 2023 CHI Conference on Human Factors in Computing Systems}}. \bibinfo{pages}{1--21}.
\newblock


\bibitem[Zimmeck et~al\mbox{.}(2019)]%
        {zimmeck2019maps}
\bibfield{author}{\bibinfo{person}{Sebastian Zimmeck}, \bibinfo{person}{Peter Story}, \bibinfo{person}{Daniel Smullen}, \bibinfo{person}{Abhilasha Ravichander}, \bibinfo{person}{Ziqi Wang}, \bibinfo{person}{Joel Reidenberg}, \bibinfo{person}{N~Cameron Russell}, {and} \bibinfo{person}{Norman Sadeh}.} \bibinfo{year}{2019}\natexlab{}.
\newblock \showarticletitle{MAPS: Scaling privacy compliance analysis to a million apps}.
\newblock \bibinfo{journal}{\emph{Proceedings on Privacy Enhancing Technologies}} \bibinfo{volume}{2019}, \bibinfo{number}{3} (\bibinfo{year}{2019}), \bibinfo{pages}{66--86}.
\newblock


\end{thebibliography}

\newpage
\appendix
\onecolumn 
\label{sec:appendix}

\section{Ground Truth Details}
\label{sec:appendix_tabs}
\begin{table*}[h!]
\footnotesize
\centering
\begin{tabular}{@{}cccccccccc@{}}
\toprule
\textbf{Company} &
  \textbf{Word Count} &
  \textbf{Sender} &
  \textbf{Subject} &
  \textbf{Recipient} &
  \textbf{Attribute} &
  \textbf{Aim} &
  \textbf{Condition} &
  \textbf{Modality} &
  \textbf{Consequence} \\ \midrule
\multicolumn{1}{c|}{Cengage~\cite{cengage2022}}       & \multicolumn{1}{c|}{3340}  & 12 & 15 & 34 & 44  & 64  & 82  & 30 & 0 \\
\multicolumn{1}{c|}{Crowdmark~\cite{crowdmark2022}}   & \multicolumn{1}{c|}{3216}  & 12 & 28 & 37 & 42  & 80  & 92  & 42 & 6 \\
\multicolumn{1}{c|}{Dropbox~\cite{dropbox2022}}       & \multicolumn{1}{c|}{2485}  & 10 & 9  & 20 & 26  & 36  & 30  & 10 & 4 \\
\multicolumn{1}{c|}{Facebook~\cite{facebook2022}}     & \multicolumn{1}{c|}{4151}  & 40 & 48 & 74 & 113 & 78  & 84  & 42 & 0 \\
\multicolumn{1}{c|}{Gradescope~\cite{gradescope2022}} & \multicolumn{1}{c|}{11431} & 18 & 20 & 39 & 69  & 104 & 152 & 28 & 2 \\
\multicolumn{1}{c|}{Honorlock~\cite{honorlock2022}}   & \multicolumn{1}{c|}{1199}  & 10 & 9  & 11 & 22  & 18  & 32  & 22 & 2 \\
\multicolumn{1}{c|}{Kultura~\cite{kultura2022}}       & \multicolumn{1}{c|}{6255}  & 15 & 29 & 54 & 63  & 96  & 164 & 52 & 4 \\
\multicolumn{1}{c|}{LinkedIn~\cite{linkedin2022}}     & \multicolumn{1}{c|}{6298}  & 37 & 58 & 80 & 111 & 110 & 174 & 22 & 0 \\
\multicolumn{1}{c|}{Matlab~\cite{matlab2022}}         & \multicolumn{1}{c|}{5580}  & 27 & 27 & 61 & 85  & 98  & 150 & 44 & 4 \\
\multicolumn{1}{c|}{Niantic~\cite{niantic2022}}       & \multicolumn{1}{c|}{5539}  & 27 & 33 & 44 & 63  & 92  & 94  & 16 & 2 \\
\multicolumn{1}{c|}{NYTimes~\cite{nytimes2022}}       & \multicolumn{1}{c|}{5000}  & 12 & 25 & 41 & 50  & 50  & 82  & 18 & 2 \\
\multicolumn{1}{c|}{Packback~\cite{packback2022}}     & \multicolumn{1}{c|}{4444}  & 11 & 14 & 25 & 35  & 40  & 94  & 18 & 4 \\
\multicolumn{1}{c|}{Panopto~\cite{panopto2022}}       & \multicolumn{1}{c|}{4167}  & 17 & 16 & 28 & 34  & 62  & 82  & 42 & 2 \\
\multicolumn{1}{c|}{Proctorio~\cite{proctorio2022}}   & \multicolumn{1}{c|}{9353}  & 28 & 24 & 61 & 89  & 124 & 140 & 54 & 2 \\
\multicolumn{1}{c|}{Stripe~\cite{stripe2022}}         & \multicolumn{1}{c|}{7460}  & 38 & 48 & 73 & 96  & 110 & 122 & 40 & 4 \\
\multicolumn{1}{c|}{Turnitin~\cite{turnitin2022}}     & \multicolumn{1}{c|}{10220} & 15 & 24 & 24 & 52  & 94  & 90  & 18 & 4 \\ \bottomrule
\end{tabular}
\caption{Number of labeled parameters in ground-truth GKC-CI annotations of 16 privacy policies (rows) from popular websites and e-learning services. All of the labels from the manually-annotated privacy policies were combined together to create the ground-truth dataset used for evaluating all models. There is no difference in which labels were used for training/testing across different models.}

\label{tab:manual-training}
\end{table*}

\section{Brat Annotation Legend}
\label{sec:bratt}
\begin{figure}[h!]
    \centering
    \includegraphics[width = 18cm]{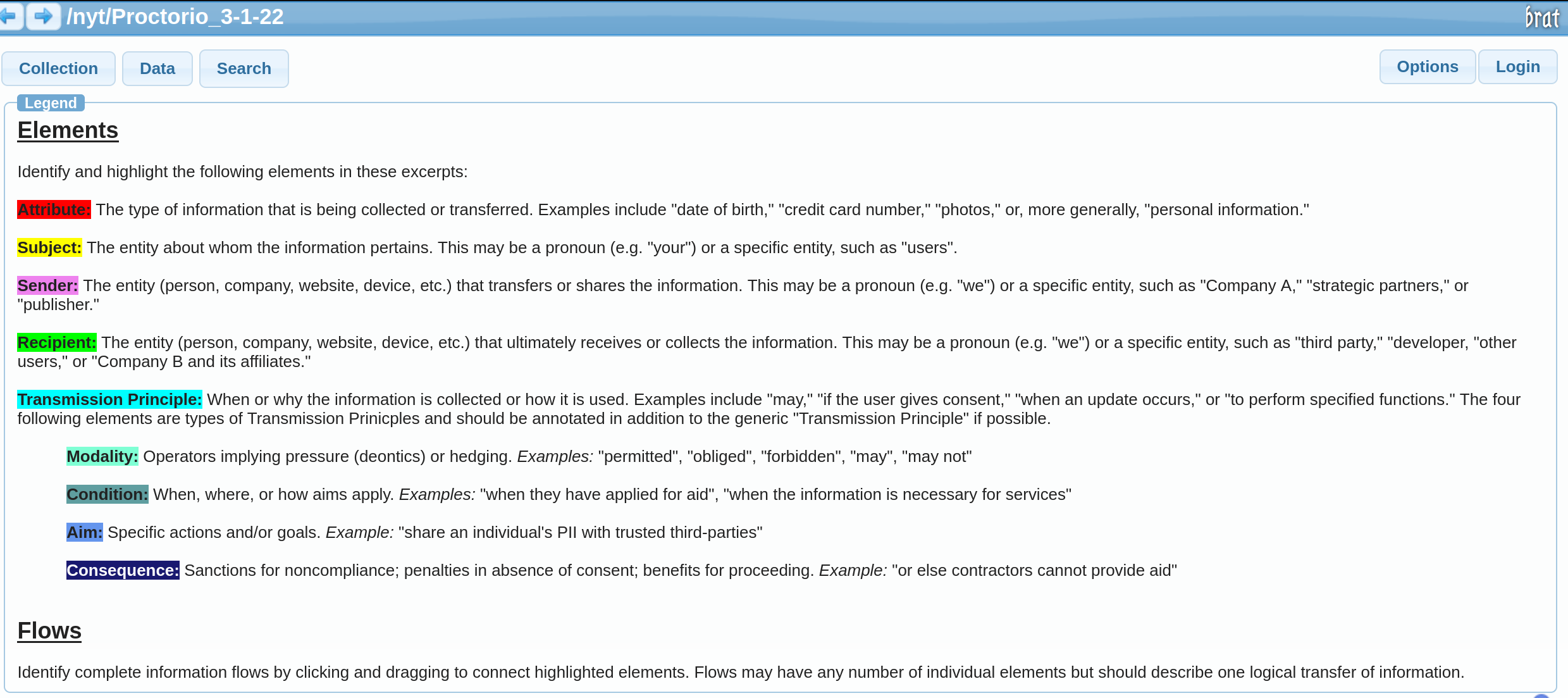}
    \caption{Legend in the customized brat annotation tool~\cite{stenetorp2012brat} for expert annotators to use as reference.}
    \label{fig:enter-label}
\end{figure}

\clearpage

\begin{figure}[h!]
\section{Performance On Sentence Versus Paragraph-Delineated Data As Epochs Vary}
\label{sec:appendix:paragraphs}
    \centering
    \includegraphics[scale=0.5]{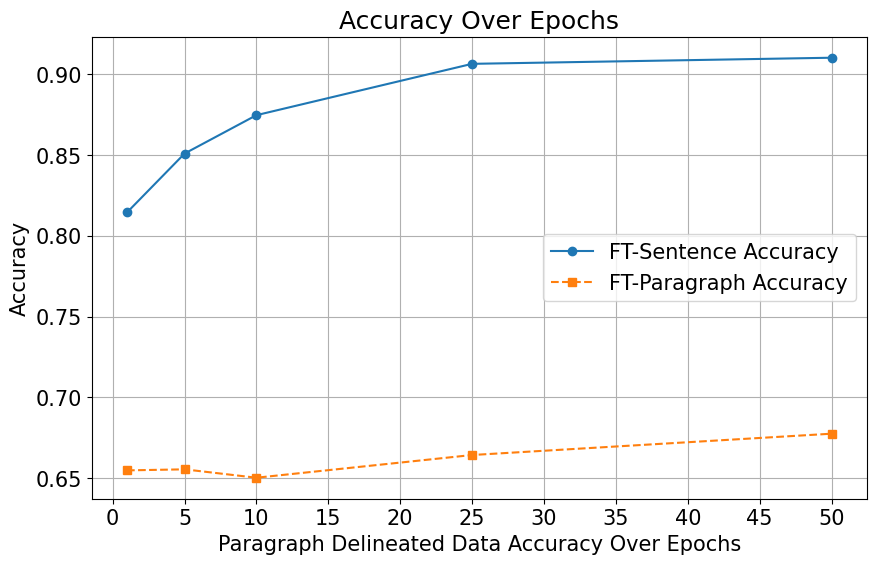}
    \caption{The solid blue line (``FT-Sentence Accuracy'') is the test-set accuracy of \texttt{GPT 3.5TPE} trained on sentence-delineated data at 1, 5, 10, 25, and 50 epochs. The dotted orange line (``FT-Paragraph Accuracy'') is the test-set accuracy of \texttt{GPT 3.5TPE} trained on paragraph-delineated data at 1, 5, 10, 25, and 50 epochs.  An accurate annotation is one that is an exact string match.}
\end{figure}

\begin{figure}[h!]
\section{Baseline Non-fine-tuned performance With N-Shot Learning as N varies}
    \label{sec:appendix:baseline}
    \centering
    \includegraphics[scale=0.5]{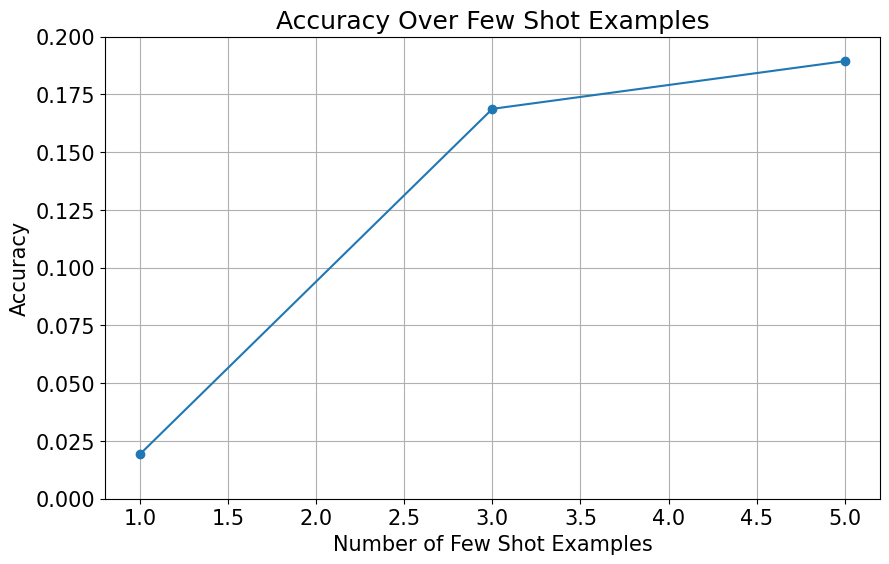}
    \caption{Test-set accuracy of non-fine-tuned \texttt{GPT 3.5TPE} with simple prompting and 1, 3 and 5 few shot examples. An accurate annotation is one that is an exact string match. Note the maximum accuracy of $<$20\%. }
    \label{fig:enter-label}
\end{figure}

\clearpage
\section{Expanded Model Names}
\label{sec:appendix:name_mapping}
\begin{table}[ht]
    \centering
    \begin{tabular}{|c|c|c|c|c|c|}
        \hline %RNN exact # of params: 66,710,688
        Model Name & Model Family & Model Size & BOS, EOS Tokens Added? & Training Objective & Epochs  \\ \hline
        RNN\_boseos\_10ep & N/A: Recurrent Neural Network & 67M & Yes & CLM & 10 \\
        llama2-7b\_boseos\_10ep & llama2 & 7B & No & CLM & 10  \\
        gpt3,5\_tpe\_10ep & GPT 3.5 Turbo & Propetary & Propetary & CLM & 10  \\
        gpt3,5\_tg\_10ep & GPT 3.5 Turbo & Propetary & Propetary & CLM & 10  \\
        gpt3,5\_t2s\_10ep & GPT 3.5 Turbo & Propetary & Propetary & CLM & 10  \\
        gpt2\_xl\_boseos\_10ep & GPT 2 & XL (1.5B) & Yes & CLM & 10  \\
        gpt2\_xl\_10ep & GPT 2 & XL (1.5B) & No & CLM & 10  \\
        gpt2\_boseos\_10ep & GPT 2 & Base (124M) & Yes & CLM & 10  \\
        gpt2\_10ep & GPT 2 & Base (124M) & No & CLM & 10  \\
        flan-t5\_seq2seq\_boseos\_1ep & Flan-T5 & Base (248M) & Yes & MLM & 1  \\
        flan-t5\_seq2seq\_1ep & Flan-T5 & Base (248M) & No & MLM & 1  \\
        flan-t5\_large\_seq2seq\_1ep & Flan-T5 & Large (783M) & No & MLM & 1  \\
        flan-t5\_large\_seq2seq\_boseos\_1ep & Flan-T5 & Large (783M) & Yes & MLM & 1  \\
        flan-t5\_large\_seq2seq\_1ep & Flan-T5 & Large (783M) & No & MLM & 1  \\ \hline
    \end{tabular}
    \caption{Model names as they correspond to their specific training/fine-tuning interventions.}
\end{table}

\begin{figure}[h!]
\section{Open-Source Model Performance as Epochs Vary}
\label{sec:appendix:epochs}
    \centering
    \includegraphics[scale=0.7]{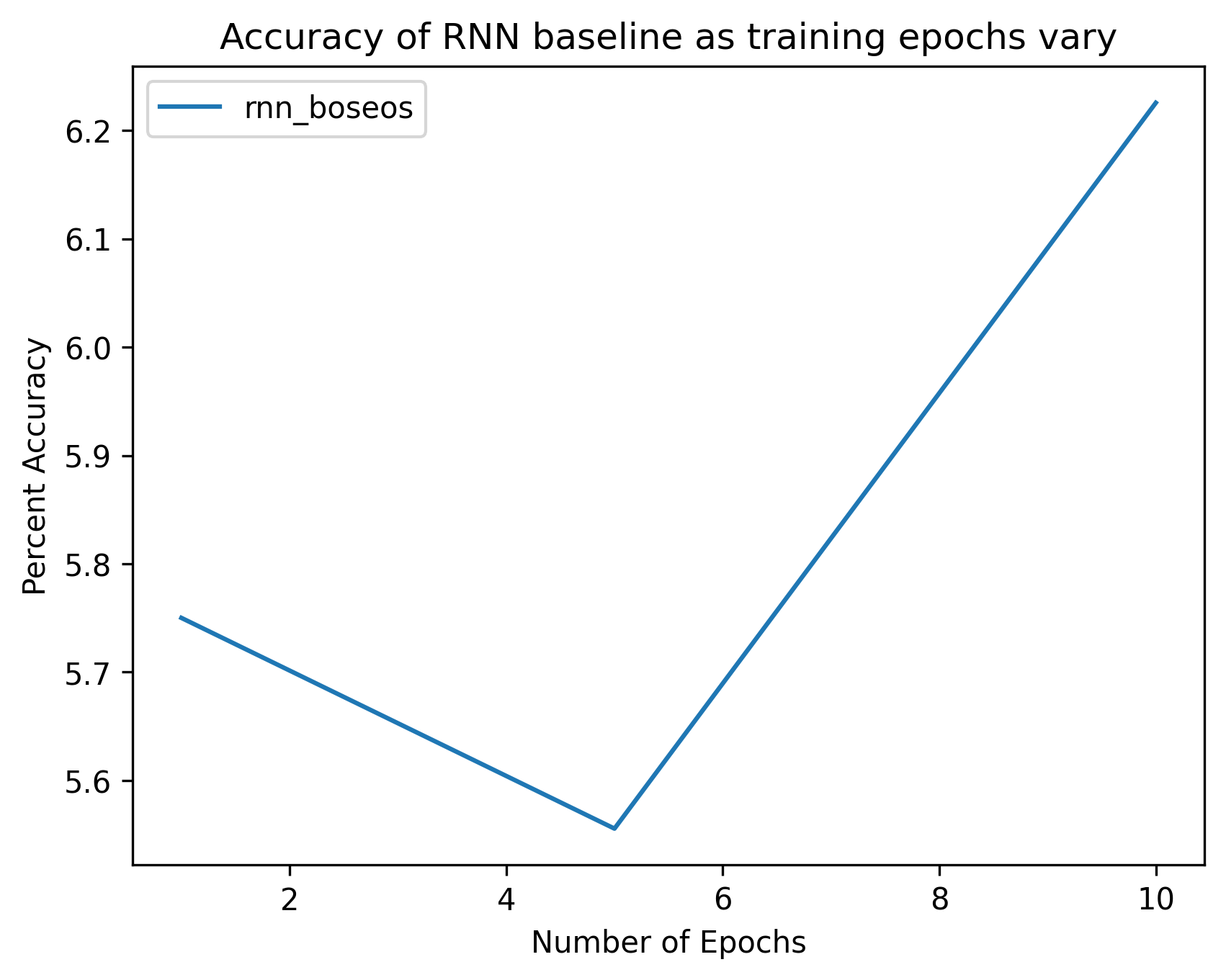}
    \caption{Test-set accuracy of our RNN at 1, 5, and 10 epochs. An accurate annotation is one that is an exact string match.}
\end{figure}

\begin{figure}[h!]
    \includegraphics[scale=0.7]{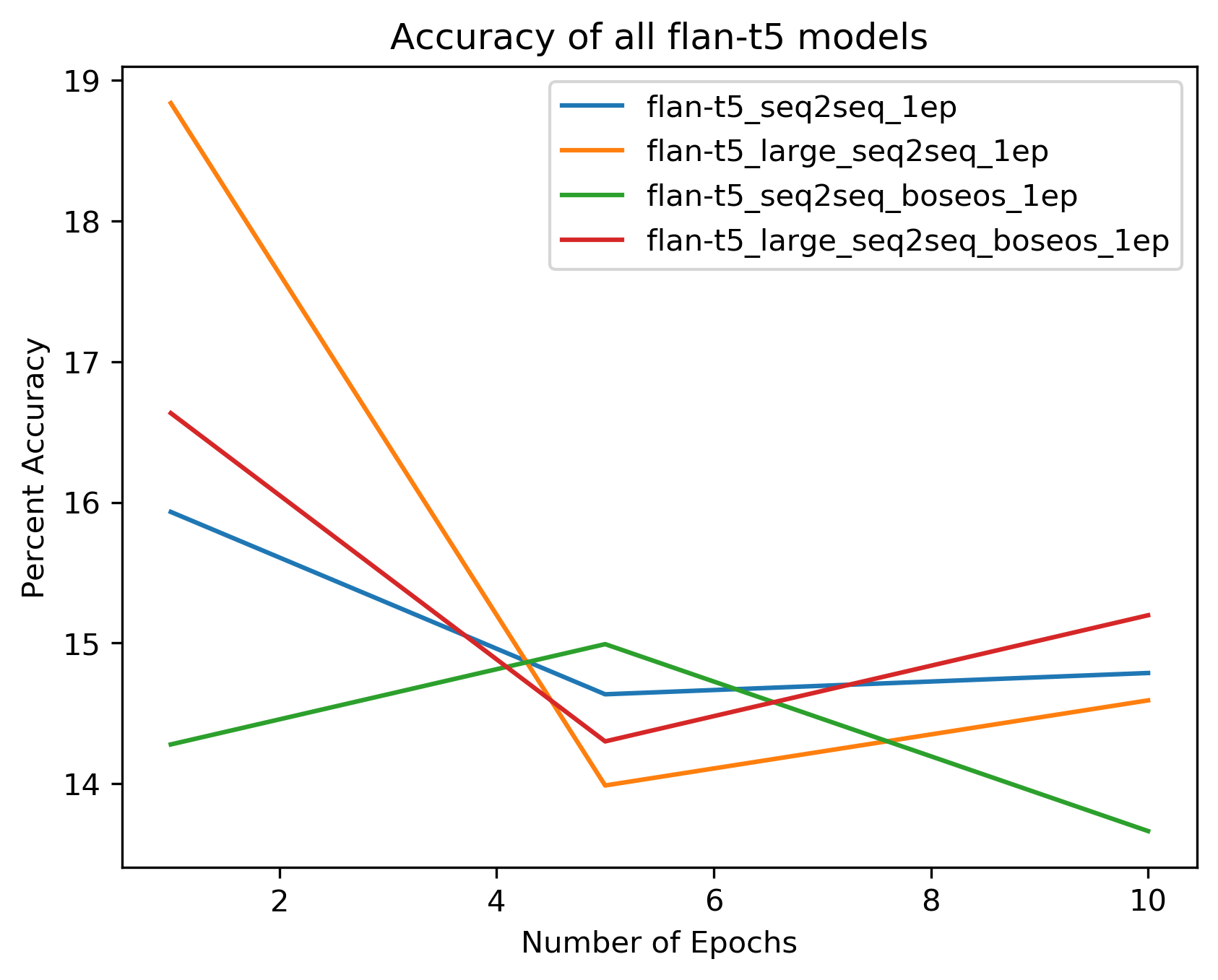}
    \caption{Test-set accuracy of Flan-T5 models at 1, 5, and 10 epochs. An accurate annotation is one that is an exact string match.}
\end{figure}

\begin{figure}[h!]
     \includegraphics[scale=0.7]{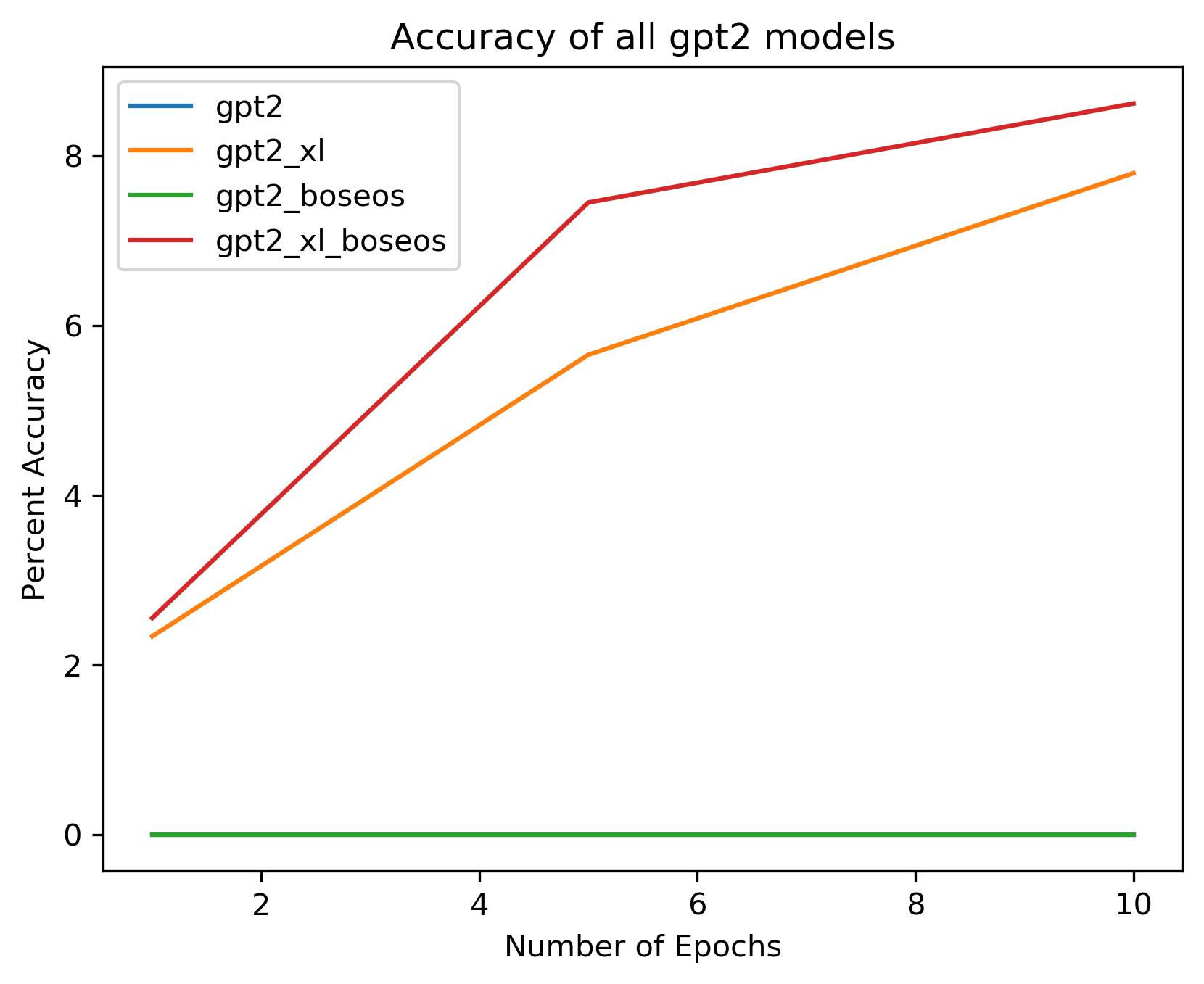}
    \caption{Test-set accuracy of GPT-2 models at 1, 5, and 10 epochs. An accurate annotation is one that is an exact string match.}
\end{figure}
   
\begin{figure}[h!]
    \includegraphics[scale=0.7]{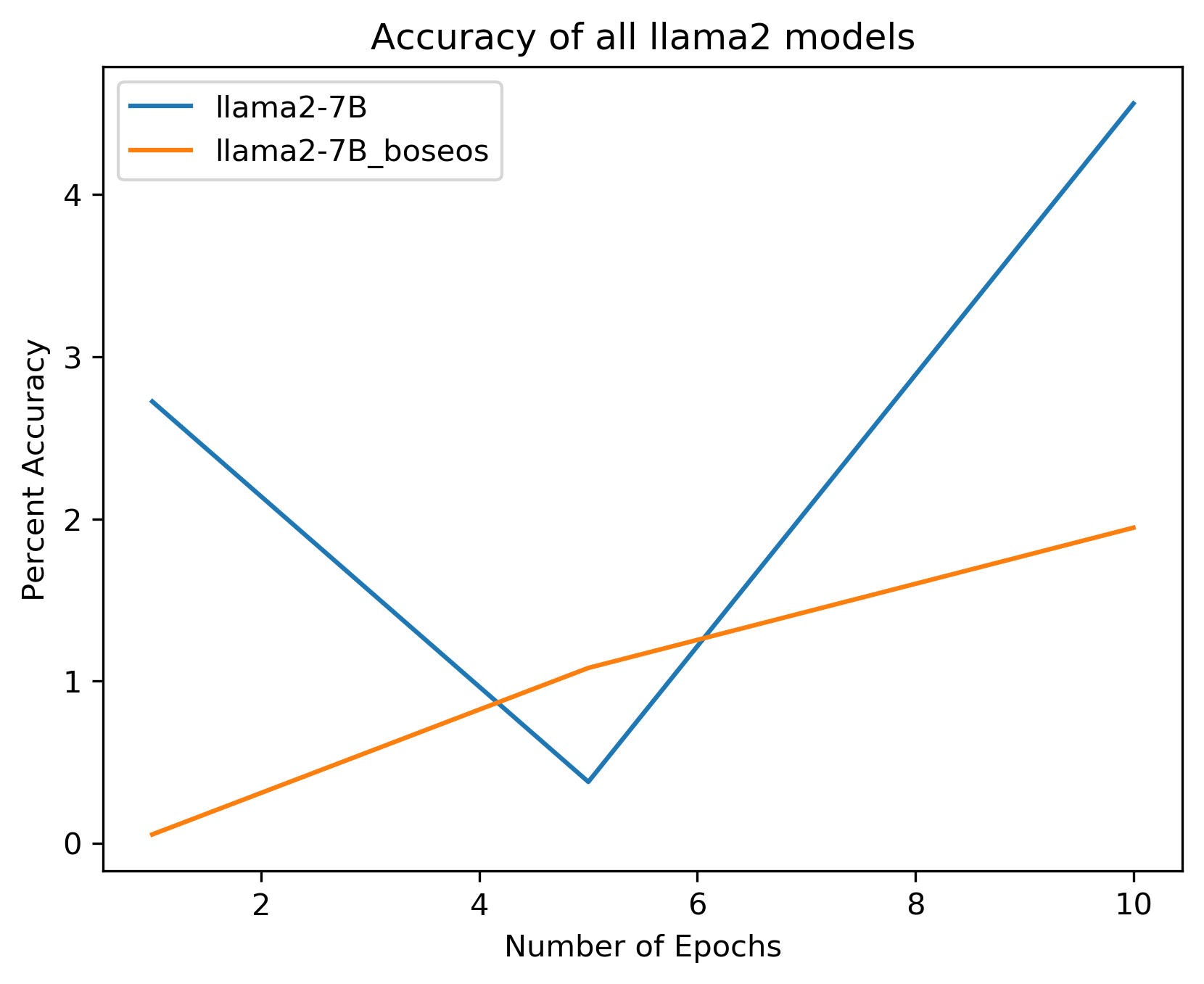}
    \caption{Test-set accuracy of Llama2 models at 1, 5, and 10 epochs. An accurate annotation is one that is an exact string match.}
\end{figure}

\clearpage

\begin{figure*}[h!]
\section{Performance Breakdown For The Three Top Performing Models}
\label{sec:appendix:top_mods_attrs}
    \centering
    \includegraphics[scale=0.65]{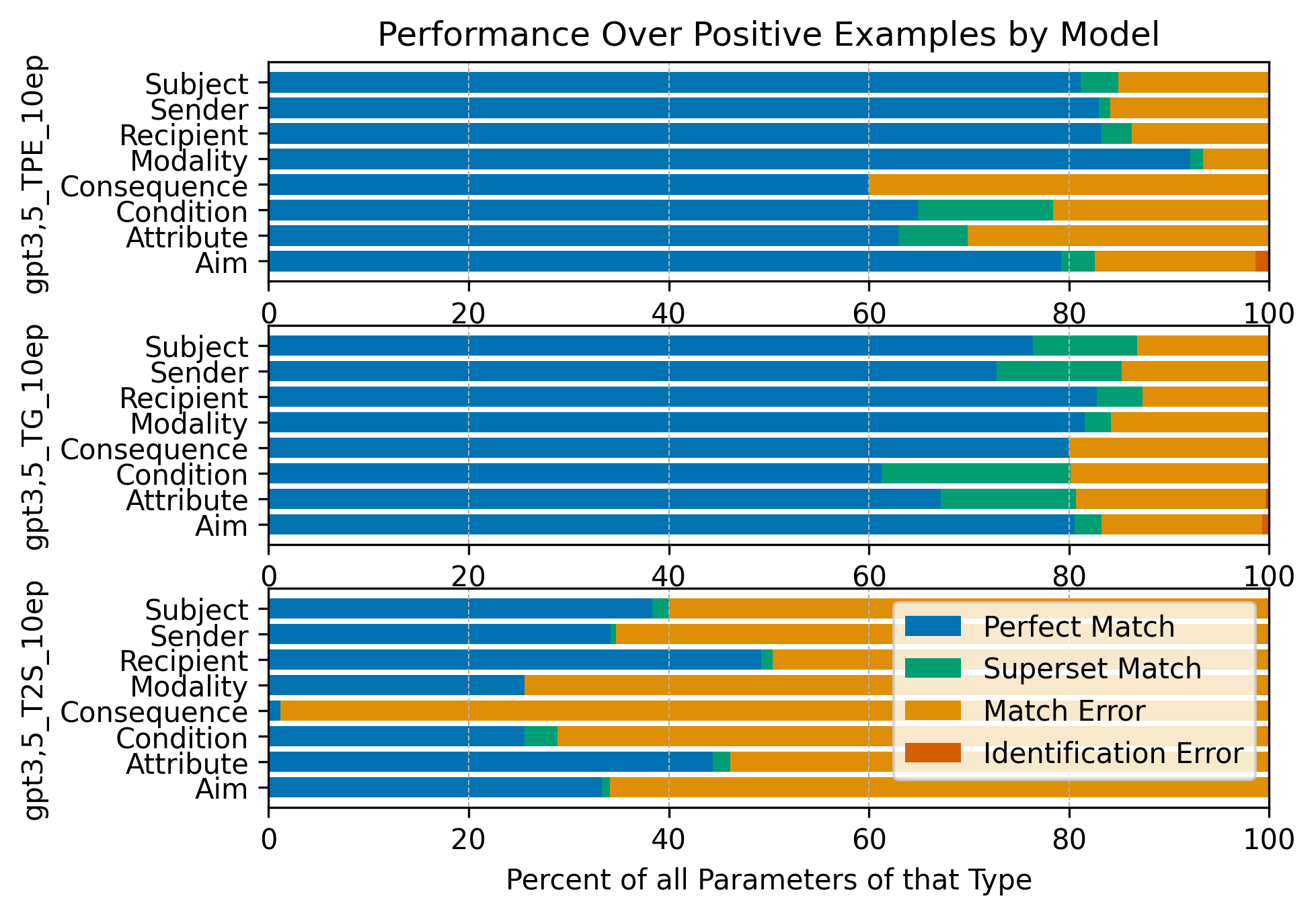}

    \caption{A comparison of our top-performing models on positive examples by GKC-CI parameter. \texttt{GPT3,5\_TG\_10ep} refers the generic GPT-3.5 Turbo model,  \texttt{GPT3,5\_TPE\_10ep} refers to the prompt-engineered version of GPT-3.5 Turbo, and \texttt{GPT3,5\_T2S\_10ep} refers to the two model GPT-3.5 Turbo system. All models were trained for 10 epochs. Note that there are only 5 "Consequence" parameters  leading to greater variance on the corresponding bars.} 
    \label{fig:top_mods_attrs}
\end{figure*}

\begin{table*}[h!]
\section{Codebook}
\label{sec:appendix:codebook}
\vspace{1em}
\footnotesize
\centering
\begin{tabular}{@{}ll@{}}
\toprule
Code                                          & Description                                                                                    \\ \midrule
\textbf{Completion Errors}           &                                                                                                \\
Completion Is Wrong                           & Completion is outright incorrect failing to provide the accurate answer.                       \\
Meaningful Subset                             & Completion partially captures the correct response but falls short of completeness.            \\
Completion Over-labeled & Completion includes correct answers but erroneously incorporates nearby words into the parameter tag. \\
\midrule
\textbf{Expert Labeling Errors} &                                                                                                \\  
Expert Labels Is Wrong               & Expert label itself is incorrect.                                                              \\
Expansive Ground Truth                        & Expert label is correct but overly broad and the completion offers a more precise response.    \\
Partial Ground Truth                          & Expert label misses a portion of the correct label, but the completion captures it accurately. \\  \midrule
\textbf{Semantic Equivalence}                 &                                                                                                \\
Semantic Equivalence    & Completion and the ground truth label differ in wording but convey equivalent semantic meanings.      \\ \bottomrule
\end{tabular}

\caption{Codebook for qualitative error analysis. Parent codes in bold. 
}
\label{tab:codebook}
\end{table*}

\clearpage

\begin{figure*}[h!]
\section{Longitudinal Data}
\label{sec:long_appendix}
    \centering
    \includegraphics[width=0.95\textwidth]{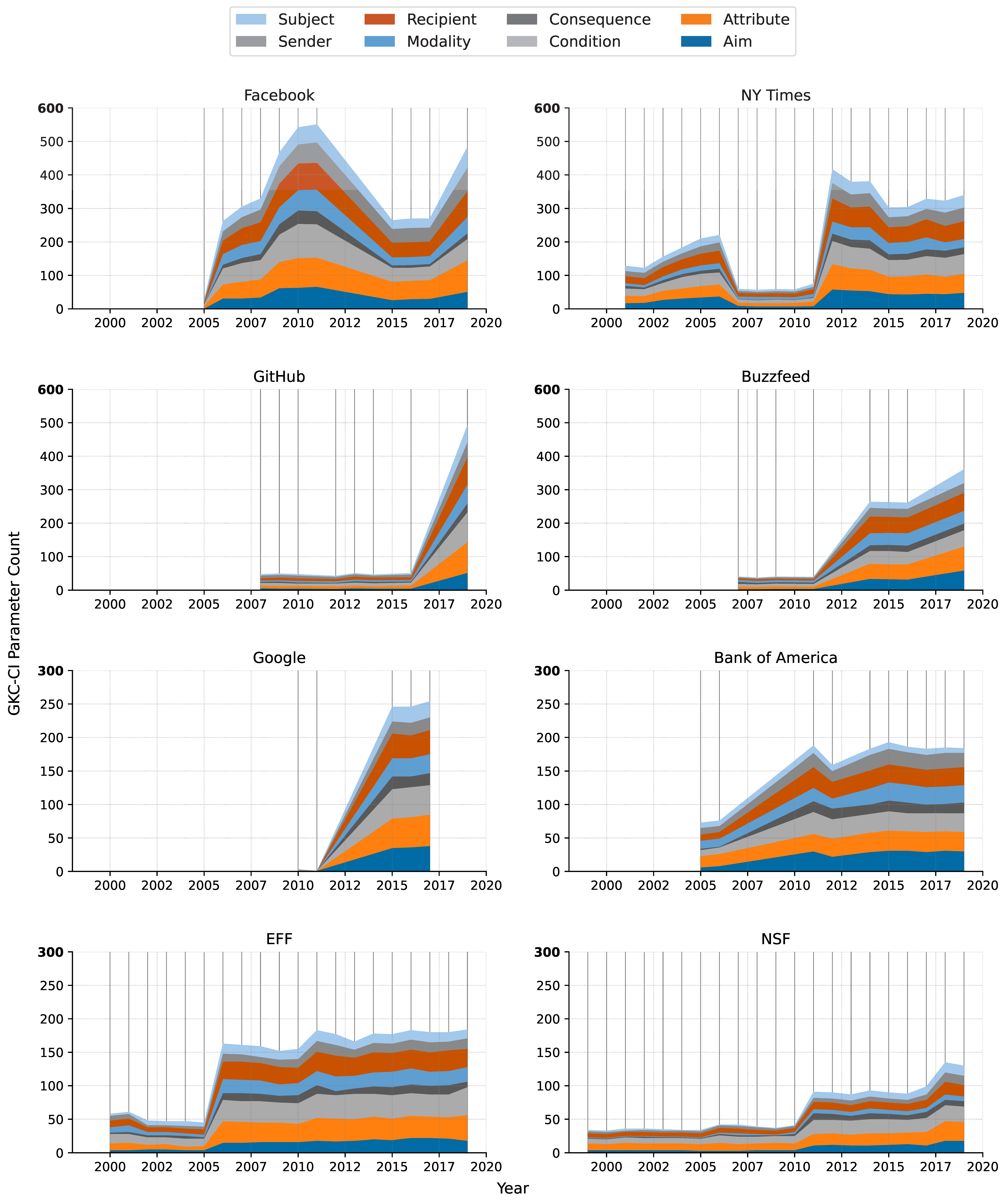}
    \caption{Number of annotated GKC-CI parameters in the privacy policies of 8 prominent companies over time. Policies from the Princeton-Leuven Longitudinal Corpus of Privacy Policies~\cite{amos2021privacy}. Bold grid lines indicate which years' policies were in the corpus and were annotated. The exact parameter counts displayed in this figure appear in Tables~\ref{tab:longitudinal-data-1}--\ref{tab:longitudinal-data-2} in Appendix~\ref{sec:long_appendix}.}
    \label{fig:longitudinal}
    %\label{fig:enter-label}
\end{figure*}

\clearpage

\begin{table}[h!]

    \centering
    \footnotesize

\begin{tabular}{lrrrrrrrrrr}
\toprule
 Company &  Year &  Aim &  Attribute &  Condition &  Consequence &  Modality &  Recipient &  Sender &  Subject &  Total \\
\midrule
Facebook &  2005 &    1 &          7 &          5 &            0 &         1 &          6 &       4 &        5 &     29 \\
Facebook &  2006 &   31 &         42 &         48 &           11 &        32 &         40 &      28 &       30 &    262 \\
Facebook &  2007 &   31 &         50 &         56 &           15 &        39 &         50 &      33 &       31 &    305 \\
Facebook &  2008 &   34 &         54 &         58 &           18 &        39 &         56 &      38 &       33 &    330 \\
Facebook &  2009 &   62 &         78 &         82 &           31 &        51 &         70 &      52 &       41 &    467 \\
Facebook &  2010 &   63 &         89 &        102 &           40 &        61 &         79 &      57 &       51 &    542 \\
Facebook &  2011 &   66 &         87 &        100 &           39 &        64 &         80 &      61 &       54 &    551 \\
\bottomrule
\end{tabular}

    \caption{Counts of annotated parameters in the privacy policies of 10 prominent companies and organizations over time \textit{(continued on next page)}.}
    \label{tab:longitudinal-data-1}
\end{table}

\begin{table}
    \centering
    \footnotesize

\begin{tabular}{lrrrrrrrrrr}
\toprule
        Company &  Year &  Aim &  Attribute &  Condition &  Consequence &  Modality &  Recipient &  Sender &  Subject &  Total \\
\midrule
Bank of America &  2005 &    6 &         17 &          9 &            2 &        12 &          9 &      10 &        8 &     73 \\
Bank of America &  2006 &    8 &         18 &         10 &            1 &        12 &         10 &       9 &        8 &     76 \\
Bank of America &  2011 &   30 &         26 &         33 &           16 &        20 &         31 &      21 &       11 &    188 \\
Bank of America &  2012 &   22 &         27 &         29 &           16 &        15 &         25 &      16 &        9 &    159 \\
Bank of America &  2014 &   29 &         29 &         28 &           14 &        24 &         27 &      23 &        9 &    183 \\
Bank of America &  2015 &   31 &         30 &         29 &           16 &        27 &         27 &      23 &       10 &    193 \\
Bank of America &  2016 &   31 &         29 &         27 &           16 &        27 &         26 &      22 &        8 &    186 \\
\bottomrule
\end{tabular}

\caption{Counts of annotated parameters in the privacy policies of 10 prominent companies and organizations over time \textit{(continued from previous page)}}
\label{tab:longitudinal-data-2}
\end{table}
\begin{table}
\section{Parameter Variance Data}
\label{sec:variance-appendix}
\centering
\tiny
\begin{tabular}{lrrrrrrrrr}
\toprule
              Website &  \% Aim &  \% Attribute &  \% Condition &  \% Consequence &  \% Modality &  \% Recipient &  \% Sender &  \% Subject &  Variance \\
\midrule
          mozilla.org &    10.0 &           4.0 &          38.0 &            10.0 &         10.0 &           6.0 &       16.0 &         6.0 &     119.7 \\
        atlassian.net &     4.2 &          12.5 &          25.0 &            33.3 &          8.3 &           8.3 &        4.2 &         4.2 &     119.0 \\
              doi.org &    23.8 &          23.8 &          14.3 &             4.8 &          9.5 &          23.8 &        0.0 &         0.0 &     109.7 \\
          mozilla.com &    10.4 &           6.2 &          35.4 &             8.3 &         10.4 &           6.2 &       16.7 &         6.2 &      98.0 \\
           apache.org &     5.9 &          23.5 &          29.4 &             5.9 &          5.9 &          11.8 &       11.8 &         5.9 &      83.4 \\
            rlcdn.com &    21.4 &          17.9 &          21.4 &             0.0 &          3.6 &          17.9 &       14.3 &         3.6 &      76.5 \\
     doubleverify.com &    25.2 &          23.6 &          10.2 &             1.6 &          7.9 &          15.0 &       11.8 &         4.7 &      71.1 \\
\bottomrule
\end{tabular}
\caption{Website privacy policies ranked by the variance of the percentages of individual parameter types out of all annotations \textit{(continued on next page)}. }
\label{tab:variance-data-1}
\end{table}

\begin{table}
    \centering
    \tiny
\begin{tabular}{lrrrrrrrrr}
\toprule
             Website &  \% Aim &  \% Attribute &  \% Condition &  \% Consequence &  \% Modality &  \% Recipient &  \% Sender &  \% Subject &  Variance \\
\midrule
        pubmatic.com &   16.08 &         15.50 &         14.04 &            3.51 &        16.67 &         15.79 &       9.65 &        8.77 &     22.17 \\
      ampproject.org &   18.23 &         18.42 &         14.36 &            5.34 &         8.84 &         14.92 &       9.58 &       10.31 &     22.17 \\
      googleapis.com &   14.34 &         18.85 &         18.44 &            6.97 &        10.66 &         14.34 &       7.79 &        8.61 &     21.98 \\
            bing.com &   16.67 &         17.65 &         15.20 &            4.90 &         9.31 &         16.67 &      10.78 &        8.82 &     21.83 \\
           3lift.com &   10.64 &         16.60 &         18.30 &            6.38 &        14.47 &         16.17 &      11.49 &        5.96 &     21.78 \\
           b-cdn.net &   17.81 &         16.44 &         15.07 &            2.74 &        12.33 &         13.70 &      10.96 &       10.96 &     21.68 \\
        applovin.com &   12.57 &         17.28 &         16.23 &            4.71 &        16.23 &         15.71 &       8.90 &        8.38 &     21.65 \\
\bottomrule
\end{tabular}
\caption{Website privacy policies ranked by the variance of the percentages of individual parameter types out of all annotations \textit{(continued from previous page)}.}
\label{tab:variance-data-2}
\end{table}
\begin{table}
\section{Parameter Density Data}
\label{sec:density-appendix}
\centering
\tiny
\begin{tabular}{lrrrrrrrrr}
\toprule
                     Website &   $\frac{\#\,\,Aims}{\#\,\, Sentences}$ &  $\frac{\#\,\,Attributes}{\#\,\, Sentences}$ &  $\frac{\#\,\,Conditions}{\#\,\, Sentences}$ &  $\frac{\#\,\,Consequences}{\#\,\, Sentences}$ &  $\frac{\#\,\,Modalities}{\#\,\, Sentences}$ &  $\frac{\#\,\,Recipients}{\#\,\, Sentences}$ &  $\frac{\#\,\,Senders}{\#\,\, Sentences}$ &  $\frac{\#\,\,Subjects}{\#\,\, Sentences}$ &   \textbf{Total \# parameters / \# Sentences} \\
\midrule
           b-cdn.net & 0.0774 &     0.0714 &     0.0655 &       0.0119 &    0.0536 &     0.0595 &  0.0476 &   0.0476 & 0.4345 \\
        facebook.com & 0.0389 &     0.0716 &     0.0473 &       0.0130 &    0.0358 &     0.0595 &  0.0534 &   0.0473 & 0.3666 \\
            bing.com & 0.0599 &     0.0634 &     0.0546 &       0.0176 &    0.0335 &     0.0599 &  0.0387 &   0.0317 & 0.3592 \\
      slideshare.net & 0.0459 &     0.0576 &     0.0596 &       0.0195 &    0.0371 &     0.0557 &  0.0410 &   0.0391 & 0.3555 \\
        linkedin.com & 0.0459 &     0.0557 &     0.0596 &       0.0205 &    0.0361 &     0.0566 &  0.0381 &   0.0381 & 0.3506 \\
             zoom.us & 0.0641 &     0.0654 &     0.0500 &       0.0179 &    0.0321 &     0.0526 &  0.0327 &   0.0353 & 0.3500 \\
      ampproject.org & 0.0635 &     0.0641 &     0.0500 &       0.0186 &    0.0308 &     0.0519 &  0.0333 &   0.0359 & 0.3481 \\
\bottomrule
\end{tabular}

\caption{Website privacy policies ranked by the total ratio of the number of annotated parameters to the number of sentences in the policy \textit{(continued on next page)}.}
\label{tab:density-data-1}
\end{table}

\begin{table}
    \centering
    \tiny
    \begin{tabular}{lrrrrrrrrr}
\toprule
                   Website &   $\frac{\#\,\,Aims}{\#\,\, Sentences}$ &  $\frac{\#\,\,Attributes}{\#\,\, Sentences}$ &  $\frac{\#\,\,Conditions}{\#\,\, Sentences}$ &  $\frac{\#\,\,Consequences}{\#\,\, Sentences}$ &  $\frac{\#\,\,Modalities}{\#\,\, Sentences}$ &  $\frac{\#\,\,Recipients}{\#\,\, Sentences}$ &  $\frac{\#\,\,Senders}{\#\,\, Sentences}$ &  $\frac{\#\,\,Subjects}{\#\,\, Sentences}$ &   \textbf{Total \# parameters / \# Sentences} \\
\midrule
           paypal.com & 0.0291 &     0.0369 &     0.0366 &       0.0100 &    0.0299 &     0.0337 &  0.0269 &   0.0251 & 0.2282 \\
                 name-services.com & 0.0391 &     0.0312 &     0.0156 &       0.0156 &    0.0312 &     0.0469 &  0.0312 &   0.0156 & 0.2266 \\
      dnsmadeeasy.com & 0.0250 &     0.0406 &     0.0344 &       0.0063 &    0.0250 &     0.0406 &  0.0219 &   0.0312 & 0.2250 \\
              tds.net & 0.0208 &     0.0417 &     0.0500 &       0.0042 &    0.0167 &     0.0292 &  0.0250 &   0.0375 & 0.2250 \\
           www.gov.uk & 0.0291 &     0.0349 &     0.0465 &       0.0174 &    0.0116 &     0.0407 &  0.0203 &   0.0233 & 0.2238 \\
              ibm.com & 0.0317 &     0.0383 &     0.0500 &       0.0083 &    0.0217 &     0.0367 &  0.0183 &   0.0183 & 0.2233 \\
         digicert.com & 0.0344 &     0.0365 &     0.0354 &       0.0083 &    0.0156 &     0.0344 &  0.0302 &   0.0229 & 0.2177 \\
\bottomrule
\end{tabular}
\caption{Website privacy policies ranked by the total ratio of the number of annotated parameters to the number of sentences in the policy \textit{(continued from previous page)}.}
\label{tab:density-data-2}
\end{table}

\clearpage

\begin{figure}[h!]
\section{Visualizer Demonstration}
\label{sec:appendix:viz}
\vspace{1em}
    \centering
    \includegraphics[width=\linewidth]{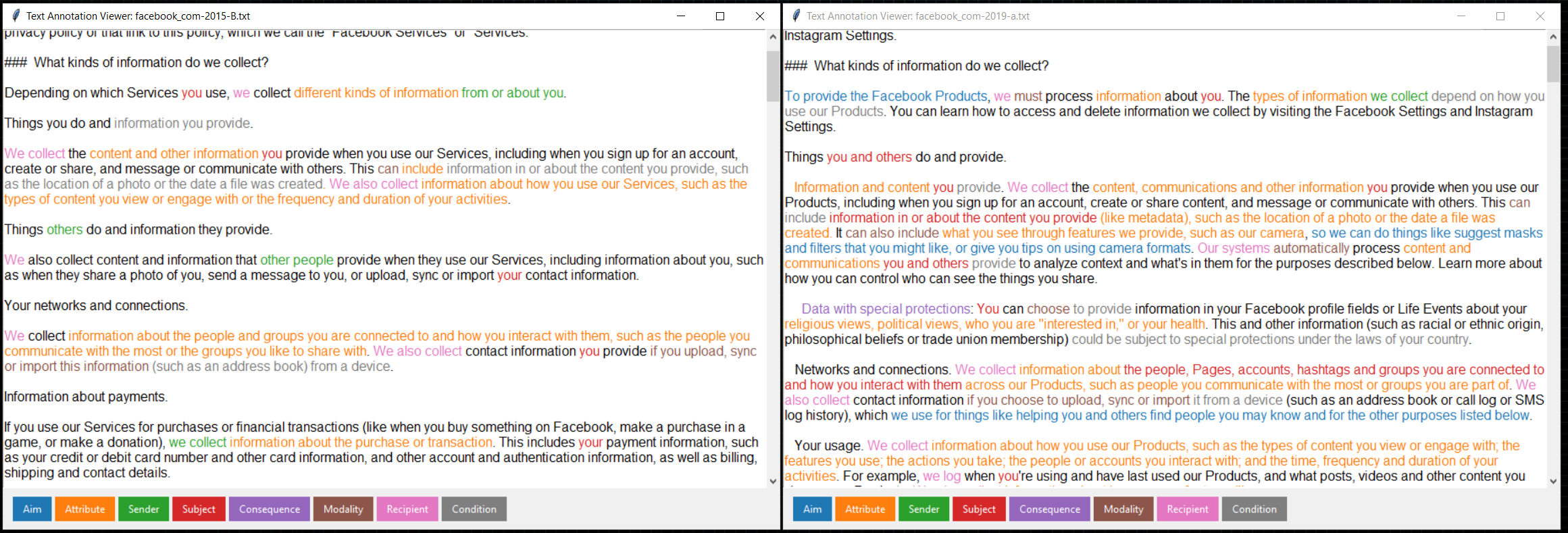}
    \caption{Side-by-side comparison of Facebook's 2015 and 2019 policies displayed using the visualizer, showing the GKC-CI annotations in the text. The legend for the color meanings is at the bottom of each GUI window. The comparison highlights information flow changes between the policies, with parameter annotations---indicated by text color---emphasizing an increase in the specificity of information flows in the 2019 policy.}
    \label{fig:enter-label}
\end{figure}

\vspace{2em}

\section{Creation of Paragraph-Level Annotations}
\label{sec:app_paragraph_lvl_annos}
To create the paragraph-level annotations, we utilize a segmentation approach tailored to capture the structural and contextual integrity of each policy text. Instead of relying on sentence-ending punctuation to define annotation boundaries, we segmented each policy into paragraphs by identifying double newline characters. This method was chosen to reflect natural paragraph divisions in the original documents, preserving the thematic and logical flow intended by the policy authors. 
Following this segmentation, we proceeded with the annotation creation and benchmarking process using the same methodology used for sentence-level annotations. 

\end{document}